\documentclass[aps,prc,twocolumn,groupedaddress,floats,amsmath]{revtex4}
\usepackage{graphicx}
\newcommand{\ktopi}{\langle K^+ \rangle / \langle \pi^+ \rangle}
\newcommand{\kmpi}{\langle K^- \rangle / \langle \pi^- \rangle}
\newcommand{\ltopi}{\langle \Lambda \rangle / \langle \pi \rangle}
\newcommand{\eden}{\varepsilon}
\newcommand{\rmax}{{\cal R}_{\rm max}}

\newcommand{\si}{\sigma}
\newcommand{\pip}{\pi^+}
\newcommand{\pim}{\pi^-}
\newcommand{\pin}{\pi^0}

\newcommand{\Kp}{K^+}
\newcommand{\Kb}{\bar K^0}
\newcommand{\Kn}{K^0}
\newcommand{\Km}{K^-}
\newcommand{\La}{\Lambda}

\newcommand{\Sp}{\Sigma^+}
\newcommand{\Sn}{\Sigma^0}
\newcommand{\Sm}{\Sigma^-}

\newcommand{\Dpp}{\Delta^{++}}
\newcommand{\Dp}{\Delta^+}
\newcommand{\Dn}{\Delta^0}
\newcommand{\Dm}{\Delta^-}

\newcommand{\rop}{\rho^+}
\newcommand{\ron}{\rho^0}
\newcommand{\rom}{\rho^-}
\newcommand{\Xm}{\Xi^-}
\newcommand{\Xn}{\Xi^0}

\newcommand{\sqs}{\sqrt{s}}

\def\be{\begin{eqnarray}}
\def\ee{\end{eqnarray}}

\begin{document}

%
%
\title{Strangeness dynamics in  heavy-ion collisions:\\
the $K/\pi$ ratios and the lifetime of a fireball}
\author{Boris Tom\'a\v sik}
\affiliation{The Niels Bohr Institute, Blegdamsvej 17,
             2100 Copenhagen \O, Denmark}
\affiliation{\'Ustav jadern\'e fyziky AV\v CR,
25068 \v Re\v z, Czech Republic}
\author{Evgeni E.\ Kolomeitsev}
\affiliation{University of Minnesota, School of Physics and Astronomy,
116 Church Street SE, Minneapolis, 55455 Minnesota, USA}
\date{December 22, 2005}
%
%
\begin{abstract}

We propose a non-equilibrium, hadronic kinetic model for
describing the relative abundance of strange particles in ultra-relativistic
heavy-ion collisions. The energy
dependence of the multiplicity ratios of charged kaons and lambdas
to pions are studied in detail. The pronounced peak in $\ktopi$ is
conjectured to be due to the decreasing lifetime of a fireball with an
increase in beam energy above 30~$A$GeV. Such a behaviour may
be the consequence of a stronger stopping at lower beam energies.
We study the dependence of the multiplicity ratios on energy and
baryon density of the fireball and its total lifetime.
\end{abstract}

\maketitle


\section{Strangeness production in nuclear collisions}

The excitation functions of multiplicity ratios of kaons to pions
are among the
most intriguing results of the energy scan at CERN's
Super-Proton-Synchrotron \cite{alt20,ktpdata}. The sharp peak of the ratio
$\ktopi$ for collisions at beam energy of 30 $A$GeV---dubbed
``the horn''---is barely reproduced by any theoretical calculation.
The statistical model expects a maximum in this region, which is due to the transition
from baryon-dominated to meson-dominated matter \cite{Cleymans:2004hj}.
That maximum is, however, much broader than the observed sharp peak.
Transport models UrQMD and HSD \cite{UrQMDHSD} overpredict $\pi^+$ production
and predict a very modest bump which is lower than the measured horn and lies
at lower beam energy. The BUU model of \cite{moselBUU} shows no peaky
structure at all. Three-fluid hydrodynamic model~\cite{toneev} reproduces
the $K^+$ and pion yields but overpredicts a multiplicity of
$K^-$. Recently, a kinetic model which assumes a phase-transition to
the deconfined phase has been proposed and was successfull in
reproducing the data~\cite{jane}.

The data can be interpreted in the framework of the Statistical Model
of the Early Stage (SMES) \cite{smes}. In this model it is assumed that
the primordial particle production follows the prescription of statistical
equilibrium. The increase of the $\ktopi$
ratio at low collision energies is a consequence of an enhanced strangeness
content due to a larger energy deposit. The sharp decrease of the ratio above
the beam energy of 30 $A$GeV
is connected with a first order phase transition between
hadronic and deconfined phases. The decreasing part of the excitation
function corresponds to the mixture of the two phases, while the flat
dependence at high energies is realized in the deconfined phase. In framework
of SMES, this observation is taken as partial evidence for deconfinement
in the nuclear collisions at threshold energy
$\sqrt{s_{NN}} \approx 7.6\, \mbox{GeV}$
($E_{\rm beam} = 30\, A\mbox{GeV}$).

At this point we would like to pose two questions which provide
the motivation for our study. Firstly, is the assumption of
chemically equilibrated primordial production justified? This is
not obvious at all. Analysis within a statistical hadronization
model shows that the limited volume in proton-proton collisions
causes suppression of strangeness once canonical statistics is
used. On the other hand, a large volume allows for the use of
grand-canonical statistics and leads to enhanced strangeness
content \cite{redlich}. But how could the incident nucleons be
influenced by volume effects in the very moment of their first
interactions? Secondly, one cannot claim evidence of deconfinement
unless \textit{all} hadronic scenarios have been safely ruled out.
We shall not touch here the question of early thermalization at
the SPS. In this paper we investigate if all hadronic
(non-equilibrium) models are safely excluded after comparison with
the data.

We propose a model in which strangeness is produced dynamically in
reactions involving the constituent hadrons of the fireball. Two
aspects are crucial for the final total amount of produced
strangeness: the energy density and the time. The energy density
is reflected in the temperature and an increase of temperature
enhances the strangeness production rate. On the other hand, at
least for a non-expanding system, if the initial amount of strangeness is
below the equilibrium value, it will certainly increase with time
until it saturates. Thus a longer lifetime also means higher
strangeness content. Based on these considerations we speculate
that the initial increase of the $\ktopi$ excitation function is
due to increasing energy density. Can the following sharp decrease
of the excitation function be due to {\em decreasing total
lifetime} of the fireball with increasing collision energy?

The assumption of decreasing total lifetime is crucial.
Sometimes, just the opposite is
expected. Due to a higher initial energy density at higher beam
energies it would take a
longer time for the fireball to cool down to the critical density when it breaks up.
Also, a larger multiplicity at higher energies assumes a larger volume at
freeze-out and thus requires longer lifetime.
These arguments, however, neglect nuclear stopping and the need to
build up longitudinal expansion. For example, the proper time
between impact and the freeze-out comes out more or less independent of
the beam energy in the transport code HSD \cite{casspriv}.
Indeed, it seems reasonable to expect that
stopping power or an ability for slowing down incident baryons is stronger
at lower energies
than at higher energies. Then, a large part of the longitudinal expansion
flow must be built up by the pressure at lower collision energies, and this takes time.
At higher collision energies, fast expansion  may be present from the very
beginning and the fireball would reach a breakup stage  earlier. We should stress
that there is no direct measurement of stopping power as it results in rapidity
distributions
{\em just after the primordial nucleon-nucleon collisions}.
Measured rapidity spectra reflect distributions at
freeze-out, i.e., after the possibly accelerated longitudinal expansion.
We know that there is some stopping
at RHIC since the net baryon spectra are peaked at a rapidity lower than
$y_{\rm beam}$ \cite{brahms-stopping}. We also know that stopping
is incomplete even down at SIS energies (few hundreds $A$MeV) as it was
demonstrated by isospin tracing in collisions
of ${}^{96}_{44}\mbox{Ru}$ and ${}^{96}_{40}\mbox{Zr}$ \cite{fopi}.
It is natural to expect that the relevant evolution scenario at AGS and SPS
energies will be somewhere between  full stopping and re-expansion
(Landau scenario) and boost-invariant non-accelerated longitudinal
expansion (Bjorken expansion).

In our model we shall adopt an ansatz for the time dependence of
energy density and baryon density of the fireball. Our calculation
is not based on any transport/cascade models or hydrodynamic
simulation. Therefore we do not directly connect to  details of
microscopic structure of the matter or the equation of state. We
gain, however, the freedom to construct as many different
evolution scenarios as we wish, and we shall use this freedom to
explore various scenarios, their impact on data, and identify
those which are allowed by data. In particular, we shall focus on
the dependence of total production of strange particles on the
lifetime of the system. We present results for multiplicity ratios
$\ktopi$, $\kmpi$, and $\ltopi$.  It will turn out that a
non-equilibrium hadronic scenario cannot be excluded by this set
of data.


\section{The model}
\label{model}

We shall only calculate {\em ratios} of yields and not the yields
themselves. Therefore, it will be sufficient to know average
densities of individual species if we assume that they come from
the same volume. (Feed-down from resonance decays is also
necessary, but we defer its discussion for later.)
Applying later this approach to kaons ($K^+$ and $K^0$),
we implicitly assume that
kaons do not decouple from the system until the overall
freeze-out. This is in line with a class of hydro-chemical models
\cite{knoll1,knoll2,broko}. This assumption will be supported by
the analysis of the kaon scattering rate and the mean free path in
Section~\ref{s:whyk}.

The evolution of kaon density in proper time is
described by the master equation which can be derived from
\begin{equation}
\label{pme}
\frac{d\rho_K}{d\tau} = \frac{d}{d\tau}\frac{N_K}{V} =
\rho_K\, \left (- \frac{1}{V}\, \frac{dV}{d\tau} \right )+ \frac{1}{V} \,
\frac{dN_K}{d\tau} \, .
\end{equation}
The first term on the right-hand-side is due to the change of a fireball volume
and includes the expansion rate. The expansion rate is the change of the density
of any conserved charge due to expansion. For example the relation
\begin{equation}
-\frac{1}{V} \frac{dV}{d\tau} = \frac{1}{\rho_B}\, \frac{d\rho_B}{d\tau}
\label{erate}
\end{equation}
holds for the baryon number. The second term on the right-hand-side of eq.~\eqref{pme}
stands for the change of a kaon number due to production or
annihilation reactions. It can be divided into gain term and loss term
\be
\frac{dN_K}{d\tau} &=& V \big(\mathcal{R}_{\rm gain}-\mathcal{R}_{\rm loss}\big)
\nonumber\\
\mathcal{R}_{\rm gain} &=& \sum_{i\, j\, X} \left \langle v_{ij} \sigma_{ij}^{KX} \right \rangle\,
\frac{\rho_i \, \rho_j}{1+\delta_{ij}}
+\rho_{K^*}\,\Gamma_{K*}
\nonumber\\
\mathcal{R}_{\rm loss} &=&
\sum_{i\, X} \left \langle v_{Ki} \sigma_{Ki}^{X} \right \rangle\,
\frac{\rho_K \, \rho_i}{1+\delta_{Ki}}\, .
\label{chem}\ee
Terms in angular brackets are the cross-sections for reactions
$ij \to KX$ or $Ki \to X$ which are multiplied by relative velocities
of incoming particles and averaged over distribution of relative
velocities. Here, $i$ and $j$ denote any single species with densities
$\rho_i$ and $\rho_j$, and $X$
stands for any number of any species. The sums run, in principle,
over all possible reaction channels. In practice they are truncated after the
most significant contributions have been included.

Using eq.~\eqref{chem}, the master equation for kaon density evolution
can be written
\be
\label{me}
\frac{d\rho_K}{d\tau} =
\rho_K\, \left (- \frac{1}{V}\, \frac{dV}{d\tau} \right )
+\mathcal{R}_{\rm gain}-\mathcal{R}_{\rm loss}
\ee
In our study we shall evolve densities of $K^+$, $K^0$, $K^{*+}$,
and $K^{*0}$ according
to this equation. We shall explain how densities for other strange
species are obtained in Section~\ref{checo}.


\subsection{Expansion dynamics}
\label{expdyn}

In order to calculate the production and annihilation rates
we need the densities of interacting species and distribution of
relative velocities. These follow from the energy density $\eden$
and the densities of conserved charges, baryon number $\rho_B$ and the
third component of isospin $\rho_3$, which is also
related to the electric charge.  We shall assume
{\em thermal} equilibrium and use $\eden,\, \rho_B$ and $\rho_3$ to
calculate temperature. The temperature determines the distribution
of relative velocities. From the time-dependence of $\rho_B$ we shall obtain
the expansion rate.

We want to be able to explore a variety of possible expansion scenarios.
We shall tune the amount of stopping which is directly connected with the
time-derivative of the density at the beginning. Less stopping leads to
larger negative time-derivative of $\rho_B$
because of the initial stage expansion.
Explored scenarios can also differ in the initial acceleration due to inner
pressure. To include these two effects in a simplest possible way, we use
quadratic dependence as an Ansatz for the initial time-dependence of
the energy density
\begin{subequations}
\label{dini}
\begin{equation}
\label{edini}
\eden(\tau) = \eden_0 ( 1 - a\tau - b\tau^2)
\end{equation}
with $\eden_0$, $a$, and $b$ being tunable parameters.
In correspondence to this we use for the initial time dependence
of conserved charge densities
\begin{equation}
\label{bdini}
\rho_i(\tau) = \rho_{i0} (1 - a \tau - b \tau^2 )^\delta
\end{equation}
\end{subequations}
where the index $i$ stands for any of the two conserved charges. The
power $\delta$ is used here as the simplest way of representing the
equation of state. For example, in case of non-interacting
{\em freely streaming} gas no work is performed by pressure and
the energy density follows the same time-dependence as baryon
density, $\delta = 1$. On the other hand, maximum pressure is
present in a {\em massless} gas, in which $\rho_B \propto \eden^{3/4}$,
$\delta = 3/4$. We shall write
\begin{equation}
\label{delta}
\delta = \frac{1}{1+\kappa}\, , \qquad 0 \le \kappa \le\frac{1}{3}\, .
\end{equation}

The dependence of Hanbury-Brown/Twiss correlation radii on the
transverse momentum of a pair  indicates that at late times
the fireball expansion follows a power-law in time
\begin{subequations}
\label{lat}
\begin{eqnarray}
\label{edlat}
\eden (\tau)&  = & \frac{\eden_0^\prime}{(\tau - \tau_0)^{\alpha/\delta}}, \\
\label{bdlat}
\rho_i(\tau)&  = & \frac{\rho_{i0}^\prime}{(\tau -
\tau_0)^{\alpha}}.
\end{eqnarray}
\end{subequations}
The power $\alpha$ is a model parameter. The value of $\alpha = 1$
corresponds to a one-dimensional boost-invariant expansion (Bjorken scenario).
Due to transverse flow, $\alpha$ acquires values larger than 1. The
parameter $\tau_0$ allows for a shift of the power-law prescription
\eqref{lat} to later times and leaves thus time for acceleration.

Ansatz for initial times \eqref{dini} is matched with the prescription
for late times \eqref{lat} at some moment $\tau_s$. We require continuous
time dependence together with its first time derivative.
In summary, the time dependence is given by
\begin{subequations}
\label{td}
\begin{eqnarray}
\label{ed}
\eden(\tau) & = &
\begin{cases}
\eden_0 ( 1 - a \tau - b \tau^2) & : \quad \tau < \tau_s \\
\frac{\eden_0^\prime}{(\tau - \tau_0)^{\alpha/\delta}} &
: \quad \tau \ge \tau_s
\end{cases},
\\
\label{bd}
\rho_i (\tau) & = &
\begin{cases}
\rho_{i0} ( 1 - a \tau - b \tau^2)^\delta & : \quad \tau < \tau_s \\
\frac{\rho_{i0}^\prime}{(\tau - \tau_0)^{\alpha}} & : \quad \tau \ge \tau_s
\end{cases}.
\end{eqnarray}
\end{subequations}
For the energy density we have parameters $\eden_0$, $a$, $b$,
$\eden_0^\prime$, $\tau_0$, $\tau_s$, $\alpha$, and $\delta$,
out of which two are constrained by matching conditions at $\tau_s$.
In addition to this, for both baryon and $I_3$ density there are
$\rho_{i0}$ and $\rho_{i0}^\prime$ which are connected by the requirement
of continuity as well.

The parameters which appear in eqs.~\eqref{td} are not directly related
to measurements and/or one may not
have good feeling for what their reasonable values should be. Therefore,
for practical application, we shall specify the time evolution in terms
of the following quantities:\\[0.5ex]
\begin{tabular}{p{.1\linewidth}p{0.87\linewidth}}
$\eden_0$ &
initial energy density,\\[0.5ex]
$\tau_T$ &
total lifetime of the fireball until chemical freeze-out after which
no change of number of strange hadrons occurs; this time must be bigger than
$\tau_s$,\\[0.5ex]
$\eden_f$ &
final energy density, \\[0.5ex]
$\rho_{Bf}$ &
final baryon density, \\[0.5ex]
$\rho_{3f}$ &
final density of the third component of isospin $I_3$, \\[0.5ex]
$\tau_0$ &
reference time for the power-law dependence in the late stage of evolution;
in principle, it is the time $(\tau_T - \tau_0)$ which would appear in
expansion velocity gradients at freeze-out; $(\tau - \tau_0)$ is also
the time measured by longitudinal HBT radius, \\[0.5ex]
$\alpha$ &
exponent of the late-stage power-law expansion; $\alpha=1$ corresponds
to one-dimensional scaling expansion, $\alpha=3$ means three-dimensional
Hubble flow \\[0.5ex]
$\delta$ &
exponent characterizing the underlying equation of state, \\[0.5ex]
$\rmax$ &
the maximum expansion rate, which is equal to the negative of maximum
decrease rate of baryon density $(-1/\rho_B)\, (d\rho_B/d\tau)$ ($\rmax$
is always positive number); the maximum rate is achieved at
$\tau_s$.
\end{tabular}\\[0.5ex]
We list in Appendix~\ref{param} relations between these physical
parameters and those of parameterizations \eqref{td}.

The main parameters, which influence on the final strangeness
yields we shall study below, are the initial energy density
$\eden_0$  and the total lifetime of the fireball $\tau_T$. The
final state parameters $\eden_f$, $\rho_{Bf}$, $\rho_{3f}$ will
be calculated from the thorough analyses of various particle
ratios in the chemical freeze-out model~\cite{becatt}. Parameters
$\tau_0$ and $\alpha$ will be chosen to agree with femtoscopy
analyses~\cite{lpsw}. We shall choose a value for $\delta$ and fix
it and $R_{\rm max}$ can be varied only in very limited range and
has little influence on the results.

The expansion rate needed in equation~\eqref{me} is obtained by
taking time-derivative of eq.~\eqref{bd}. Using eq.~\eqref{erate}
we have
\begin{equation}
\label{fer}
-\frac{1}{V}\, \frac{dV}{d\tau} =
\begin{cases}
- \frac{\delta (a+2b\tau)}{1-a\tau-b\tau^2} & : \quad \tau < \tau_s \\
-\frac{\alpha}{\tau-\tau_0} & : \quad \tau \ge \tau_s
\end{cases}\, .
\end{equation}

\begin{figure}
\centerline{\includegraphics[width=4.5cm]{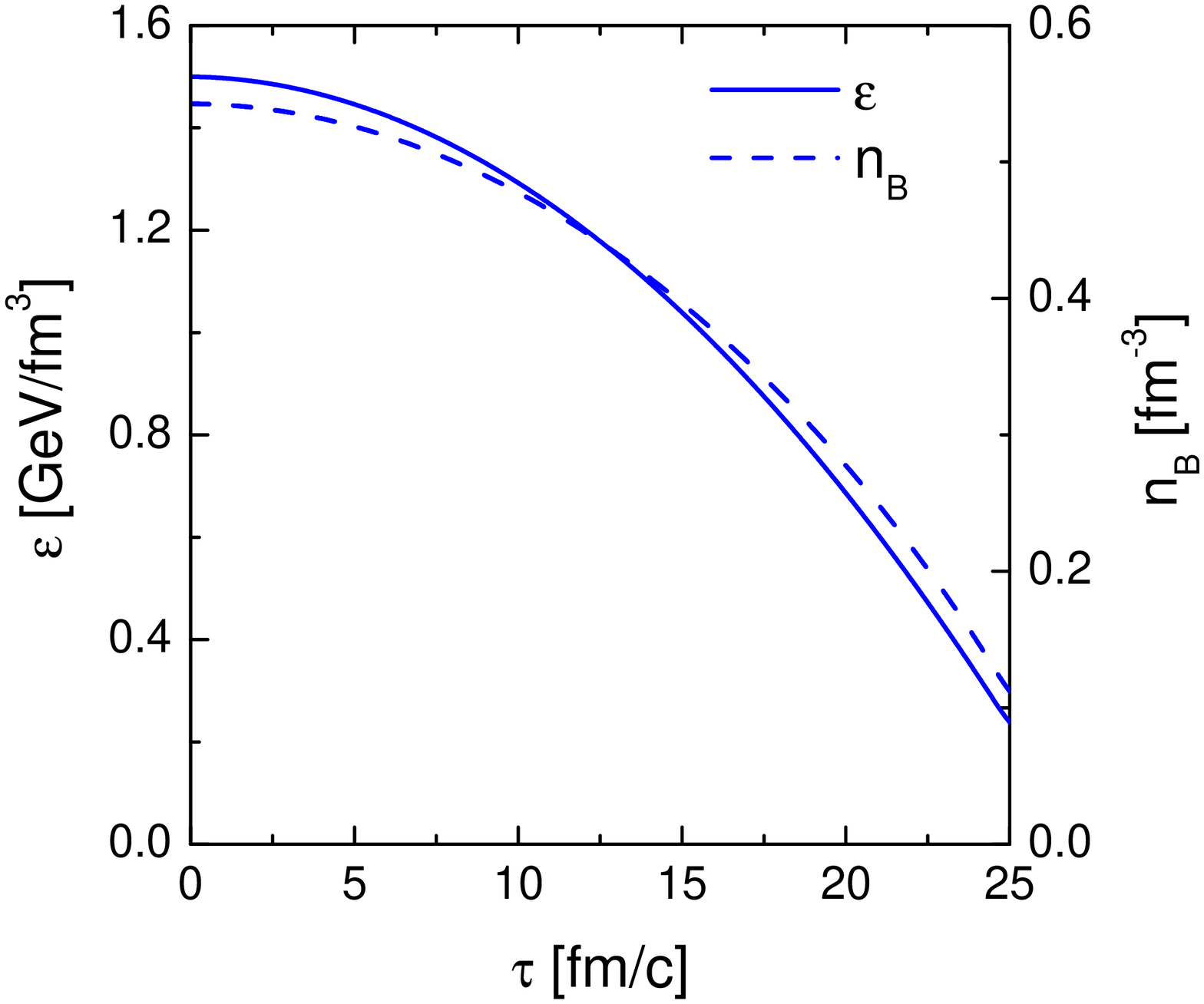}\,\,\,
\includegraphics[width=3.9cm]{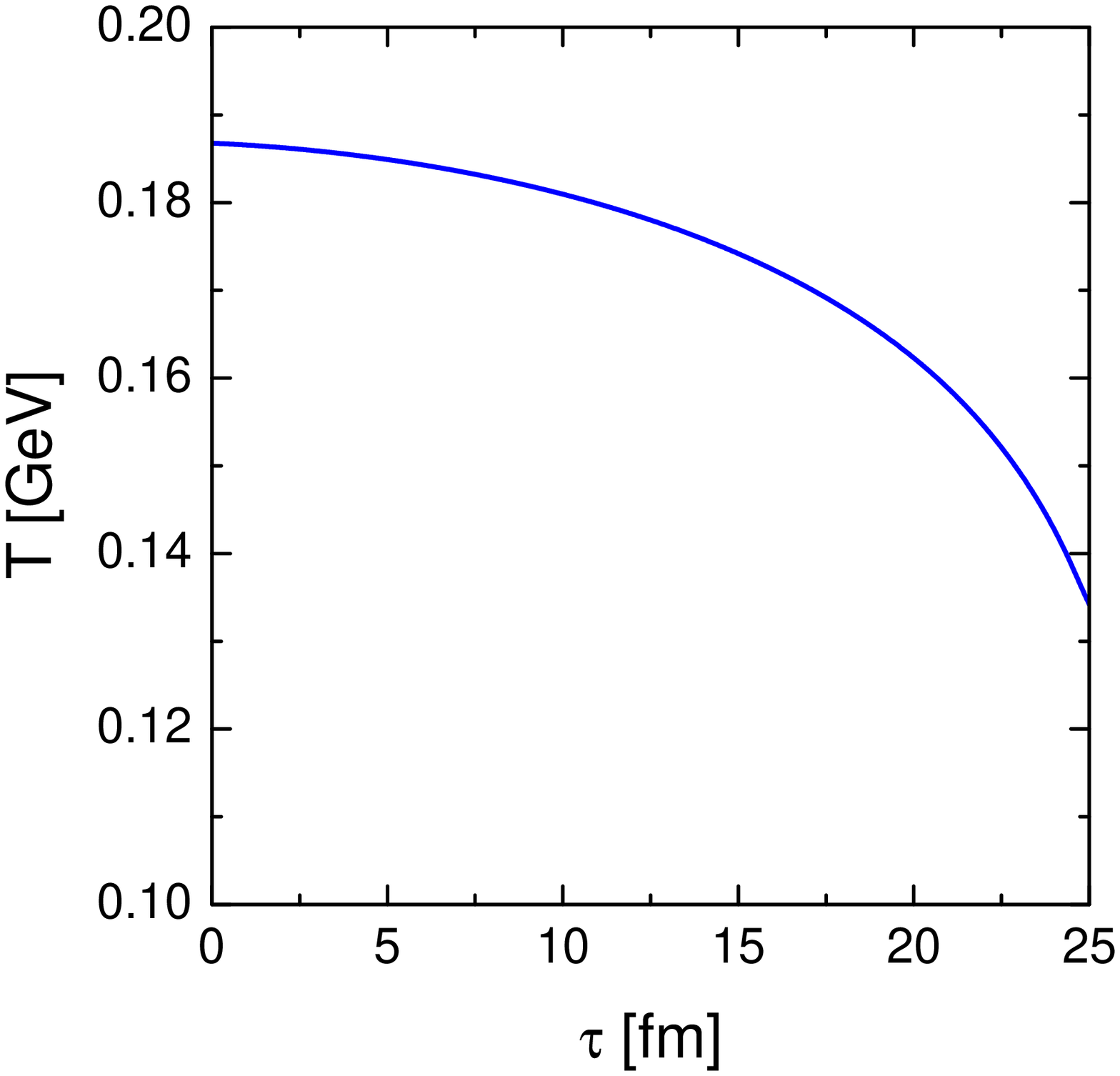}}
\caption{\label{f:eden}
Left panel: the energy density (left scale) and the baryon
density (right scale) as a function of time for a scenario that
reproduces the data from 30 $A$GeV collisions. Right panel:
the corresponding time dependence of the temperature.
}
\end{figure}
For illustration we plot in Fig.~\ref{f:eden}
the time dependence of the energy density and baryon density and the temperature.
The curves correspond to the scenario which reproduces
experimental data on $\ktopi$, $\kmpi$, and $\ltopi$ ratios for
30~$A$GeV collisions as it is shown in Fig.~\ref{f:cdat} with parameters
listed in Tables \ref{fs} and \ref{t:epar}. Time evolutions in other
scenarios are qualitatively very similar to the one presented here.
It turns out that the shape of the curves is almost specified by the
initial and final energy density and the lifetime.
In fact, most of the time the fireball spends in the regime
of accelerated expansion while the power law is realized only
towards the end. This is driven by the requirement of a rather low
initial energy density. Since the final density is fixed by the data, a
longer power law tail leads readily to very
high initial energy densities.
We emphasize that the proposed evolution does not contradict to
observations related to the freeze-out stage of a fireball
expansion reconstructed via single-particle spectra and femtoscopy.


\subsection{Chemical composition and reactions}
\label{checo}

For the sake of averaging over relative velocities, we shall assume
that the momenta are distributed according to Boltzmann distribution
\begin{equation}
n_i(p) \propto \exp \left ( - \frac{\sqrt{m_i^2 + p^2}}{T} \right )\, .
\end{equation}
Thus we make an assumption of thermal equilibrium, though chemically the
system will be treated as non-equilibrated. The averaged cross section
is then obtained as \cite{koform}
\begin{widetext}
\begin{equation}
\langle v_{ij} \sigma_{ij}^{X} \rangle = \\
\frac{\int_{\sqrt{s_0}}^{\infty} dx\, \sigma_{ij}^{X}(x)\,
K_1\left ( \frac{x}{T} \right )\,
\left [ x^2 - (m_i + m_j)^2\right]\left[ x^2 - (m_i-m_j)^2 \right ]}{%
4\, m_i^2\, m_j^2\, T\, K_2(m_i/T)\, K_2(m_j/T)}
\label{sa}
\end{equation}
\end{widetext}
where $K_i$'s are the modified Bessel functions and
$\sqrt{s_0} = \max ( m_i+m_j,\, \sum_{\rm final} m_a )$ is the reaction
threshold.

The following reaction channels producing kaons were taken into
account
\be\begin{tabular}{lllclllclll}
$\pi N     $ & $\leftrightarrow$ & $K Y      $ && $\pi N       $ & $\to$ & $NK\bar K   $ && $N\Delta      $ & $\to$ & $NNK\bar K$ \\
$\pi \Delta$ & $\leftrightarrow$ & $KY       $ && $\pi\Delta   $ & $\to$ & $NK\bar K   $ && $\Delta\Delta $ & $\to$ & $NNK\bar K$ \\
$\pi Y     $ & $\leftrightarrow$ & $K\Xi     $ && $NN          $ & $\to$ & $KNY        $ && $NN           $ & $\to$ & $NNK\bar K$ \\
$\pi\pi    $ & $\leftrightarrow$ & $K\bar K  $ && $N \Delta    $ & $\to$ & $NYK        $ &&                 &       &             \\
$\pi \rho  $ & $\leftrightarrow$ & $K\bar K  $ && $N \Delta    $ & $\to$ & $\Delta KY  $ &&                 &       &             \\
$\pi \rho  $ & $\leftrightarrow$ & $K\bar K^*$ && $\Delta\Delta$ & $\to$ & $\Delta YK  $ &&                 &       &             \\
$\rho\rho  $ & $\leftrightarrow$ & $K\bar K  $ && $NN          $ & $\to$ & $\Delta KY  $ &&                 &       &             \\
$K^*       $ & $\leftrightarrow$ & $K\pi     $ &\phantom{xx}&                &       &               &\phantom{xx}&                 &       &             \\
\end{tabular}
\ee
In order to keep the detailed balance we also included the
inverse reactions for all channels with two particles in final
state. Those processes with three and more final state particles
are rather suppressed due to high thresholds and smaller phase
space. By the same reason we neglect strange antibaryons. The
error we thus introduce is small, because the matter is rather
baryon-dominated in the investigated energy domain. The ratio
$\bar \La /\La$ is about 10\% at the highest SPS energy, so our
discrepancy will be at most of this order. This is acceptable for
a schematic model like the present one.

Only production of  $ K=(K^+,K0)$ and $K^*=(K^{*+},K^{*0})$ mesons will be
calculated from master equation \eqref{me} by including the
above reaction channels.
All the explicitly calculated species contain strange antiquark.
They are produced being accompanied by strange and multistrange
baryons, and $\bar K$ and
$\bar K^*$ mesons containing a strange quark.
In principle, we could trace
density evolutions of those species kinetically.
In practice, however, the reactions which do not create strange quarks
but rather rearrange them in different hadrons are very quick.
Therefore with a good approximation we can assume that all
species containing strange quarks
are in {\em relative chemical equilibrium}.
In our calculations we include
$K^-$, $\bar K^0$, $K^{*-}$, $\bar K^{*0}$, $\La$, $\Sigma$,
$\La(1405)$, $\La(1520)$, $\Sigma(1385)$, $\Xi$, and $\Omega$.

For the {\em non-strange} sector we assume that all species
are chemically equilibrated. We included into calculations
all non-strange mesons with masses below 1.5~GeV and baryons
up to 2~GeV.

Since we follow evolution of every isospin state separately, we do not use
isospin-averaged parameterizations for the cross sections. Appropriate
cross section for every isospin channel is used instead. We list the
used parameterizations in Appendix \ref{xsec}.

For the calculation of densities of individual species and the
evaluation of the average in eq.~\eqref{sa} we thus need the temperature $T$,
and chemical potentials $\mu_B$ and $\mu_3$ connected with the baryon number
and the third component of isospin, respectively. We also need phase-space
occupation factors for kaons $\gamma_{K^+}$, $\gamma_{K^0}$,
$\gamma_{K^{*+}}$, and $\gamma_{K^{*0}}$,  as well as one common
factor for species with $S<0$, $\gamma_S$. On the other hand, the time
evolution is formulated in terms of $\eden$, $\rho_B$, $\rho_3$, and
kinetically calculated $\rho_{K^+}$, $\rho_{K^0}$, $\rho_{K^{*+}}$, and
$\rho_{K^{*0}}$. We thus need to write down relations connecting these two sets
of quantities. First, we note that the overall strangeness neutrality
dictates that the density of strange quarks
\begin{equation}
\rho_S \equiv \sum_{i,\, S<0} |S_i|\, \rho_i =
\rho_{K^+} + \rho_{K^0} + \rho_{K^{*+}} + \rho_{K^{*0}}
\label{ns}
\end{equation}
where the sum should run over all species containing strange quark,
though practically we include only those mentioned above.
Then, in Boltzmann approximation, we have the following relations
between thermodynamic quantities and densities
\begin{subequations}
\label{tdr}
\begin{eqnarray}
\eden & = & \frac{1}{2\pi^2}\,  \sum_{i,S=0} g_i\, \lambda_i\,
m_i^2\, T^2
\\ \nonumber  && \qquad \qquad
\left \{ \frac{m_i}{T} K_1\left(\frac{m_i}{T}\right ) +
3K_2\left(\frac{m_i}{T}\right ) \right \}
\\ &&
+ \frac{1}{2\pi^2}\, \sum_{i,S<0} g_i\, \lambda_i\, \gamma_S^{|S_i|}\,
m_i^2\, T^2
\nonumber \\ && \qquad \qquad
\left \{ \frac{m_i}{T} K_1\left(\frac{m_i}{T}\right ) +
3K_2\left(\frac{m_i}{T}\right ) \right \}
 \nonumber \\&&
 + \frac{1}{2\pi^2}\, \sum_{i,{\rm kaons}} g_i\, \lambda_i\, \gamma_i\,
m_i^2\, T^2
\nonumber \\ && \qquad \qquad
\left \{ \frac{m_i}{T} K_1\left(\frac{m_i}{T}\right ) +
3K_2\left(\frac{m_i}{T}\right ) \right \} \nonumber
\end{eqnarray}
\begin{eqnarray}
\rho_B & = & \frac{1}{2\pi^2}\,  \sum_{i,S=0} B_i\, g_i\, \lambda_i\,
m_i^2 \, T\, K_2\left ( \frac{m_i}{T} \right ) \\
& & + \frac{1}{2\pi^2}\,  \sum_{i,S<0} B_i\, g_i\, \lambda_i\,
\gamma_S^{|S_i|}\,
m_i^2 \, T\, K_2\left ( \frac{m_i}{T} \right )\nonumber
\end{eqnarray}
\begin{eqnarray}
\tilde \rho_3 & = & \frac{1}{2\pi^2} \,
\sum_{i,S=0} I_{3,i}\, g_i\, \lambda_i\,
m_i^2 \, T\, K_2\left ( \frac{m_i}{T} \right ) \\
& & + \frac{1}{2\pi^2}\,  \sum_{i,S<0} I_{3,i}\, g_i\, \lambda_i\,
\gamma_S^{|S_i|}\,
m_i^2 \, T\, K_2\left ( \frac{m_i}{T} \right ) \nonumber\\
& & + \frac{1}{2\pi^2}\,  \sum_{i,{\rm kaons}} I_{3,i}\, g_i\, \lambda_i\,
\gamma_{i} \,
m_i^2 \, T\, K_2\left ( \frac{m_i}{T} \right ) \nonumber
\end{eqnarray}
\begin{equation}
\rho_S  =  \frac{1}{2\pi^2}\, \sum_{i,S<0} |S_i|\, g_i\, \lambda_i\,
\gamma_S^{|S_i|}\, m_i^2\, T\, K_2\left ( \frac{m_i}{T} \right )
\end{equation}
and for all kaon species $ $
\be
\nonumber
\rho_i = \frac{1}{2\pi^2}\, \exp\left(\frac{I_{3,i}\mu_3}{T}\right )\,
\gamma_i\, m_i^2\, T \, K_2\left ( \frac{m_i}{T} \right )\, ,
\\ \label{kdenT}
i=K^+,\, K^0,\, K^{*+},\, K^{*0}\,.
\ee
\end{subequations}
These relations can be inverted numerically and $T$, $\mu_B$, $\mu_3$,
$\gamma_S$, and the four $\gamma_i$'s can be determined.

In equations \eqref{tdr},
$g_i$ is the spin degeneracy of species $i$ (different isospin states
are accounted for separately). The fugacity
\begin{equation}
\lambda_i = \exp\left ( (B_i\, \mu_B + I_{3,i}\, \mu_3)/T \right )\, ,
\end{equation}
$B_i$, $I_{3,i}$, and $S_i$ are baryon number, third component of isospin,
and strangeness, respectively.

\subsection{Kaons as a part of the system}
\label{s:whyk}

Kaons have rather small cross sections for interactions  in
baryon-rich matter and therefore their scattering rate (mean number
of collisions per unit of time) is not large
and the mean free path is long. In our simulations we calculate the scattering
rate in a similar way as the annihilation rate in eq.~\eqref{chem}
\begin{equation}
\label{kscatt}
{\cal R}_{\rm scatt} = \sum_i \langle v_{Ki} \sigma_{Ki}^{\rm tot} \rangle\,
\frac{\rho_i}{1 + \delta_{Ki}}\,  .
\end{equation}
In determining
${\cal R}_{\rm scatt}$ we take all processes that are included in kaon
annihilation and add the elastic scattering off protons and neutrons
with the cross sections given by eqs.~\eqref{kpxs} and \eqref{knxs}.

\begin{figure*}
\centerline{\includegraphics[width=4.8cm]{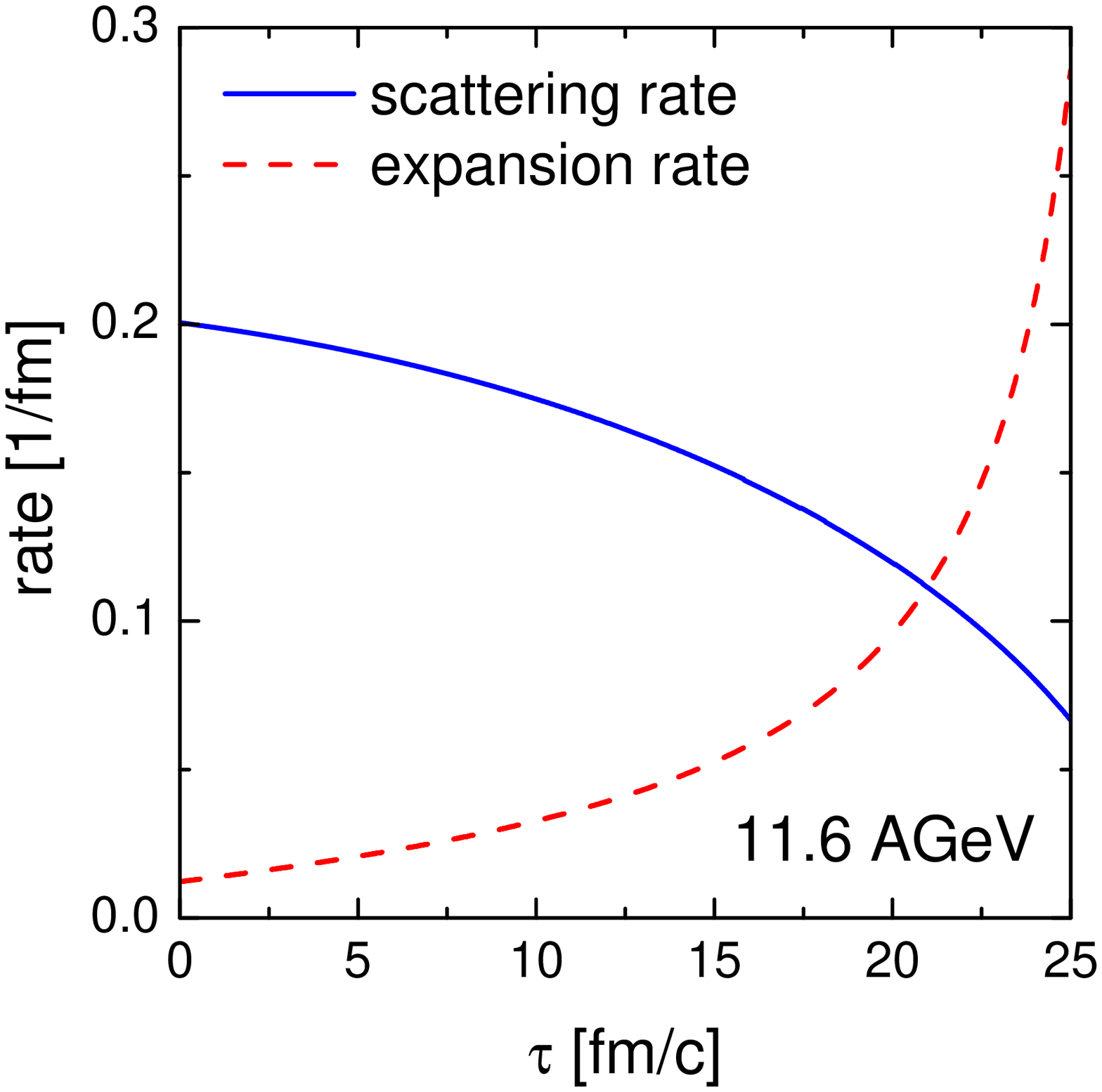}\quad
\includegraphics[width=4.8cm]{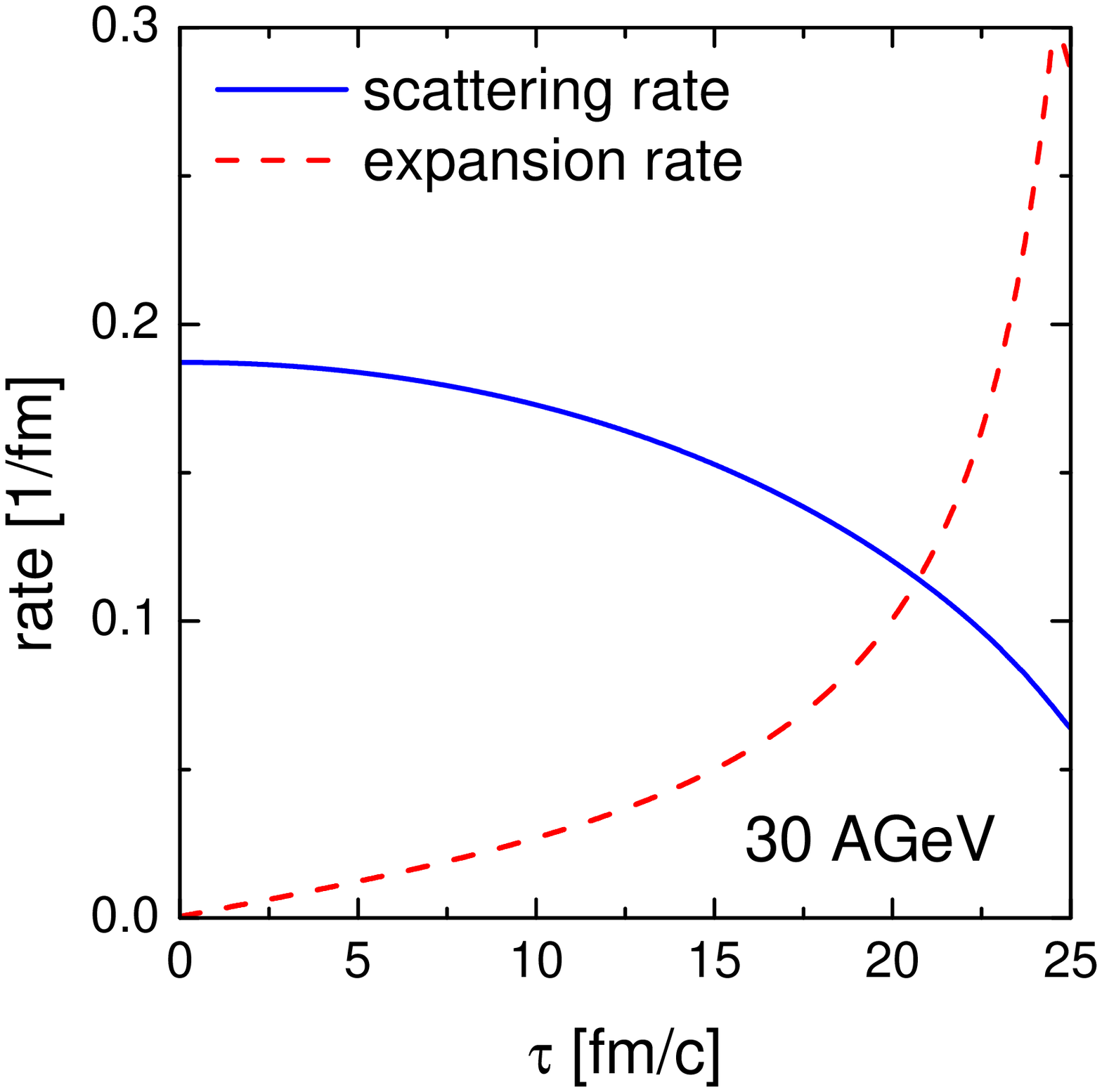}\quad
\includegraphics[width=4.8cm]{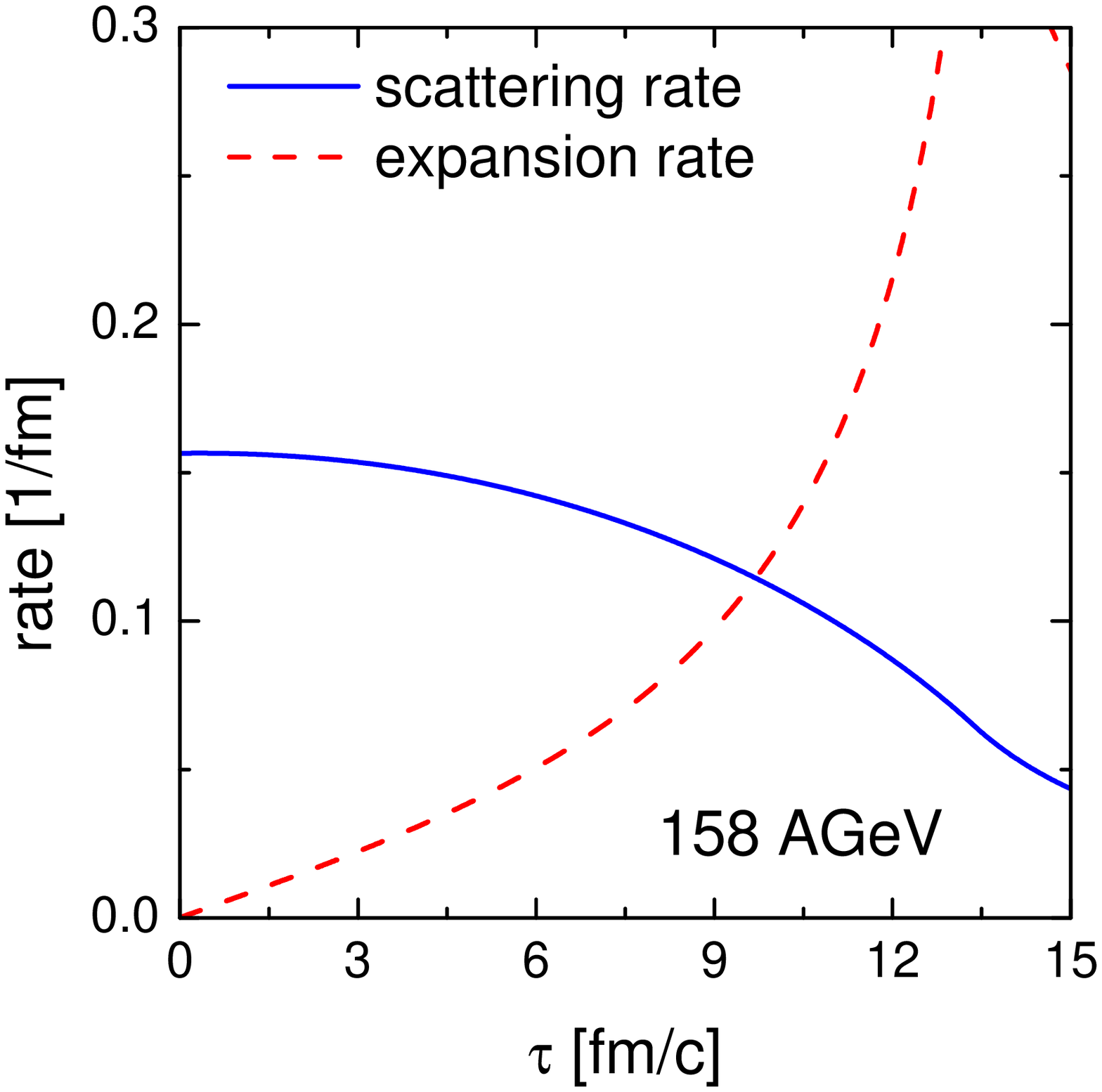}}
\caption{\label{f:kdec}
Comparison of the kaon scattering rate calculated according to
Eq.~\eqref{kscatt} and the expansion rate introduced in Eq.~\eqref{erate}
as a function of time
for three scenarios which fit the data at: 11.6~$A$GeV (left),
30~$A$GeV (middle), and 158~$A$GeV (right).
}
\end{figure*}

Kaon scattering rates for three representative
expansion scenarios are plotted in Fig.~\ref{f:kdec}.
They are typically of the order 0.1~fm$^{-1}$; somewhat larger at higher density
in the beginning and smaller at later times. With thermal kaon velocity
of about $0.7\,c$ this leads to a mean free path of 7~fm. Even if the kaon
moved with light velocity, like some primordially produced kaons,
the mean free path would still come to about 10~fm. This length is
comparable to the size of a fireball.
Would the kaon mean free path  be much bigger than the size of a
fireball, one could assume kaons to escape from the system
right after the production event without any further scattering.
Oppositely, a mean free path much smaller than the fireball size would indicate
that kaons rescatter intensively and stay in thermal equilibrium with fireball medium.
Reality is somewhere in between these two limits.
In any case, they do not simply escape from the fireball.
In our type
of model we can only make two extreme assumptions: decouple the produced
kaons from the system {\em completely} or treat them {\em all} as a part
of the system. Due to a chance of absorption the latter appears closer
to reality.
For practical reason, in calculation of kaon scattering and annihilation
rates thermally equilibrated velocity distributions have been assumed.

Let us also note that the inverse slope parameters $T^*$ of
kaon spectra, together with pions and protons seem to follow the empirical
prescription $T^* = T + m\langle v_t \rangle^2$, where $\langle v_t \rangle^2$
is average transverse expansion velocity \cite{na57}.
This suggests that all these species freeze-out kinematically at the
same time.

The influence of our simplification is twofold. Firstly, we may
overestimate the amount of kaons that are annihilated. Secondly,
the energy which the kaons carry is not taken out of the system as it
would be so if kaons decouple.

\begin{figure*}
\centerline{\includegraphics[width=5cm]{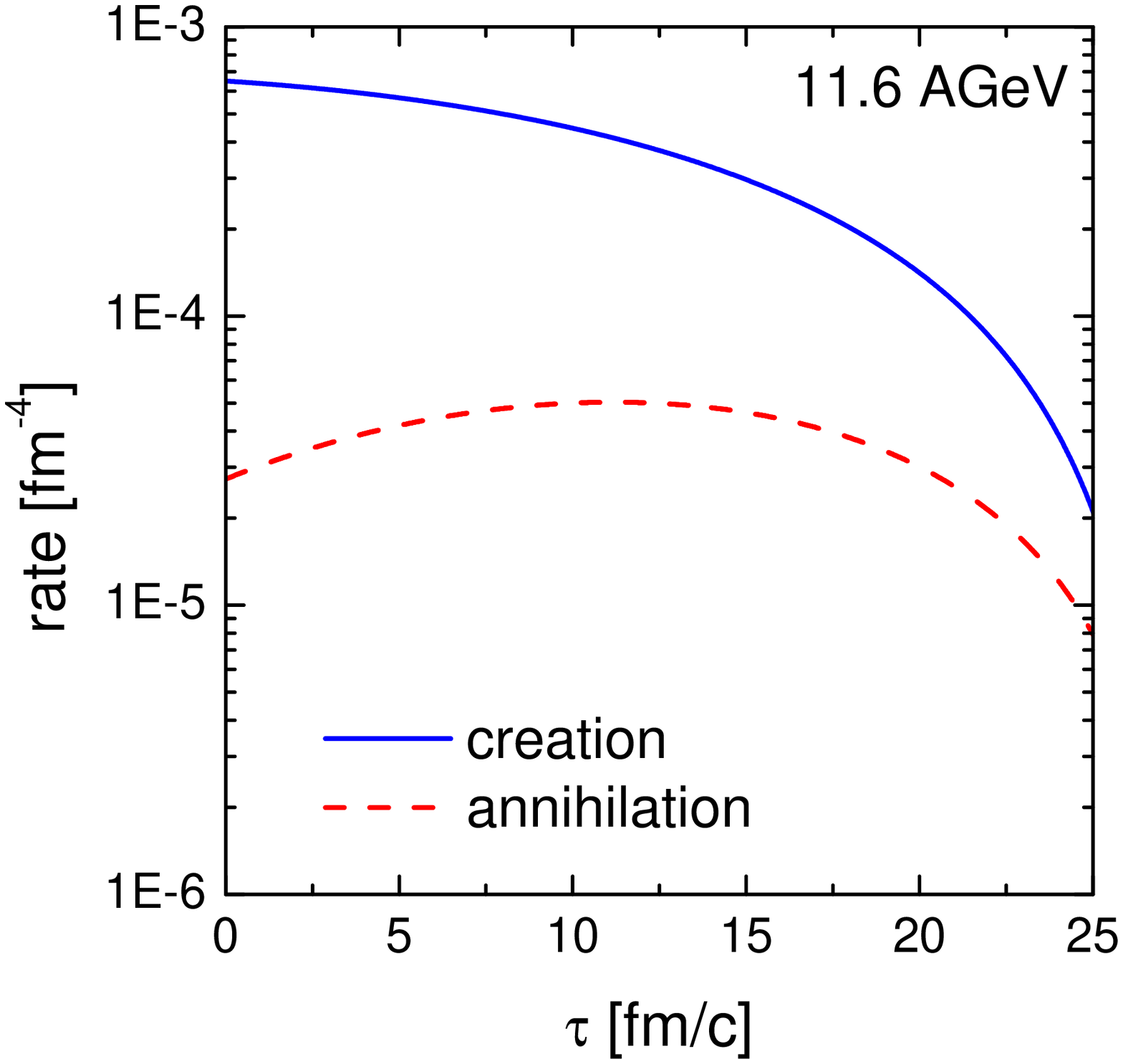}
\includegraphics[width=5cm]{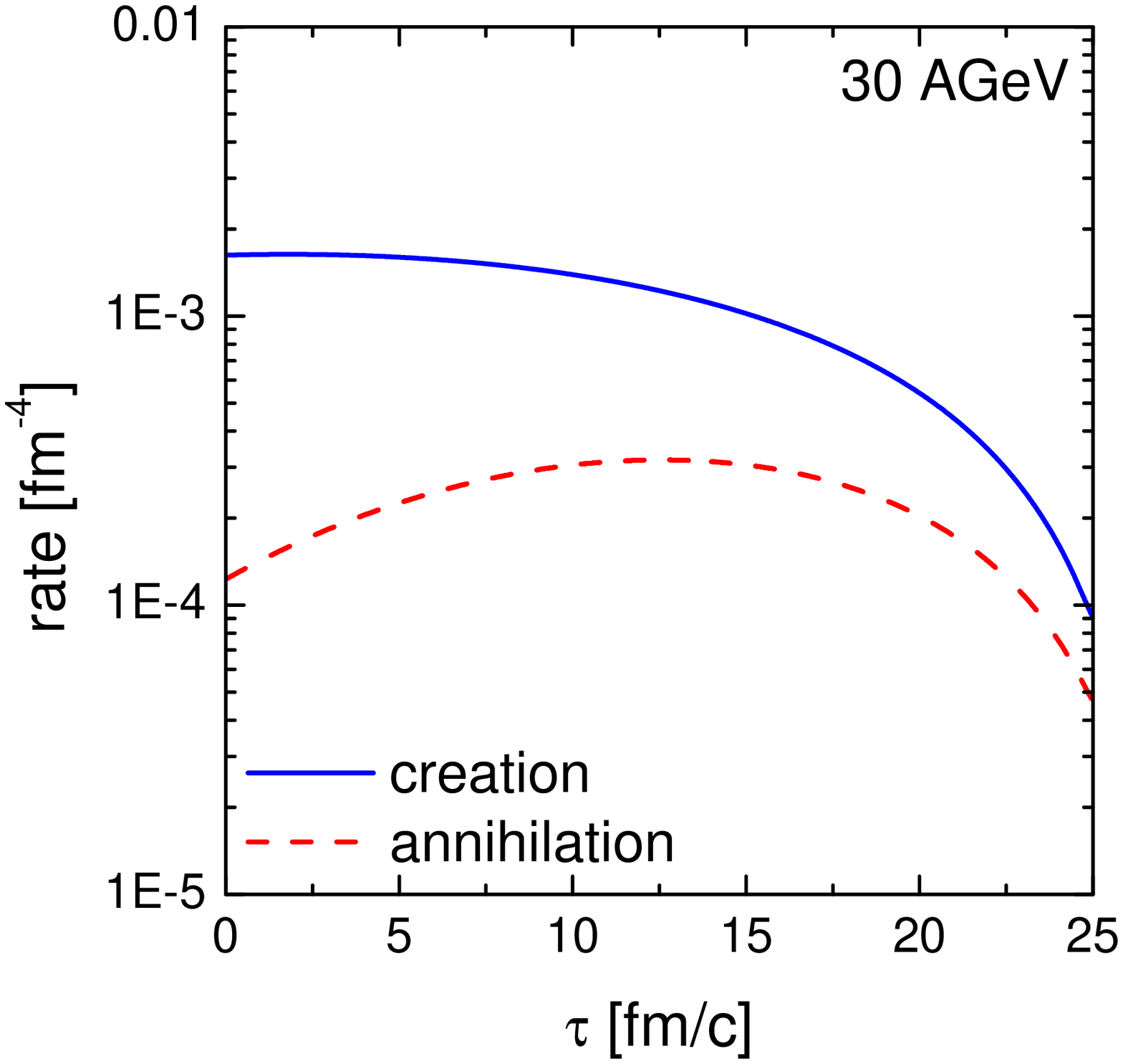}
\includegraphics[width=5cm]{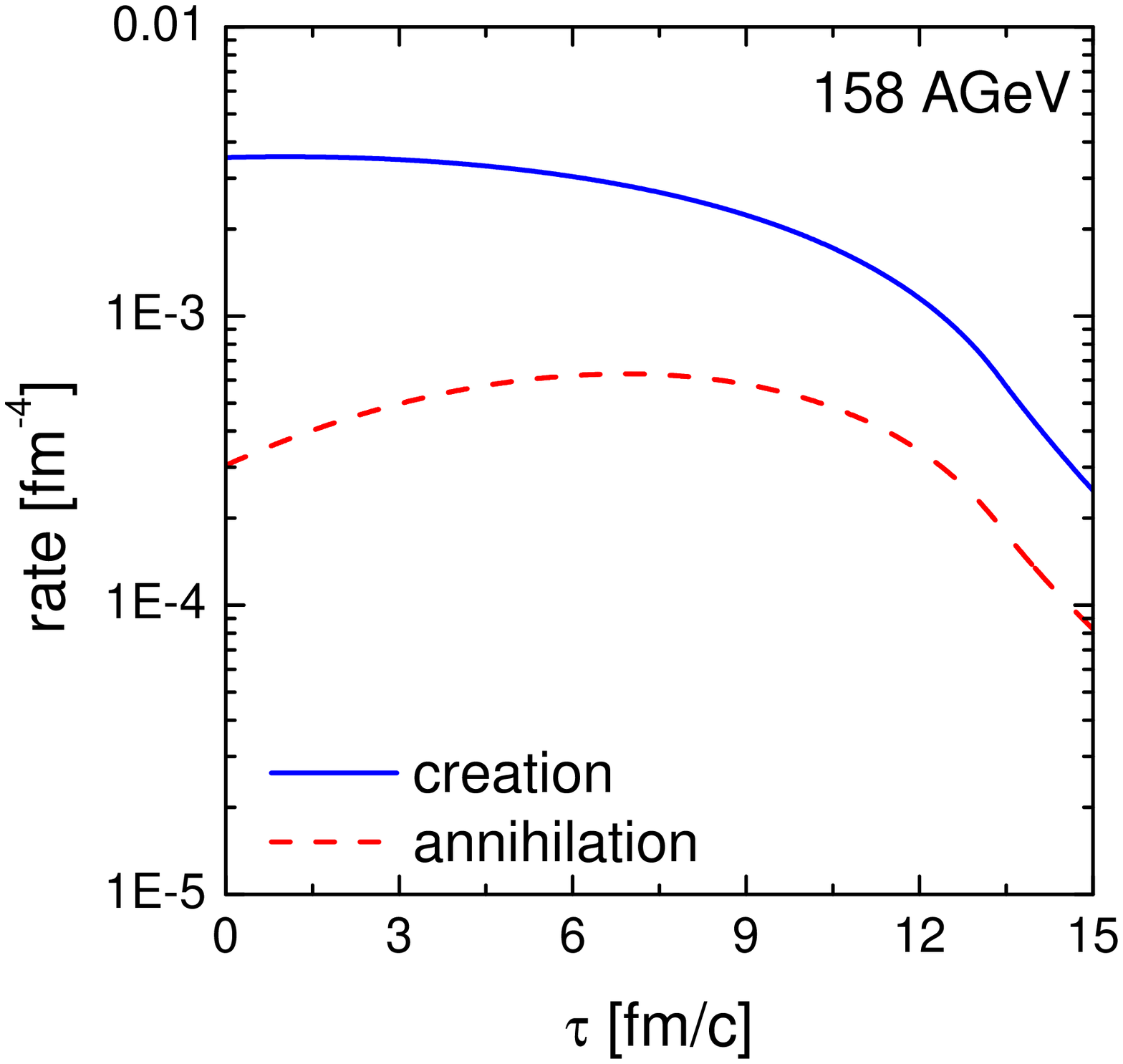}}
\caption{\label{f:kca}
Comparison of total kaon creation and annihilation rates
as a function of time
for three scenarios which fit the data at: 11.6~$A$GeV (left),
30~$A$GeV (middle), and 158~$A$GeV (right).
}
\end{figure*}

The overestimation of kaon annihilation would introduce only small
error if the annihilation rate is small compared to the production
rate. The rates are plotted for three typical scenarios
in Fig.~\ref{f:kca} (same scenarios as in Fig.~\ref{f:kdec}).
The annihilation rate grows since the kaon abundance increases.
Originally it is smaller than the creation rate by an order of magnitude,
at late times by a factor of 2. From the fact that up to 50\% of final state
direct kaons are produced primordially and
not due to kaon creation modelled
here (see Fig.~\ref{f:kden}),
we conclude that a decrease of the (small) annihilation rate would not
change our results too much. At most, it would slightly shorten the lifetimes
necessary to describe the data.

\begin{figure*}
\centerline{\includegraphics[width=5cm]{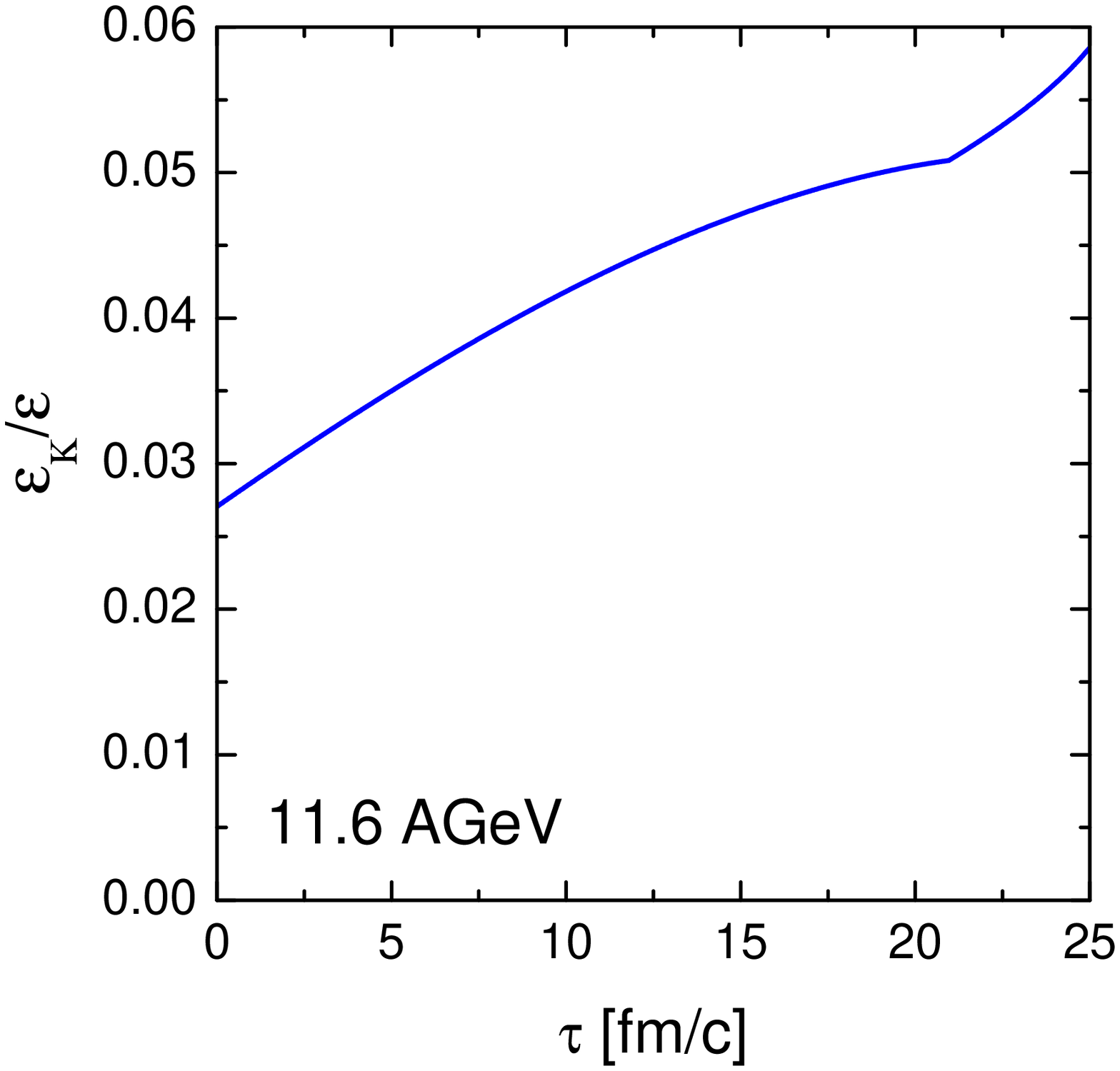}
\includegraphics[width=5cm]{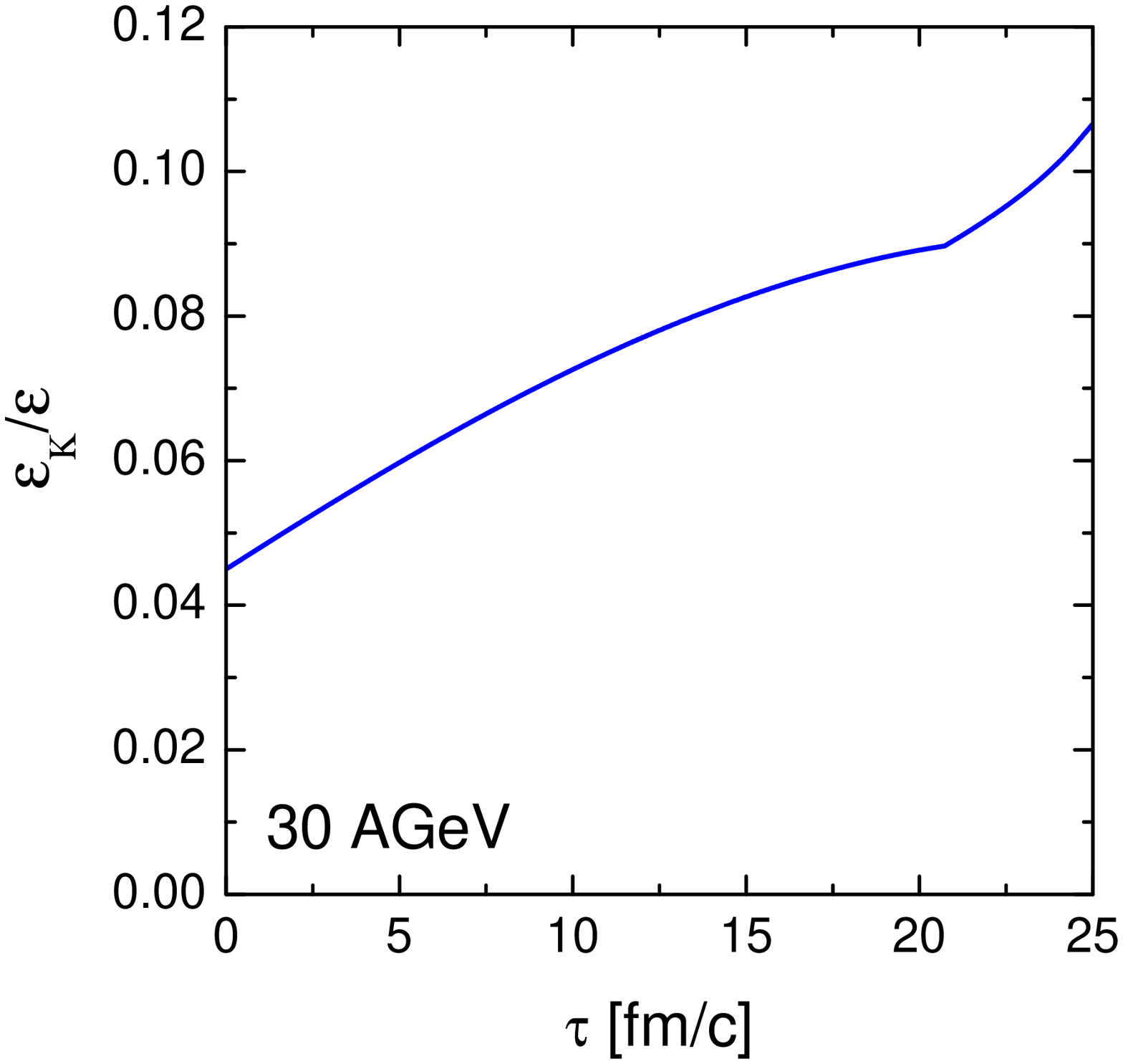}
\includegraphics[width=5cm]{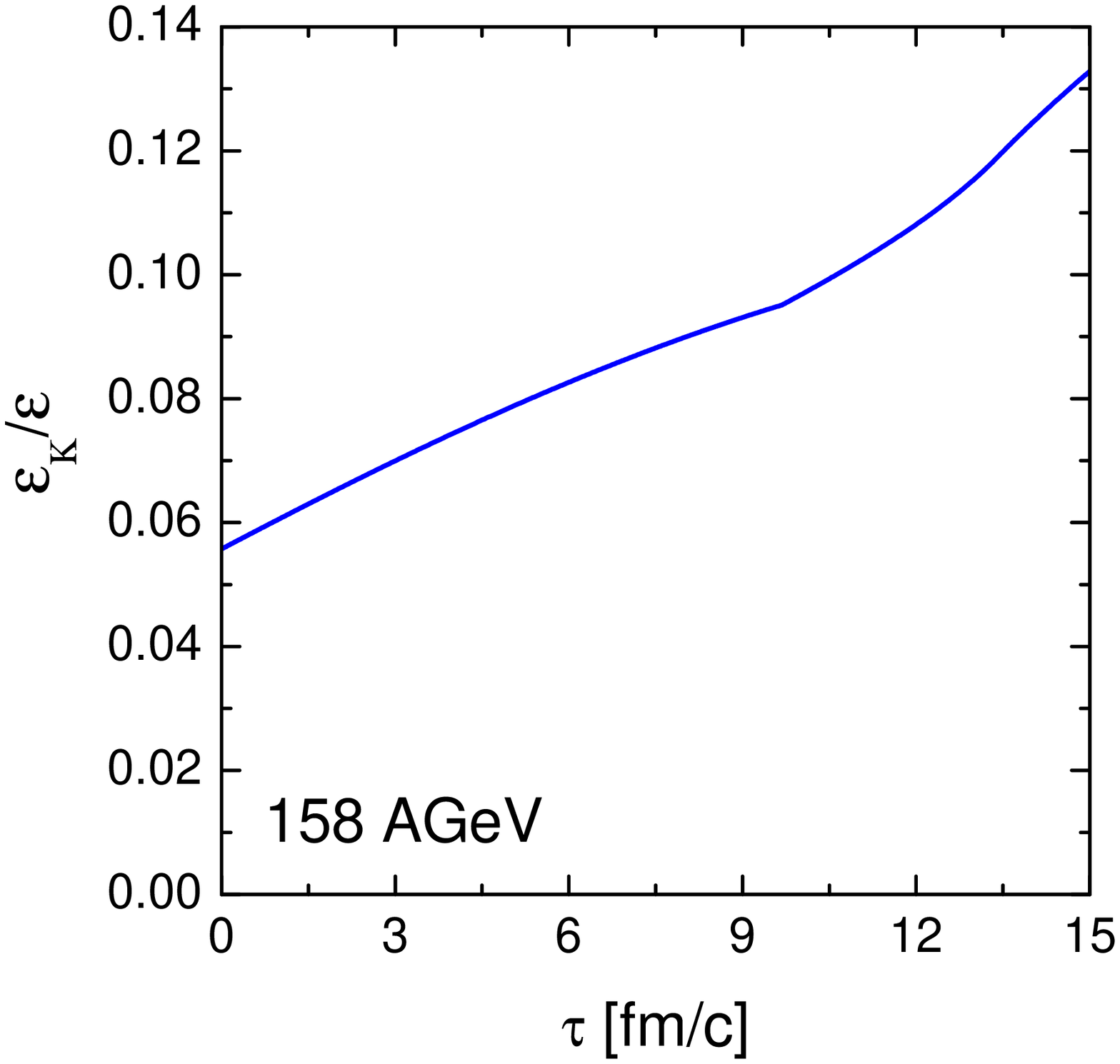}}
\caption{\label{f:kepart}
Ratios of the energy contained in kaons to the total energy of the
system as functions of time
for three scenarios which fit the data at: 11.6~$A$GeV (left),
30~$A$GeV (middle), and 158~$A$GeV (right).
}
\end{figure*}

In order to estimate a possible error in the total energy of the fireball
we plot in Fig.~\ref{f:kepart} the ratio of the energy carried by thermal kaons
to the total energy of the system.
At the beginning, at most 6\% of energy is contained in kaons. This is the
part of energy we should take away if we treat primordial kaons as
non-interacting. Since energy
density goes roughly with the fourth power of temperature, $T$
would not change much and the production rates would stay practically
unchanged. In general, subtracting a few per cent from the energy
density would have small influence on chemical composition because at the
same time we would not include kaons and the available energy would
be distributed among smaller number of degrees of freedom, thus changing
the temperature only little.

It is possible to refine our treatment of kaons as a part of the
fireball by comparing the kaon scattering rate to the expansion
rate of the fireball, cf. Fig.~\ref{f:kdec}.
At the beginning of fireball evolution, expansion rate is lower than the
scattering rate, thus we say that kaons are thermally
equilibrated. When the expansion rate becomes comparable or bigger than
the scattering rate, this means that the density of fireball drops
considerably before a kaon had a chance to scatter. Then kaons are
assumed to decouple from thermal equilibrium. Decoupling
is a gradual process but we cannot treat it so in the framework of
our model. We say that the kaons decouple at once when
\begin{equation}
\label{ktherm}
-\frac{1}{\rho_B}\, \frac{d\rho_B}{d\tau} \ge \theta \,{\cal R}_{\rm scatt}\, .
\end{equation}
When this condition becomes fulfilled, we fix the temperature for
kaons in eq.~\eqref{kdenT}. We thus assume that the kaons keep a
higher temperature than other species, because they decoupled
from the thermal bath earlier. By default, the constant $\theta$ is
set to 1, though no exact value can be specified by theory \cite{bgz,fow}.
Variations of $\theta$ turn out to have negligible influence on the
results.

In Fig.~\ref{f:kdec} we demonstrate that the decoupling according
to this prescription happens only towards the end of expansion.

\subsection{A summary of the evolution procedure}

For an easier overview, we summarize the algorithm for time evolution.
Suppose, at time $\tau$
we know all quantities: $\eden$, $\rho_B$, $\rho_3$, all kaon densities,
and thermodynamic quantities $T$, $\mu_B$, $\mu_3$, $\gamma_S$, and all
$\gamma_i$'s. We proceed to time $\tau+d\tau$ in following steps:
\begin{enumerate}
\item
Calculate new densities of $K^+$, $K^0$, $K^{*+}$, $K^{*0}$ from
eq.~\eqref{me}.
\item
Calculate total kaon scattering rate ${\cal R}_{\rm scatt}$ from eq.~\eqref{kscatt}.
\item
Obtain $\eden$, $\rho_B$, $\rho_3$ from prescription
\eqref{td}. Also obtain expansion rate from eq.~\eqref{fer}.
\item
Determine $\rho_S$ from eq.~\eqref{ns}.
\item
Decide by the use of inequality~\eqref{ktherm}
if kaons are thermalized according.
\item
Obtain $T$, $\mu_B$, $\mu_3$, $\gamma_S$, and $\gamma_i$'s
from numerically inverting
relations \eqref{tdr}. If kaons are not thermalized, fix their temperature
in eq.~\eqref{kdenT}.
\item
Calculate the density of any desired species via
\begin{equation}
\rho_i = \frac{1}{2\pi^2}\, g_i \, \lambda_i \, m_i^2 \, T \,
K_2 \left ( \frac{m_i}{T} \right )
\end{equation}
for non-strange species, and
\begin{equation}
\rho_i = \frac{1}{2\pi^2} \, g_i\, \lambda_i \, \gamma_S^{|S_i|}\, m_i^2\, T\,
K_2 \left ( \frac{m_i}{T} \right )
\end{equation}
for species with $S<0$.
\item
Continue to next time step by going to step 1.
\end{enumerate}

In the outlined setup of the model there are different handles to
tune the final $K^+$ density and those of $K^-$ and $\La$. All these,
normalized by pion densities, will be compared to data. Since the production
of $K^+$ is calculated explicitly, its density depends on temperature
and time. If we can reproduce the $\ktopi$ ratio, this means that we
have gotten the total strangeness production right. The strange
quarks are then distributed among $K^-$, $\La$, and other species
according to the temperature and chemical potentials. Thus the key to
simultaneous fit to $\kmpi$ and $\ltopi$ ratios is the correct value
of the temperature.


\subsection{Final state and feed-down from resonance decays}
\label{fistate}

Chemical composition of the final state in Au+Au collisions
at beam energy 11.6~$A$GeV \cite{b23,b24,b25,b26,bcksr} and Pb+Pb
collisions at beam energies 30 AGeV \cite{alt20}, 40, 80, and 158
$A$GeV \cite{ktpdata} have been analyzed in the framework of the
statistical hadronization model complemented by strangeness
suppression factor $\bar \gamma_S$
\footnote{Note that notation $\gamma_S$ is used in
\cite{becatt} for what we call $\bar \gamma_S$ here. We change the notation
in order not to mix up with our occupation factor for only $S<0$ species}
by Becattini and collaborators
\cite{becatt}. We use their results on temperature, chemical potentials
and $\bar \gamma_S$ in order to characterize the state of chemical freeze-out
which we aim for.
\begin{table}
\caption{\label{fs}
Thermodynamic quantities and densities characterising the chemical
freeze-out state in collisions at different energies. Values of
$T$, $\mu_B$, $\mu_3$, $\mu_S$, and $\bar\gamma_S$ from
\cite{becatt} serve as an input in reconstructing the final state values
of $\epsilon_f$, $\rho_{B\,f}$ and $\rho_{3\,f}$.
}
\begin{tabular}{c|ccccc}
\hline\hline
$E_{\rm beam}$ [$A$GeV] & 11.6 & 30 & 40 & 80 & 158 \\
\hline
$T$ [MeV] & 118.1 & 139.0 & 147.6 & 153.7 & 157.8
\\
$\mu_B$ [MeV] & 549.1 & 423.1 & 375.1 & 293.6 & 243.7
\\
$\mu_3$ [MeV] & --11.8 & --11.0 & --10.3 & --8.2 & --7.1
\\
$\mu_S$ [MeV] & 117.5 & 99.1 & 90.5 & 69.7 & 59.2
\\
$\bar \gamma_S$ & 0.652 & 0.938 & 0.757 & 0.73 & 0.843
\\ \hline
$\eden_f$ [GeV/fm$^3$] & 0.132 & 0.173 & 0.203 & 0.194 & 0.198
\\
$\rho_{Bf}$ [fm$^{-3}$]& 0.086 & 0.087 & 0.091 & 0.068 & 0.058
\\
$\rho_{3f}$ [fm$^{-3}$]& --0.0083 & --0.0097 & --0.0102 & --0.0079 & --0.0070
\\ \hline\hline
\end{tabular}
\end{table}
Thus obtain the final state value of the energy density,
$B$ and $I_3$ densities (Table~\ref{fs}). These are
three of the parameters which specify the time evolution. Note that
even fixing the final state densities does not guarantee that we arrive
at the correct final chemical composition because final temperature
depends on the amount of produced strangeness which is calculated
kinetically. On the other hand, it is important to realize that if
we end up in a state with the correct amount of kaons and the correct
final energy and baryon densities guaranteed by construction,
then we also reproduce the temperature as inferred in chemical freeze-out
fits in \cite{becatt} and therefore {\em all} ratios of abundances, i.e.
also those involving only non-strange species. If we then compare calculated
density of one of the species to the {\em total} measured multiplicity
of that species we could infer the volume and thus multiplicities of all
other species.

A significant portion of final state pions stems from decays
of resonances. These must be included when obtaining ratios
of multiplicities. This is done by determining the number of
resonances and multiplying by the average number of pions produced
in a decay of one resonance. Also, there is feed-down to kaon production
from $K^*$ decays and to the number of $\La$'s from other
strange baryons. A list of all contributions is provided in
Appendix \ref{feeddown}.


\subsection{Initial state}

Now we have to specify the initial conditions for the evolution
equation \eqref{me}.
Note that strangeness is also produced in primordial collisions of incident nucleons.
The amount will be extrapolated from strangeness abundance in nucleon-nucleon
interactions. We shall equate the ratio of $K^+$ density and the density
of negative hadrons to the ratios of multiplicities
$\langle K^+\rangle_{NN}/\langle h^-\rangle_{NN}$
extrapolated from nucleon-nucleon collisions. The values are summarized
in Table~\ref{t:inis}.
\begin{table}
\caption{\label{t:inis}
Mean multiplicity $\langle K^+ \rangle_{NN}$ from nucleon-nucleon
collisions and ratios
$\langle K^+\rangle_{NN}/\langle h^-\rangle_{NN}$ extrapolated
to a mixture of protons and neutrons corresponding to Au+Au or
Pb+Pb collisions at indicated beam energies.
}
\begin{tabular}{c|cc} \hline\hline
& $\langle K^+ \rangle_{NN}$ &
$\frac{\langle K^+ \rangle_{NN}}{\langle h^-\rangle_{NN}}$ \\
\hline
Au+Au @ 11.6 AGeV & $0.0607\pm 0.0252$ & 0.0581 \\
Pb+Pb @ 30 AGeV   & $0.143\pm 0.027$ & 0.0821 \\
Pb+Pb @ 40 AGeV   & $0.163\pm  0.036$ & 0.0842 \\
Pb+Pb @ 80 AGeV   & $0.218\pm 0.059$ & 0.0880 \\
Pb+Pb @ 158 AGeV  & $0.272\pm0.056$ & 0.0883 \\
\hline\hline
\end{tabular}
\end{table}
Density of $K^0$ is then determined as $\rho_{K^0} = \rho_{K^+} \exp (-\mu_3/T)$.
No $K^*$'s are assumed in the initial state. Species with $S<0$ must
balance the total strangeness to 0 and are set into {\em relative}
chemical equilibrium. We describe the choice of the initial state in
detail in Appendix~\ref{istate}.


\section{Results and discussion}

We calculated the resulting $\ktopi$, $\kmpi$ and $\ltopi$ ratios
for many different evolution scenarios for the beam energies
11.6~$A$GeV (Au+Au), and 30, 40, 80, 158~$A$GeV (Pb+Pb).
Thus we explore the region of the peak in $\ktopi$.

At every beam energy the investigated scenarios differ by
the initial energy density and total lifetime of the fireball.
The final configurations listed in Table~\ref{fs} are fixed from
the analysis~\cite{becatt}.
The range of initial energy densities goes up to 3.0~GeV/fm$^3$.
This seems to surpass the critical value for deconfinement indicated
by lattice calculations. We have two comments here, however: first,
the system under investigation is out of equilibrium and it
is a question whether results of equilibrium statistical physics can
be applied here. The relaxation time is not known. Second,
an uncertainty in determination of the critical temperature
was reported at the Quark Matter 2005 conference \cite{katz}
leading to a value higher than one previously used.

The parameter $\tau_0$ is chosen  such that it is smaller than the total
lifetime by 7~fm/$c$, $\tau_0=\tau_T-7{\rm fm}/c$.
Such a value is motivated by
measurement of the longitudinal HBT radius $R_l$, which---in a Bjorken
scenario---indicates that the {\em thermal} freeze-out happens about
8-9 fm/$c$ after the reference $\tau_0$. As we deal in our study
with {\em chemical} freeze-out, we set $\tau_T-\tau_0$ a little smaller.

Since we know from transverse momentum spectra and
HBT radius anlyses that there is considerable transverse expansion, we fix
$\alpha = 2$. The equation of state is set by a modest value of
$\kappa = 1/6$, or $\delta = 6/7$.

The last parameter to specify is $\rmax$. If we fix the initial and
final densities and the total lifetime in which the fireball evolves
between these two states, there is only a limited range of values
the maximum expansion rate can assume. For every pair of chosen initial
energy density and lifetime we explored the two limiting values of
$\rmax$. It turned out that the differences due to selection of
$\rmax$ in the studied ratios were of the order 10\%.

\begin{figure}
\centerline{\includegraphics[width=0.475\textwidth]{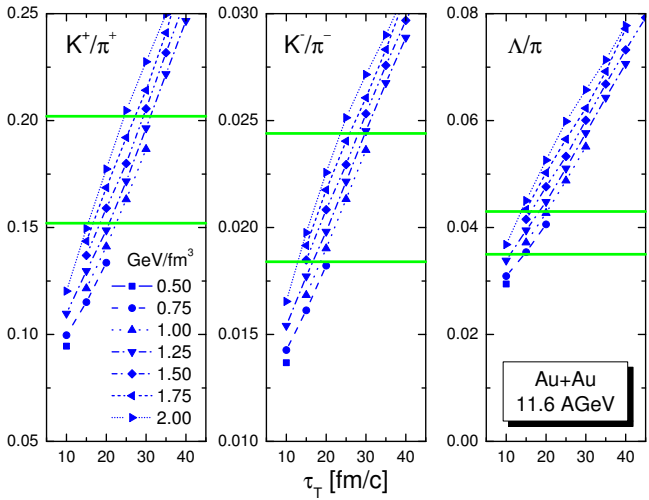}}
\caption{\label{f:rAGS}
Ratios $\ktopi$, $\kmpi$, and $\ltopi$ as a function of the total
lifetime of the system, calculated for Au+Au collisions at
beam energy of 11.6~$A$GeV. Different curves correspond to different
initial energy densities. Horizontal lines indicate 1$\sigma$
intervals around the measured data.
}
\end{figure}

\begin{figure}
\centerline{\includegraphics[width=0.475\textwidth]{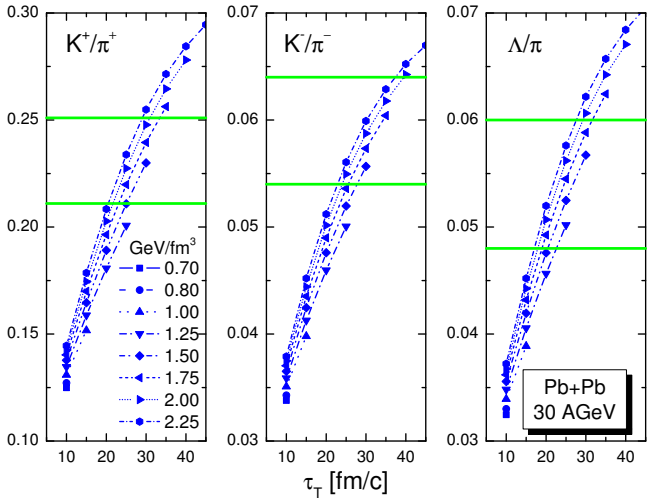}}
\caption{\label{f:r30}
Ratios $\ktopi$, $\kmpi$, and $\ltopi$ as a function of the total
lifetime of the system, calculated for Pb+Pb collisions at
beam energy of 30~$A$GeV. Different curves correspond to different
initial energy densities. Horizontal lines indicate 1$\sigma$
intervals around the measured data.
}
\end{figure}

\begin{figure}
\centerline{\includegraphics[width=0.475\textwidth]{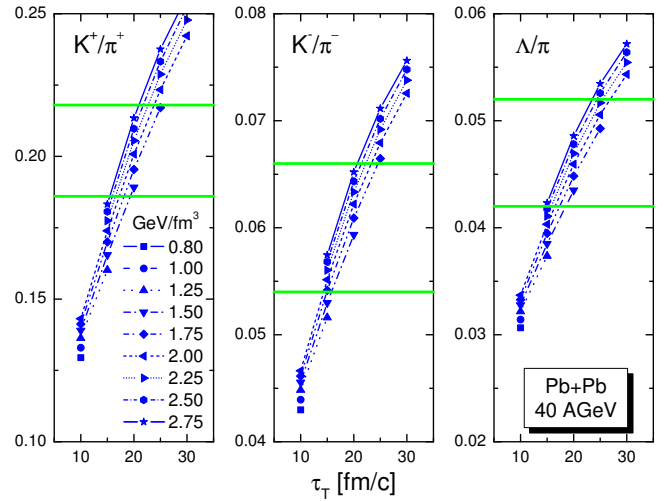}}
\caption{\label{f:r40}
Ratios $\ktopi$, $\kmpi$, and $\ltopi$ as a function of the total
lifetime of the system, calculated for Pb+Pb collisions at
beam energy of 40~$A$GeV. Different curves correspond to different
initial energy densities. Horizontal lines indicate 1$\sigma$
intervals around the measured data.
}
\end{figure}

\begin{figure}
\centerline{\includegraphics[width=0.475\textwidth]{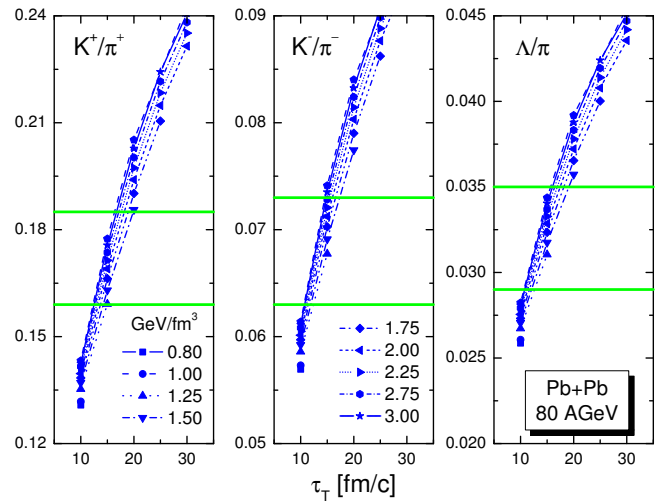}}
\caption{\label{f:r80}
Ratios $\ktopi$, $\kmpi$, and $\ltopi$ as a function of the total
lifetime of the system, calculated for Pb+Pb collisions at
beam energy of 80~$A$GeV. Different curves correspond to different
initial energy densities. Horizontal lines indicate 1$\sigma$
intervals around the measured data.
}
\end{figure}

\begin{figure}
\centerline{\includegraphics[width=0.475\textwidth]{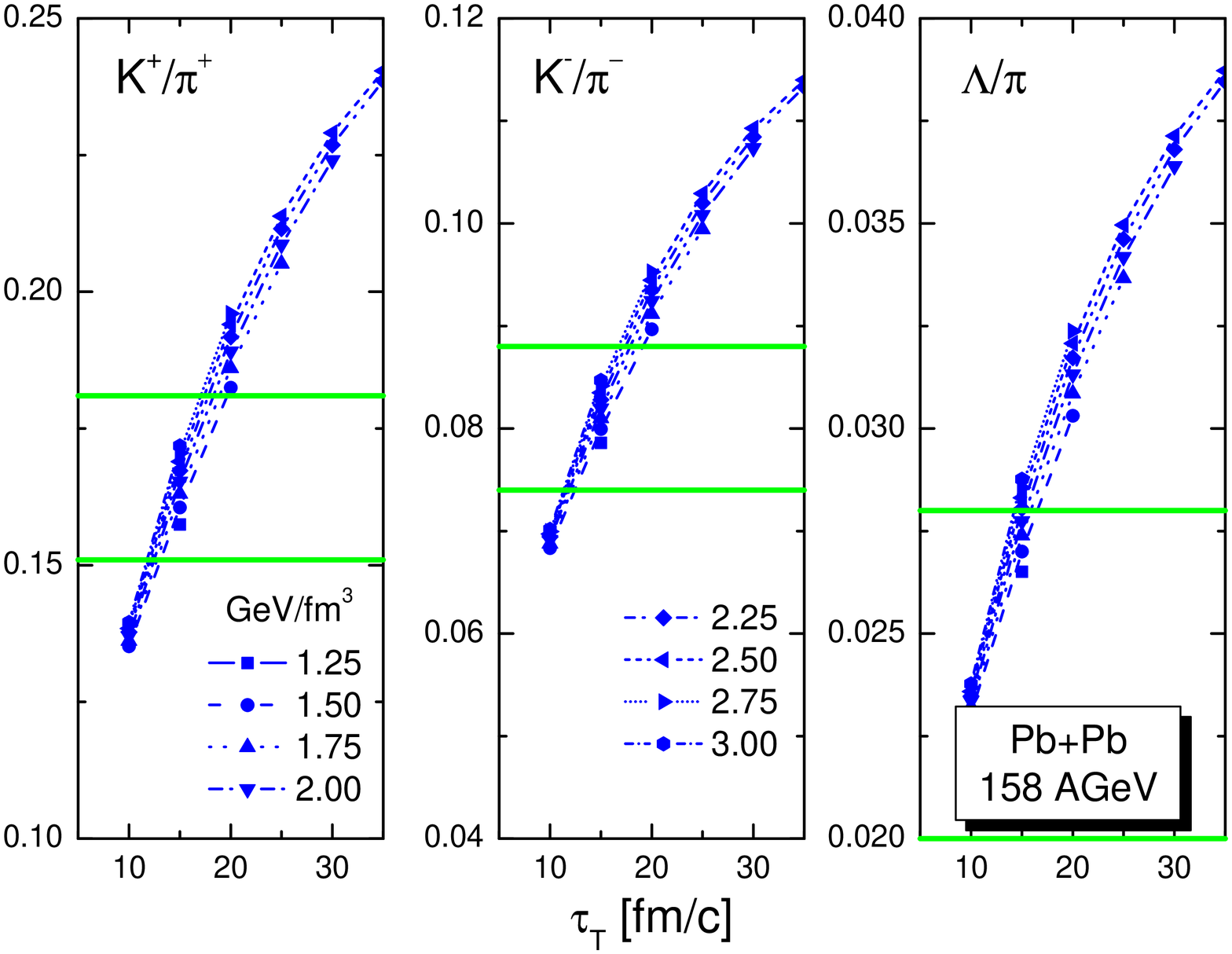}}
\caption{\label{f:r158}
Ratios $\ktopi$, $\kmpi$, and $\ltopi$ as a function of the total
lifetime of the system, calculated for Pb+Pb collisions at
beam energy of 158~$A$GeV. Different curves correspond to different
initial energy densities. Horizontal lines indicate 1$\sigma$
intervals around the measured data.
}
\end{figure}

The results are plotted in Figs.~\ref{f:rAGS}-\ref{f:r158}.
The shown results for energies 30~$A$GeV and upward were
obtained for the highest possible $\rmax$. If scenarios with
slowest possible $\rmax$ are used, the lifetime needed to reach the
same result is prolonged by 1-2 fm/$c$. For the lowest studied energy
we used the lowest $\rmax$ scenarios.

In Figs.~\ref{f:rAGS}-\ref{f:r158} we clearly observe that the resulting multiplicity ratios depend crucially
on the total lifetime of the fireball. Dependence on the initial
energy density is less important. The latter
is more strongly pronounced at lower energies, where the baryon density is higher.
Since the dependence on the energy density is so weak, we do not
expect a major change in the results if parameterization for
time dependence of the energy density was changed.

\begin{figure}
\centerline{\includegraphics[width=0.4\textwidth]{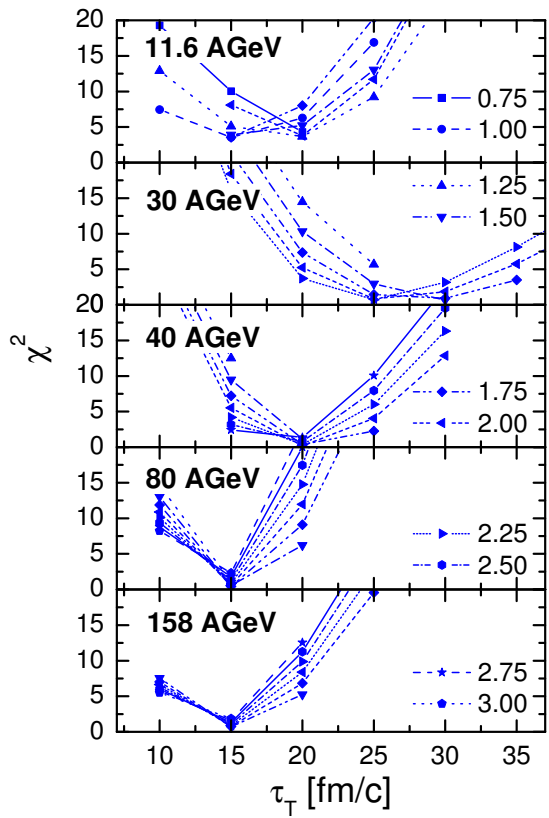}}
\caption{\label{f:sum}
The quality-of-fit measure $\chi^2$, as defined in eq.~\eqref{e:chi2}
as a function of the total lifetime. Different panels correspond to
different beam energies and various curves to different initial energy
densities.
}
\end{figure}

In order to quantify the agreement between our calculations and
experimental data, for  every examined scenario we calculate
the standard $\chi^2$ measure
\begin{equation}
\chi^2 = \sum_i \frac{(\mbox{theory}_i - \mbox{data}_i)^2}{\mbox{error}_i^2}
\, ,
\label{e:chi2}
\end{equation}
where the sum runs over the three measured data points for every beam
energy: $\ktopi$, $\kmpi$, $\ltopi$. The resulting $\chi^2$ values are
summarized in Fig.~\ref{f:sum}.
For beam energies above 30~$A$GeV the best agreement with data is indeed
obtained if the lifetime {\em decreases} as a function of beam energy.
At the highest AGS energy, our model fits data best if the lifetime is
shorter than at 30~$A$GeV. In some sense we have translated the
non-trivial excitation function of $\ktopi$ into the dependence of
the lifetime on the collision energy.
A maximum of the lifetime at 30~$A$GeV {\em could} be an
indicator of some change in evolution dynamics, e.g. a soft point
in the equation of state leading to a low pressure gradient
and a prolonged lifetime.
However, we can not insist on such conclusion here, since
comparison with data cannot clearly exclude that
a lifetime at beam energy 11.6 $A$GeV is as long as at 30~$A$GeV.
A slight change in the dependence of $\tau_T$ on the collision
energy, could be expected if other time evolution parameterizations is used.

\begin{figure*}
\centerline{\includegraphics[width=12cm]{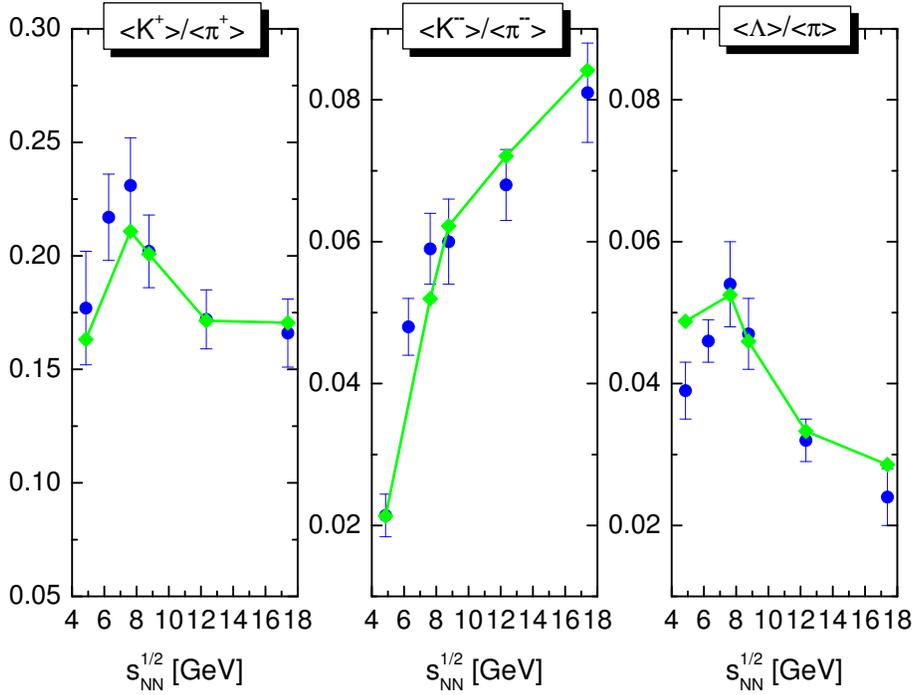}}
\caption{\label{f:cdat} Comparison of our results with data
\cite{alt20,ktpdata}. Calculations were performed with parameters
listed in Tables \ref{fs} and \ref{t:epar}. }
\end{figure*}

In Table~\ref{t:epar} we present the most favorable sets of the
evolution parameters. We choose those sets where the fireball
lifetime does not increase with an increasing beam energy.
Corresponding final densities are given in Table~\ref{fs}.
Our results for the ratios $\ktopi$, $\kmpi$, and $\ltopi$ are
shown in Fig.~\ref{f:cdat}.
\begin{table}
\caption{\label{t:epar}
Parameters of the simulations which lead to results compared to data in
Fig.~\ref{f:cdat}. In the lower two rows we compare the chemical freeze-out
temperature $T$ obtained by Becattini {\em et al.} \cite{becatt} with
the teperature $T_f$ we obtain in our simulations.
}
\begin{tabular}{c|ccccc}
\hline\hline
$E_{\rm beam}$ [$A$GeV] & 11.6 & 30 & 40 & 80 & 158
\\ \hline
$\eden_0$ [GeV/fm$^3$] & 1 & 1.5 & 2 & 2.25 & 2.75
\\
$\tau_T$ [fm/$c$] & 25 & 25 & 20 & 15 & 15
\\
$\rmax$ [fm$^{-1}$] & 0.286 & 0.305 & 0.333 & 0.379 & 0.374
\\
\hline
$T$ [MeV] & 118.1 & 139.0 & 147.6 & 153.7 & 157.8
\\
$T_f$ [MeV] & 114.7 & 134.1 & 143.3 & 149.3 & 153.6
\\ \hline\hline
\end{tabular}
\end{table}
In  Table~\ref{t:epar} we see that the final state temperature
agrees rather well with the results of chemical freeze-out fits
by Becattini {\em et al.} \cite{becatt}.
Note that we did not perform a calculation at the  beam energy 20~$A$GeV
due to lack of final state analysis from
this collision energy in \cite{becatt}.

Table~\ref{t:epar} and Fig.~\ref{f:cdat} are the main results of
our study. In the subsequent subsections we illustrate
the time evolution of various densities and rates.


\subsection{The evolution}

\begin{figure*}
\centerline{\includegraphics[width=5.2cm]{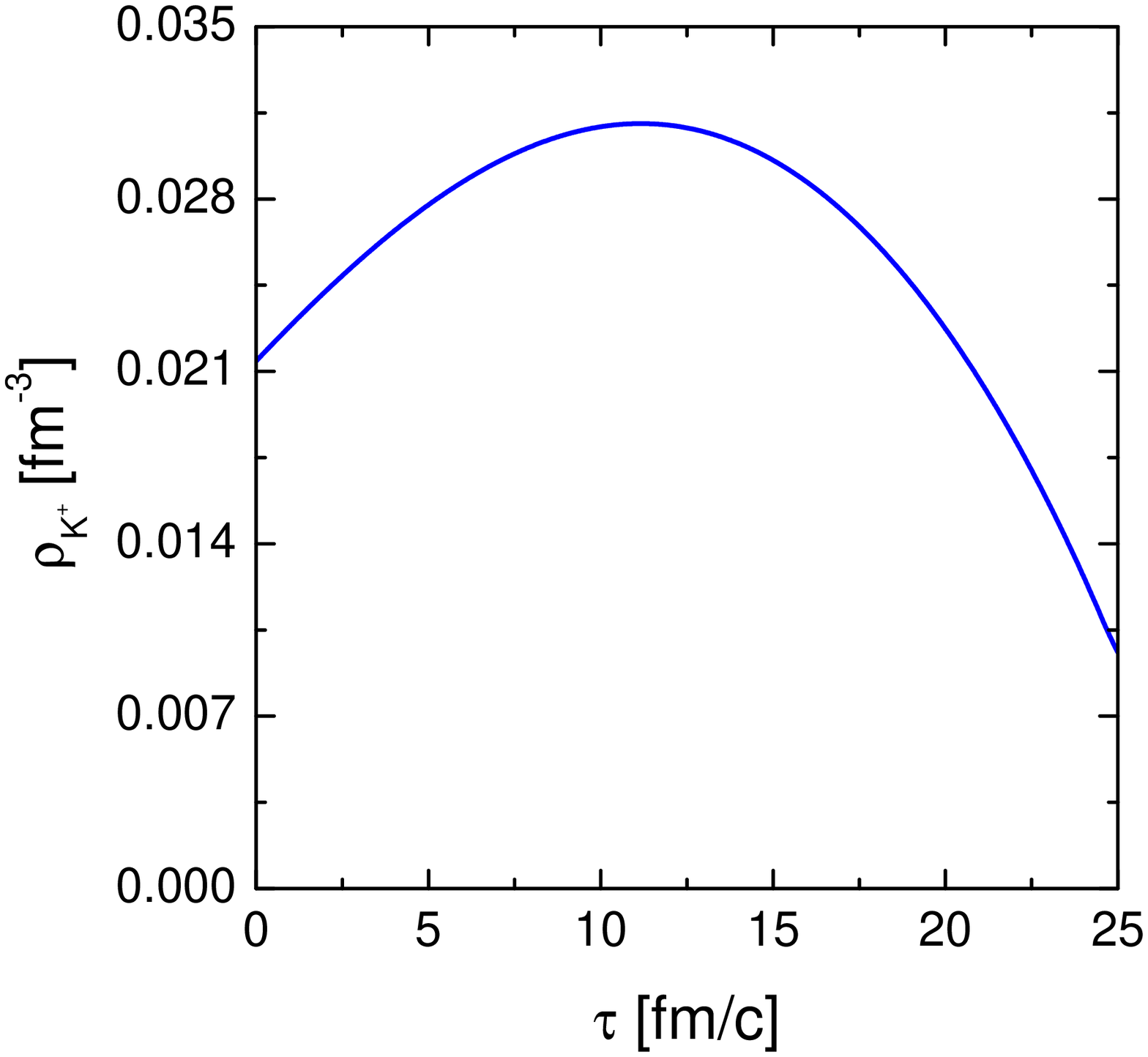}
\includegraphics[width=5.1cm]{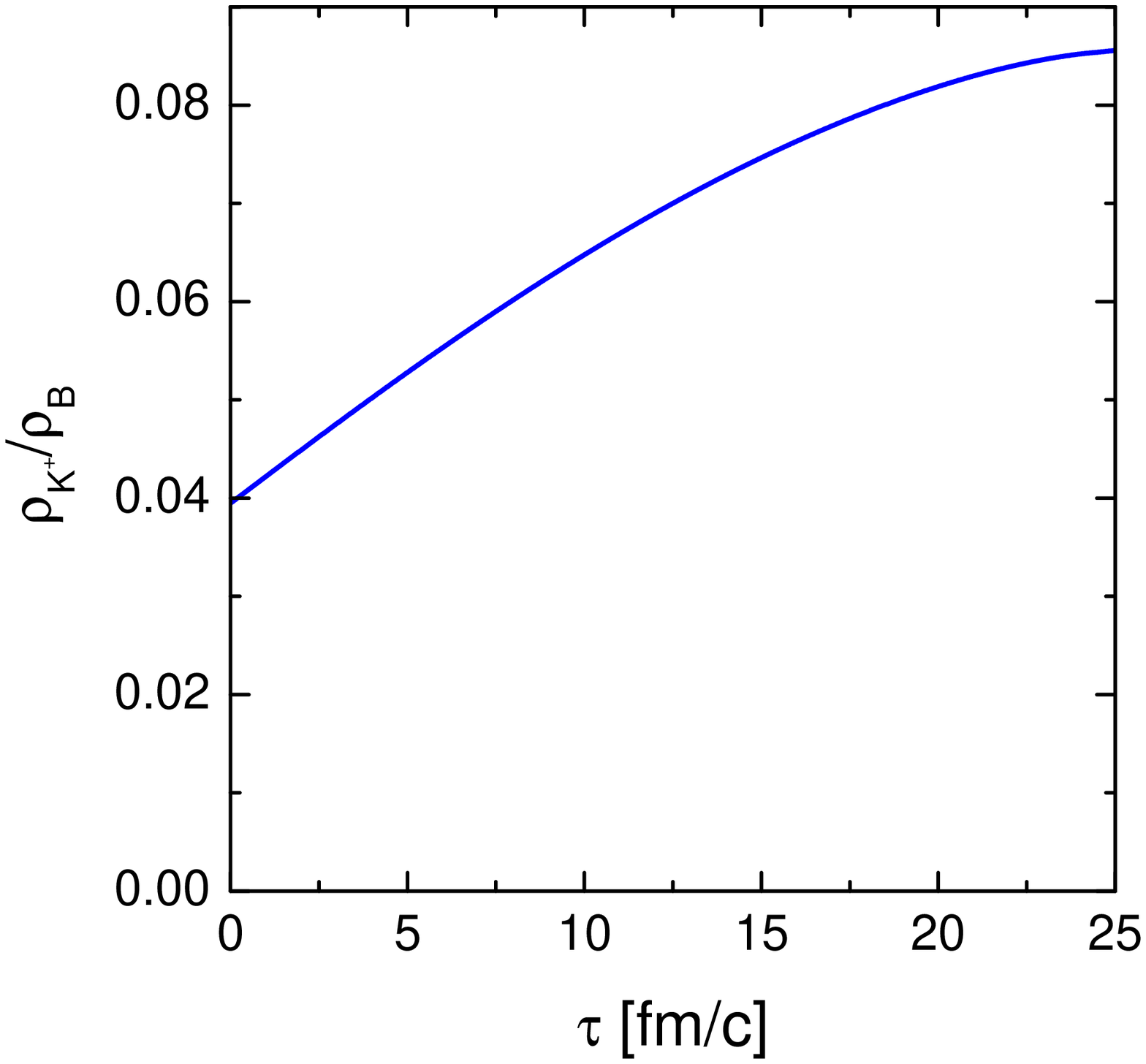}
\includegraphics[width=5cm]{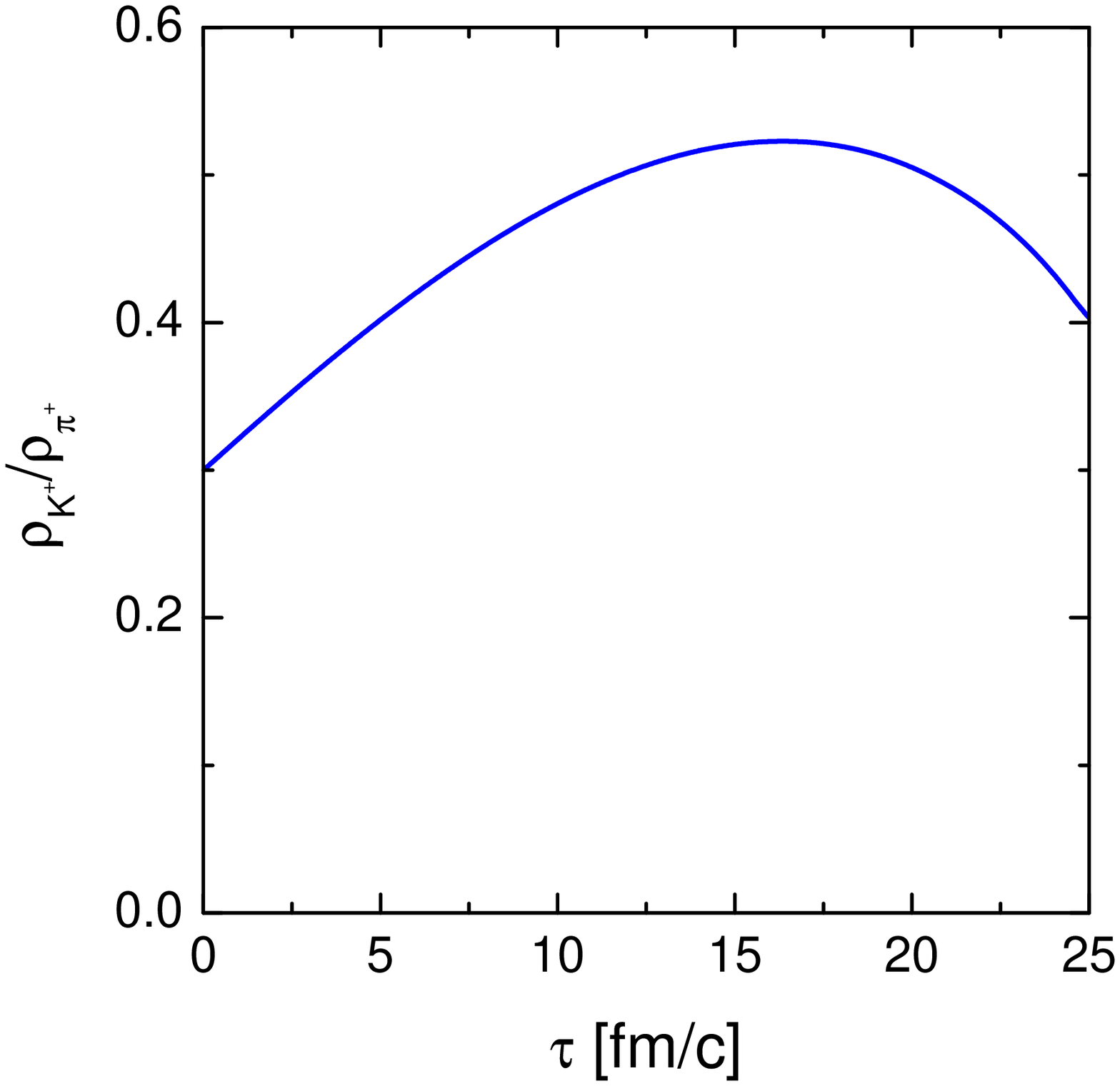}}
\caption{\label{f:kden}
Left panel: evolution of the kaon densities. Middle panel: evolution
of the kaon density divided by the baryon number density. Right panel:
evolution of the ratio $\rho_{K^+}/\rho_{\pi^+}$.}
\end{figure*}

Evolution of kaon density in Pb+Pb collisions at 30~$A$GeV
is illustrated in Fig.~\ref{f:kden} (left panel).
The growth of $\rho_{K^+}$ is soon overrun by the decrease due
to expansion.
To cancel the expansion effect we have to normalize
the kaon density by the baryon density. The latter,
being density of a conserved charge, changes due to expansion only.

The ratio $\rho_{K^+}/\rho_B$ (middle panel in Fig.~\ref{f:kden})
shows a steady increase in time. The ratio
of $\rho_{K^+}/\rho_{\pi^+}$ (right panel in Fig.~\ref{f:kden}) finishes with  rather high value which
does not correspond to the measured $\ktopi$. The final result which does
reproduce the data includes, however, feed-down from resonance decays
which contribute largely to pion production.

\begin{figure}
\parbox{42mm}{\includegraphics[width=4.2cm]{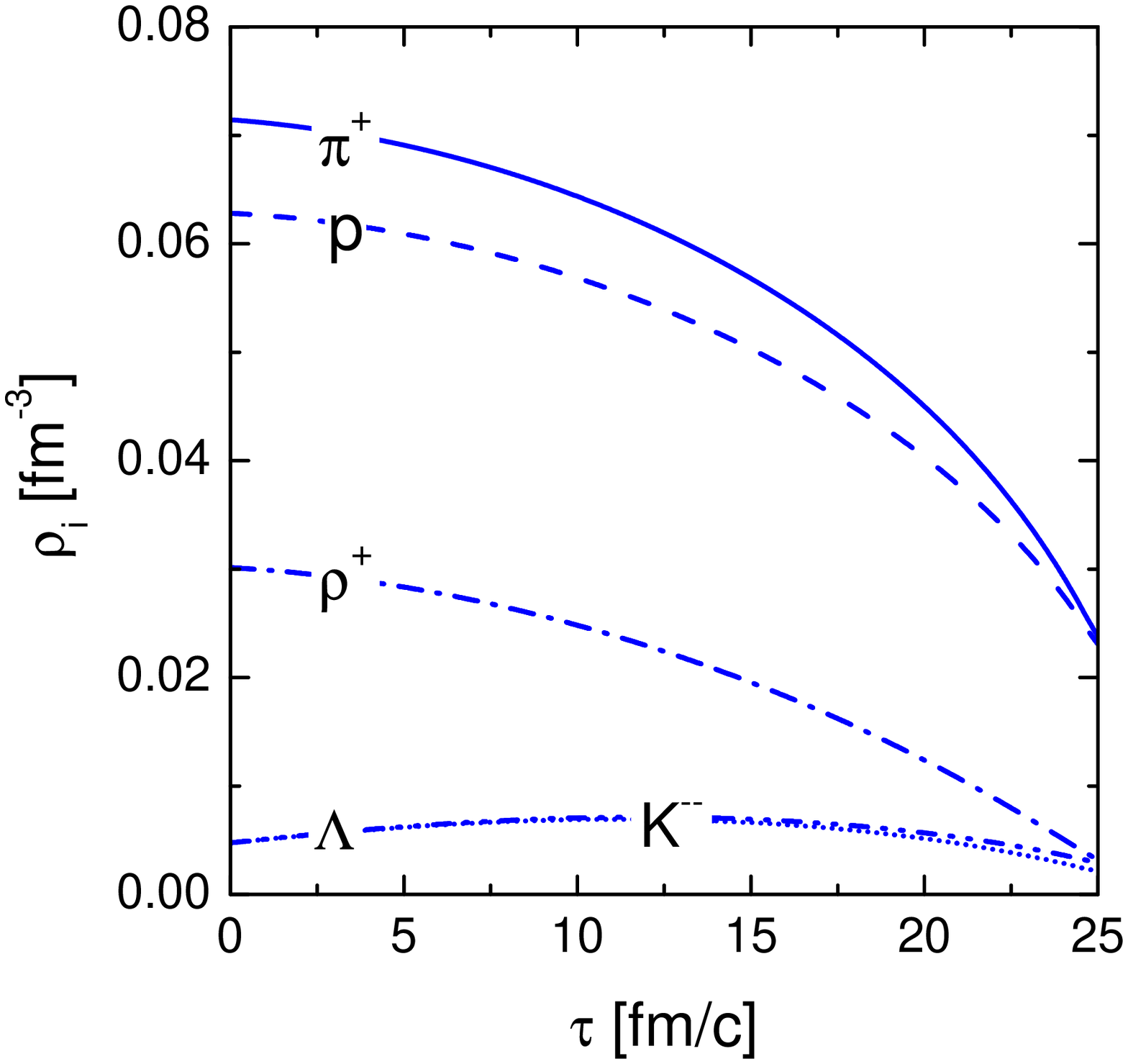}}
\parbox{42mm}{\includegraphics[width=4.2cm]{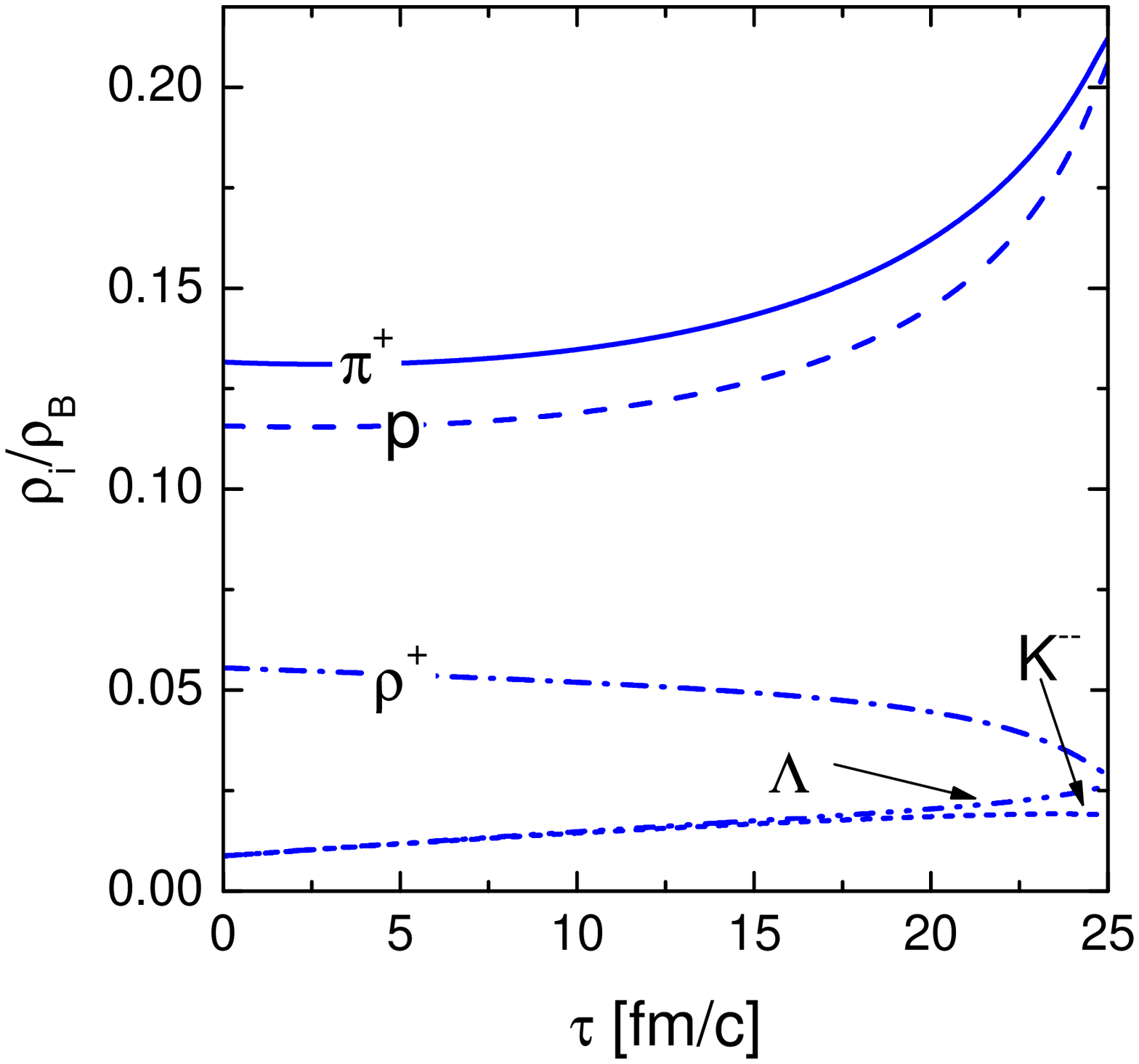}}
\caption{\label{f:oden}
Densities of $\pi^+$, $p$, $\rho^+$, $\Lambda$, and $K^-$ as functions
of time for the scenario which reproduces 30 $A$GeV data. In the lower
panel the densities are normalized to the baryon density.
}
\end{figure}

In Fig.~\ref{f:oden} we show the density evolution for some other species.
When the densities are normalized to baryon density in order to get rid
of the effect of expansion we clearly see that due to cooling the
relative abundance of pions and nucleons as the lightest mesons and baryons
increases with respect to other species.
Normalized $K^-$ and  $\Lambda$ densities rise continuously in
time.

\subsection{The rates}
\label{s:rates}

\begin{figure*}
\centerline{\includegraphics[width=5cm]{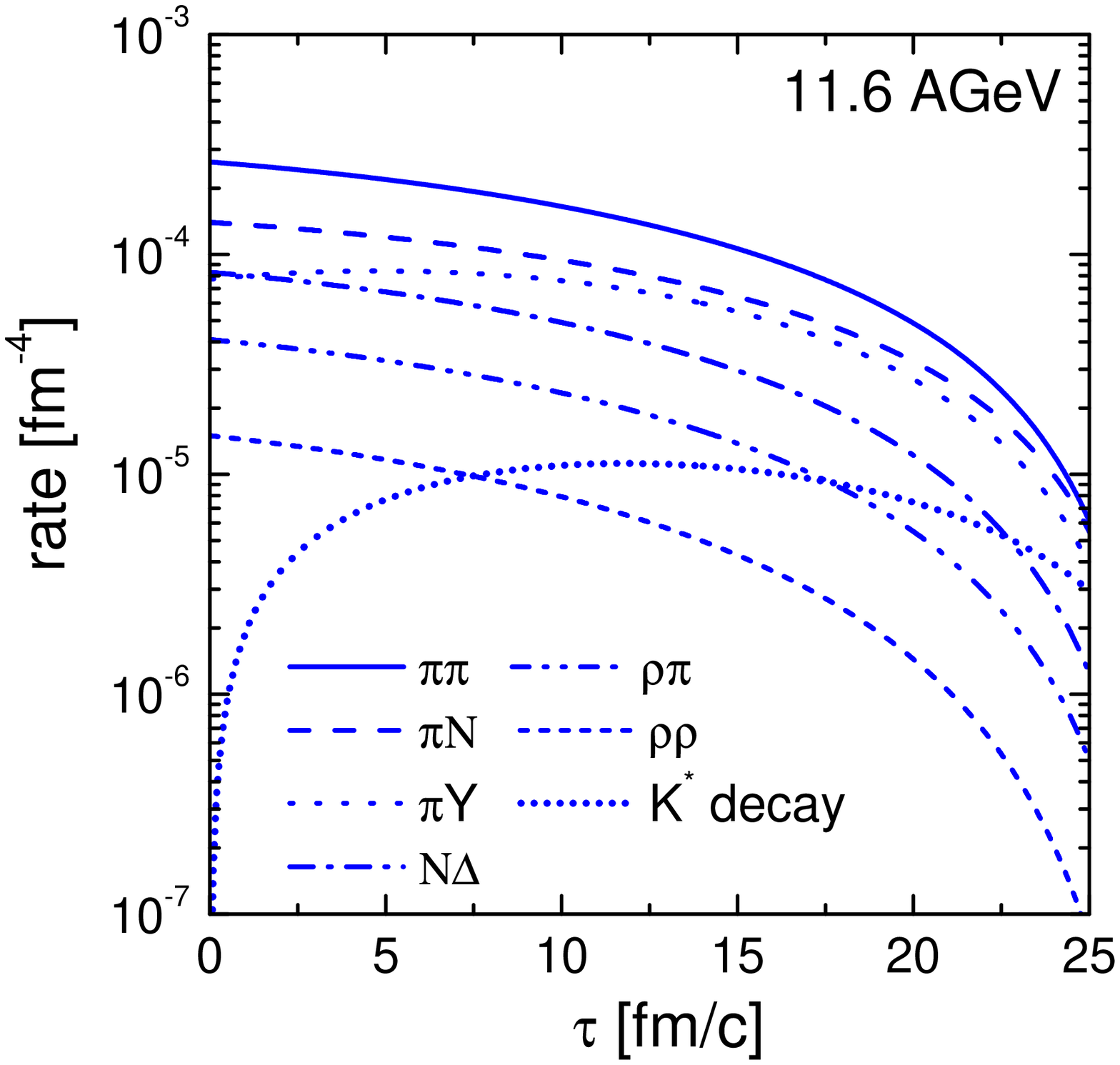}
\includegraphics[width=5cm]{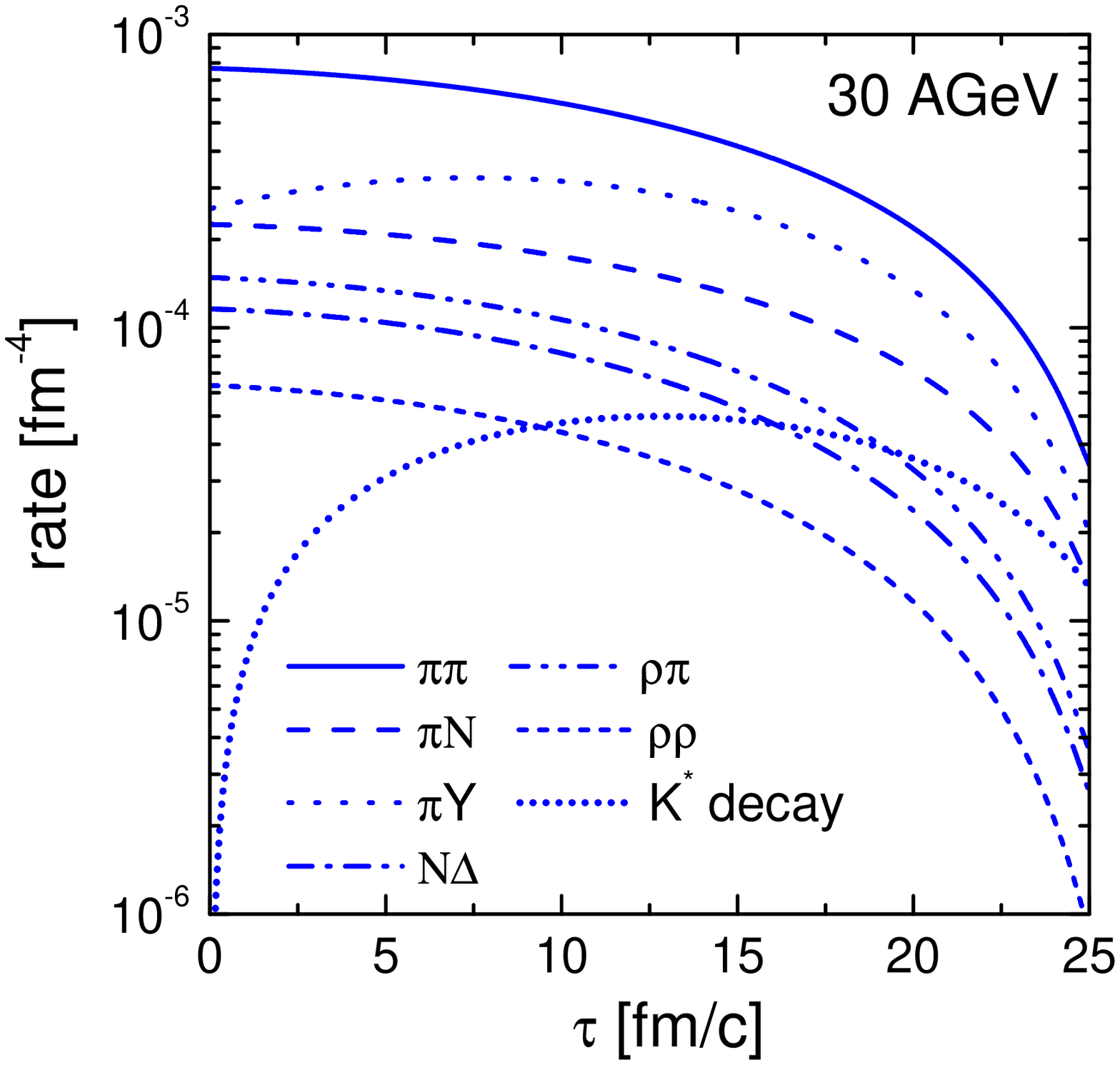}
\includegraphics[width=5cm]{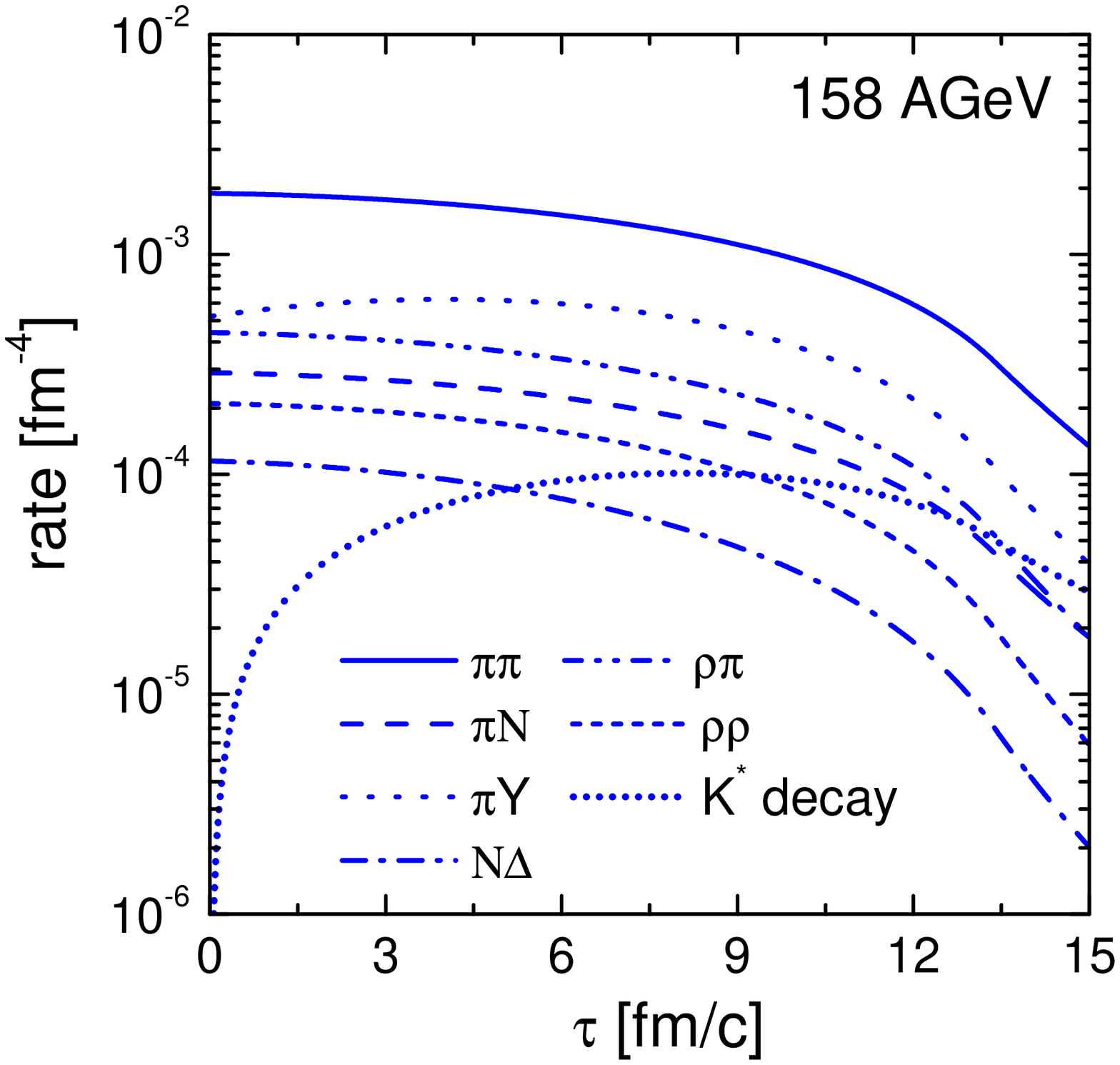}}
\caption{\label{f:allrat}
Production rates of $\Kp$ due to selected processes as a function
of time for three scenarios which fit the data at: 11.6~$A$GeV (left),
30~$A$GeV (middle), and 158~$A$GeV.
}
\end{figure*}

Kaons are produced in various reactions. In Fig.~\ref{f:allrat}
we show the production rates of different reactions as functions
of time.
The three shown examples refer to scenarios which reproduce data
for 11.6, 30, and 158 $A$GeV in Fig.~\ref{f:cdat}.
As expected, due to an increase of the energy density the
production rates grow with increasing beam energy. In all cases, the
dominant contribution is due to $\pi\pi$ reactions. A very important
contribution comes from reactions of pions with hyperons. Recall
that the corresponding cross section is not known experimentally and
we chose a constant matrix element for this reaction. This introduces
some uncertainty into our quantitative results. Note however that
since the contribution from $\pi Y$ reactions has almost always the same
relative importance in comparison to $\pi\pi$ reactions, we do not
expect any {\em qualitative} change of our results under modification
of $\pi Y$ cross sections.

There is a clear change when moving from lower, baryon-dominated, energies
toward higher, meson-dominated energies. While at 11.6~$A$GeV the
second largest contribution comes from $\pi N$ reactions, the contribution
from this type of reactions decreases with increasing beam energy and
actually becomes even lower than $\pi \rho$ contribution at 158~$A$GeV.

\begin{figure*}
\centerline{\includegraphics[width=5cm]{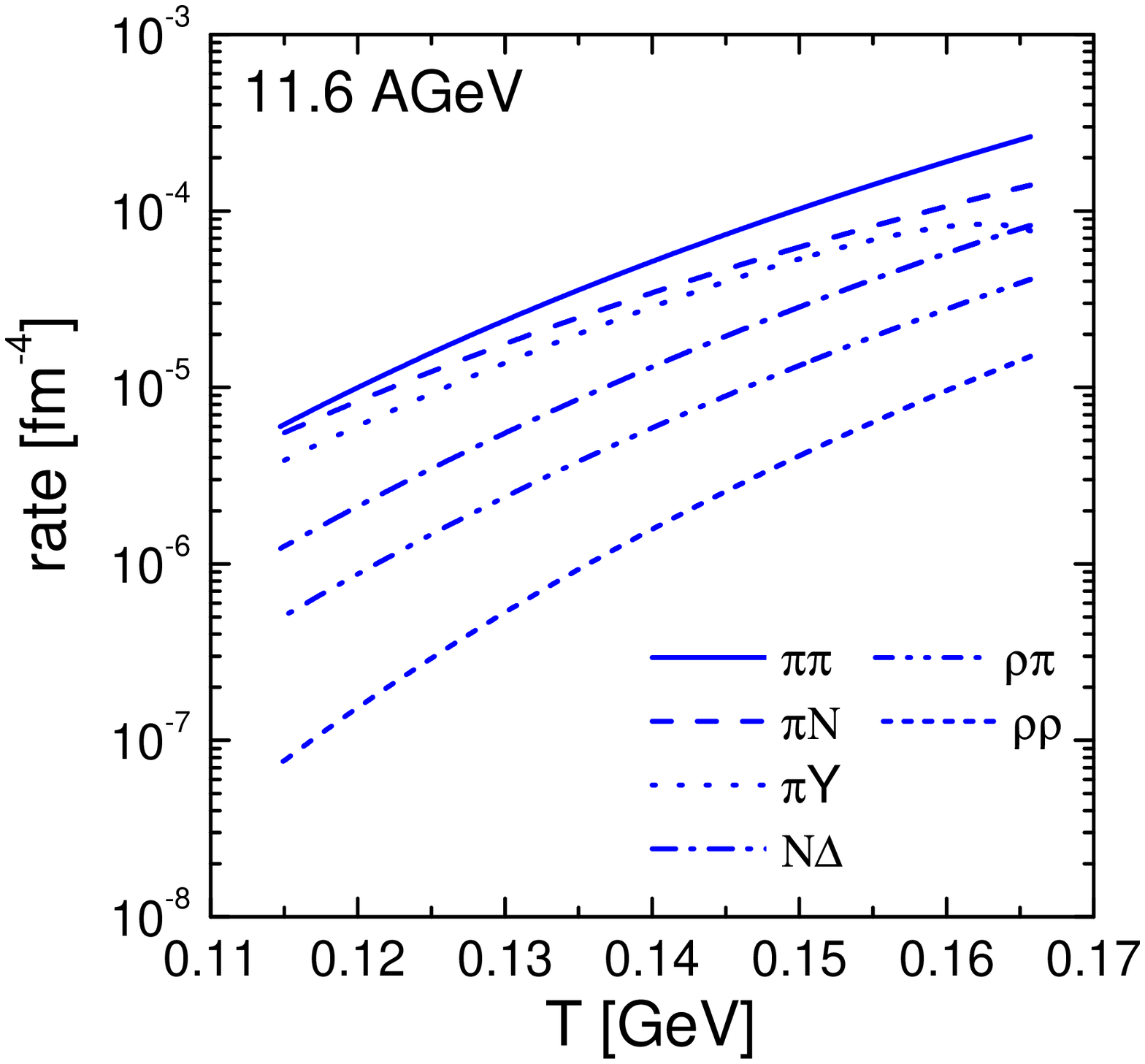}
\includegraphics[width=5cm]{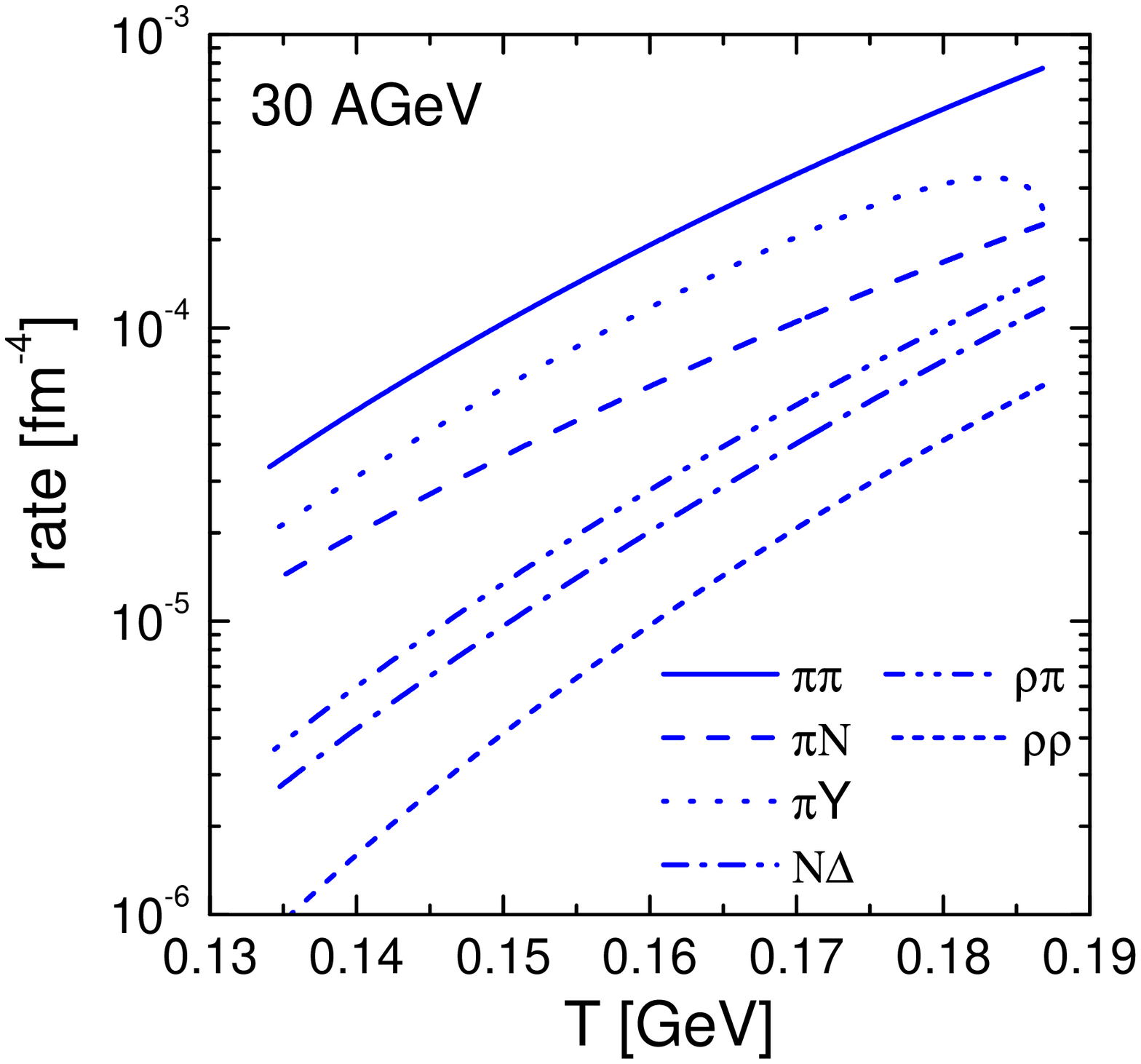}
\includegraphics[width=5cm]{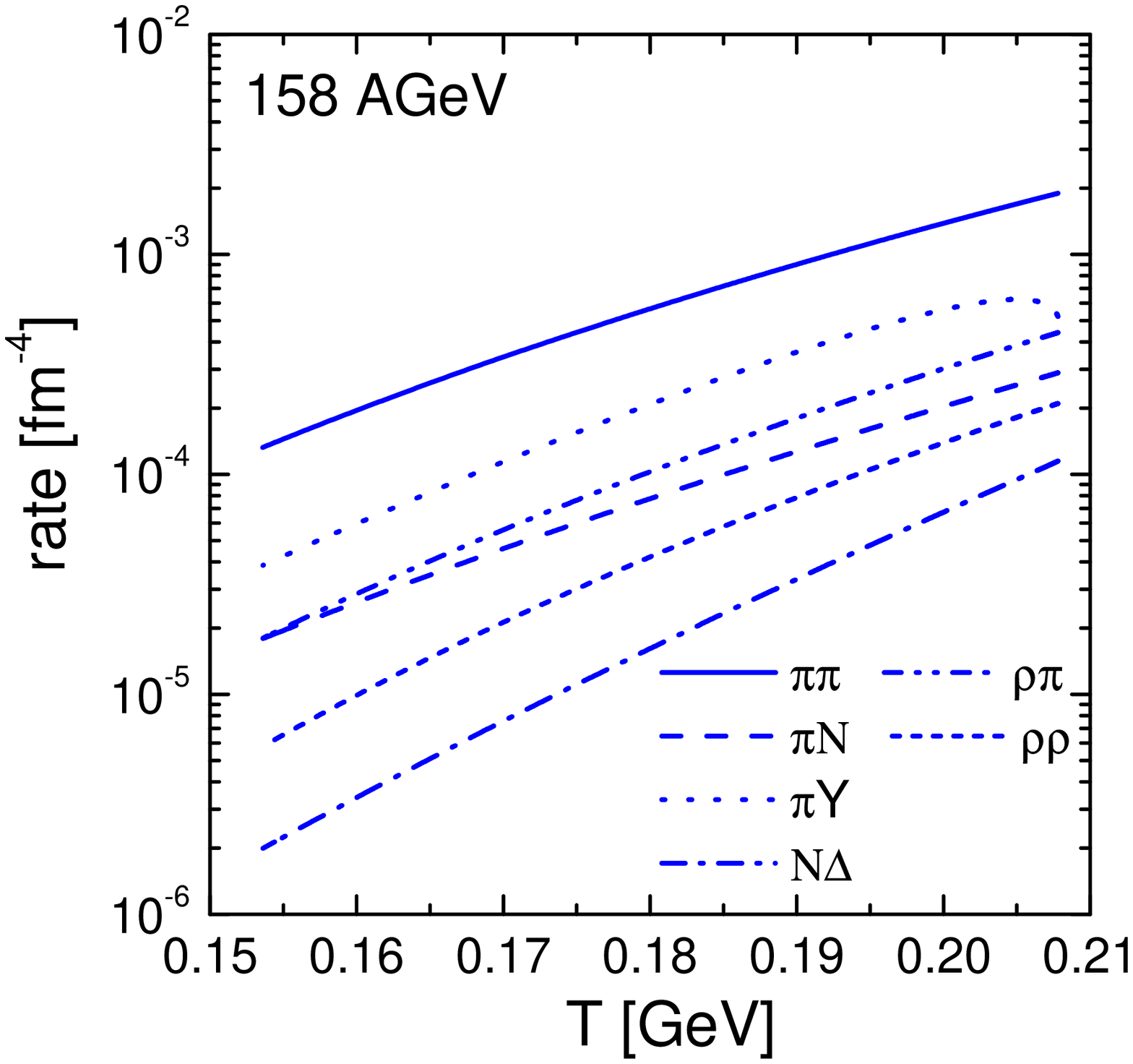}}
\caption{\label{f:trat}
Production rates of $\Kp$ from selected reaction processes as a function
of temperature for three scenarios which fit the data at: 11.6~$A$GeV (left),
30~$A$GeV (middle), and 158~$A$GeV.
}
\end{figure*}

We study the meson-meson production rates together with $\pi N$, $\pi Y$ and $N\Delta$ rate
as a function of temperature in Fig.~\ref{f:trat}.
As mentioned, the difference between the three different panels in
in baryon density. The $\pi \pi$ rate becomes relatively more and
more important. As it is outlined in Appendices~\ref{rhorho} and
\ref{pirho}, we took a great care in determining the $\pi\rho$ and
$\rho\rho$ cross sections. Note that we do not treat all
meson-meson cross section as being the same, in contrast to
\cite{HSD,mosel}. The resulting suppression in $\pi\rho$ channel
is due to $p$-wave suppression of the reaction $\pi\rho\to \phi\to
K \bar K$ and a higher threshold in the channel $\pi\rho \to K
\bar K^*$. The later cross section was determined from the
scattering amplitudes following from the covariant solution of the
Bethe-Salpeter equation~\cite{LK} with the  leading chiral
order Weinberg-Tomozawa term as an interaction kernel. The
untarization effects are found to be important leading to a
smaller cross section than what would be inferred from $\pi\pi$
cross section. The $\rho\rho$ rate is suppressed thermally
due to large mass of the $\rho$ mesons, and the resulting contribution
is very low.



\section{Conclusions}
\label{conc}

In the introduction we motivated our study by the question whether
it is possible to rule out any hadronic interpretation of the
observed excitation function of $\ktopi$, $\kmpi$, and $\ltopi$
ratios. We proposed a non-equilibrium hadronic model which is able
to reproduce the data by virtue of varying the total lifespan of a
fireball at different energies.

Thus one has to do more in trying to exclude this kind of models.
Another indications of non-trivial effects in the region of the
$\ktopi$ peak is a step in kaon spectra inverse slopes, and a
change in the pion yield per participant. The model must be tested
on these observables, as well.

Our findings are based on the assumption that the lifetime of a
fireball becomes shorter as one moves to higher beam energies. We
want to stress that most of the information we have about the
space-time evolution of a fireball comes from {\em hadronic}
spectra and correlations, which are formed {\em at the freeze-out}
stage of fireball evolution. This is just the final state to which
many possible evolution scenarios may lead. One of the future
tasks should be a careful check how hadronic observables
calculated within our model fit the data. This is rather involved
project. Although we were lead by the data in formulating our
model, we refrained from such detailed and careful fits here.

The whole duration  of fireball evolution is reflected in
penetrating probes like dilepton spectra. Their measurement was
improved dramatically in the last years~\cite{ceres,ceres40,na60}.
It seems to be crucial, therefore, to check our model assumption
about the lifetime against the dilepton spectra.

Our approach addressed here only a part of the whole problem by
asking how do different fireball evolution scenarios influence the
data. We ignored the question how a specific evolution scenario
follows from the microscopic properties of the created matter?
This question must be, of course, addressed in the future if the
model survives all experimental tests. On the other hand, if we
succeed to exclude the model on a set of data, there will be no
need to answer this question. This approach might be simpler to
begin with.

\acknowledgments
We appreciate stimulating discussions with
members of theory groups at University of Frankfurt and Giessen
and at GSI, Darmstadt. We thank Francesco Becattini for sharing
with us the values of chemical potential connected with electric
charge which are not listed in \cite{becatt}. The work of BT was
supported by the Marie Curie Intra-European Fellowship within the
6th European Community Framework Programme. The work of EEK was
supported by the US Department of Energy under contract No.
DE-FG02-87ER40328.

%
%

\appendix

\section{Parameters of the time dependence ansatz}
\label{param}

Here we express the parameters of eqs.~\eqref{td} in terms of
$\eden_0$, $\tau_T$, $\eden_f$, $\rho_{Bf}$, $\rho_{3f}$, $\tau_0$,
$\alpha$, and $\rmax$, which were introduced in Section  \ref{expdyn}.

The easiest to obtain are $\eden_0^\prime$, $\rho_{B0}^\prime$,
$\rho_{30}^\prime$
\begin{subequations}
\begin{eqnarray}
\eden_0^\prime & = & \eden_f \, (\tau_T - \tau_0)^{\alpha/\delta} \\
\rho_{B0}^\prime & = & \rho_{Bf}\, (\tau_T - \tau_0)^{\alpha} \\
\rho_{30}^\prime & = & \rho_{3f}\, (\tau_T - \tau_0)^{\alpha}
\end{eqnarray}
\end{subequations}
Since the expansion rate at time $\tau_s$ must be $\rmax$, from
eq.~\eqref{fer} we see
\begin{equation}
\tau_s = \frac{\alpha}{\rmax} + \tau_0\, .
\end{equation}
Parameters $a$ and $b$ are obtained from matching at time $\tau_s$
\begin{subequations}
\begin{eqnarray}
\eden_0 (1 - a\tau_s - b\tau_s^2) & = &
\frac{\eden_0^\prime}{(\tau_s - \tau_0)^{\alpha/\delta}} \\
\eden_0 (1 + 2b\tau_s) & = &
\frac{\alpha\, \eden_0^\prime}{\delta (\tau_s - \tau_0 )^{\alpha/\delta -1}}
\end{eqnarray}
\end{subequations}
which lead to
\begin{subequations}
\begin{eqnarray}
\nonumber
a & = & \frac{2}{\tau_s} \\
& & - \frac{\eden_f}{\eden_0}\,
\left ( \frac{2}{\tau_s} +
\frac{\alpha}{\delta (\tau_s -\tau_0)} \right )\,
\left ( \frac{\tau_T - \tau_0}{\tau_s-\tau_0}\right )^{\alpha/\delta}\\
\nonumber
b & = & \frac{\eden_f}{\eden_0}\, \left ( \frac{1}{\tau_s^2} +
\frac{\alpha}{\delta \tau_s (\tau_s - \tau_0)}\right )
\left ( \frac{\tau_T - \tau_0}{\tau_s - \tau_0}\right )^{\alpha/\delta}\\
& &
- \frac{1}{\tau_s^2}\, .
\end{eqnarray}
\end{subequations}

For $B$ and $I_3$ densities we obtain from matching at $\tau_s$
\begin{subequations}
\begin{eqnarray}
\rho_{B0} & = & \frac{\rho_{Bf}}{(1-a\tau_s - b \tau_s^2)^\delta}\,
\left ( \frac{\tau_T - \tau_0}{\tau_s - \tau_0} \right )^\alpha \\
\rho_{30} & = & \frac{\rho_{3f}}{(1-a\tau_s - b \tau_s^2)^\delta}\,
\left ( \frac{\tau_T - \tau_0}{\tau_s - \tau_0} \right )^\alpha \, .
\end{eqnarray}
\end{subequations}

In Table~\ref{t:evpar} we present the values of parameters
$a$, $b$  and $\tau_s$ corresponding to the energy density,
fireball lifetime and $\mathcal{R}_{\rm max}$ given in
Table~\ref{t:epar}

\begin{table}
\caption{\label{t:evpar}
Parameters of the time dependence~\eqref{td} corresponding to the energy density,
fireball lifetime and $\mathcal{R}_{\rm max}$ given in
Table~\ref{t:epar}.
}
\begin{tabular}{c|ccccc}
\hline\hline
$E_{\rm beam}$ [$A$GeV] & 11.6 & 30 & 40 & 80 & 158
\\ \hline
$a\times 10^{4}$ [fm$^{-1}$] & $142.6$ & $5.482$ & $4.963$
& $10.77$ & $8.448\cdot 10^{-2}$
\\
$b\times 10^{4}$ [fm$^{-2}$] & $7.752$ & $13.29$ & $21.55$
& $41.60$ & $44.70$
\\
$\tau_s$ [fm/$c$] &25.00  & 24.56  & 19.01 & 13.28  & 12.98  \\
\hline\hline
\end{tabular}
\end{table}


\section{Cross sections for kaon production}
\label{xsec}

In this section we specify the cross sections of all reactions
with strangeness production which are relevant at the collision
energies under consideration.
We disregard a possible modification of cross sections in dense
and hot nuclear matter and accept their vacuum forms.
This approximation complies with our assumption above that
properties of all particles do not change in medium either.

For processes with two particles in the final state the inverse
reactions are included in the calculation, as well.
The cross section for the inverse
reactions follows from phase-space considerations as
\be
\sigma_{34\to 12}(\sqrt{s}) &=& \frac{(2J_3+1)(2J_4+1)}{(2J_1+1)(2J_2+1)}\,
\frac{p^2_{\rm cm} (s,m_1,m_2)}{%
p^2_{\rm cm} (s,m_3,m_4)}\,
\nonumber\\
&\times&
\sigma_{12\to 34}(\sqrt{s})\, ,
\label{ixs}
\ee
where $J_i$ and $m_i$ are spins and masses of the participating
species,
and $p_{\rm cm}$ is the center-of-mass momentum defined as
\be
p^2_{\rm cm} (s,m_a,m_b)&=&{\frac{[s - (m_1^2 + m_2^2)]^2 - 4\, m_1^2 \, m_2^2}{
4\,s}}\,.
\label{pcm}
\ee

In all parameterizations below, energy units will be GeV and cross sections
are measured in millibarns.


\begin{widetext}
\subsection{Reactions of $\pi + N$}
\label{piN}

Cross sections for $\pi  N \to \Kp  \La$ are all related
by isospin symmetry \cite{kujon} to $\pin n \to K^0 \La$,
which we take from \cite{tsau}
\be
2\si(\pin  p \to \Kp  \La)  &=&
\si(\pip  n \to \Kp  \La)  =  \si(\pim p \to K^0 \La)=  2\si(\pin n \to \Kn \La)
=\frac{0.007665 (\sqs - 1.613)^{0.1341}}{(\sqs - 1.720)^2 +
0.007826}\,.
\label{i1}
\ee
In Fig.~\ref{fig:pin2yk} we compare this parameterization of the
$\pip p \to K^0 \La$ reaction with the available experimental data.

\begin{figure*}
\parbox{5cm}{\includegraphics[width=5cm]{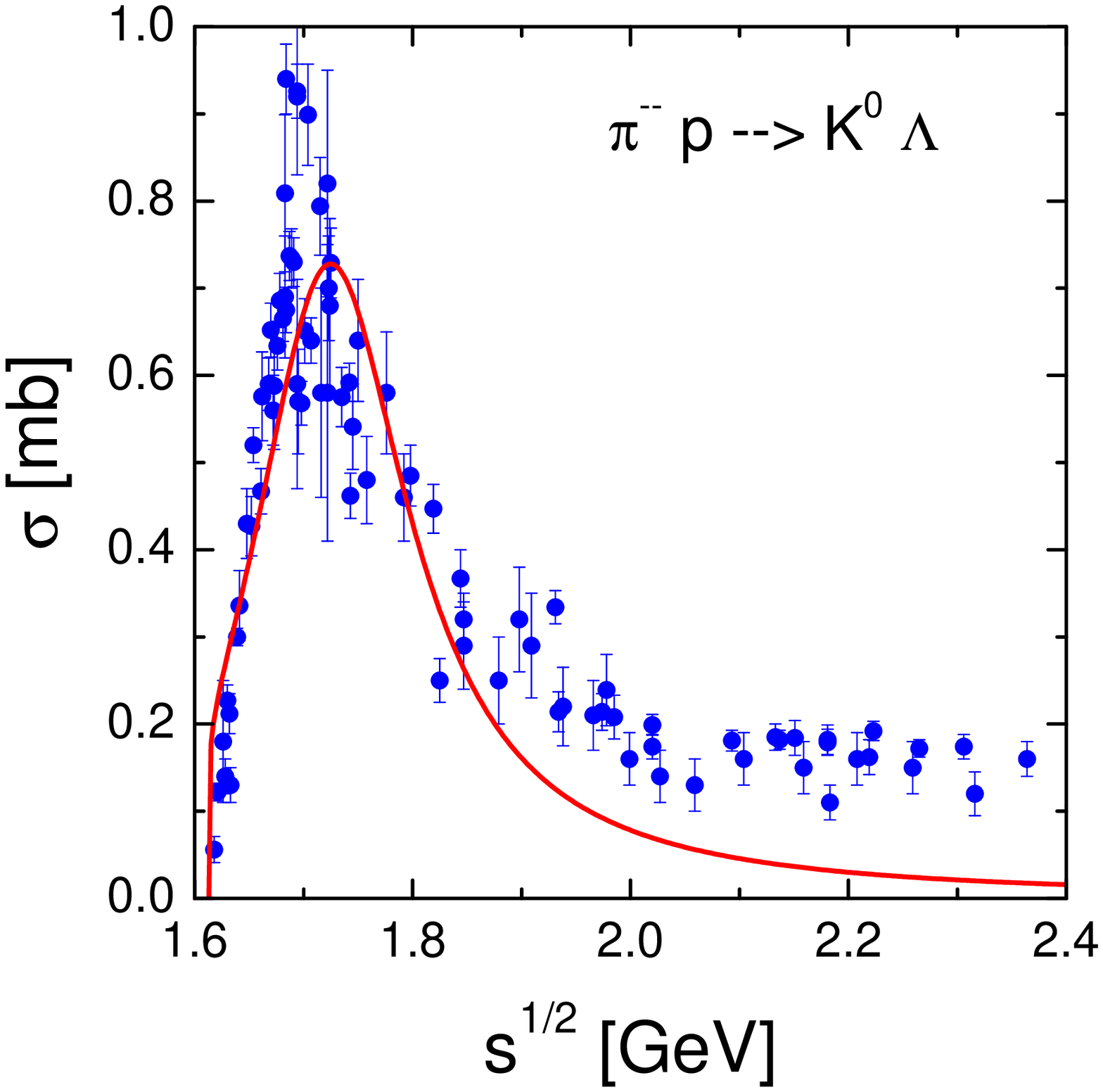}}\,
\parbox{5cm}{\includegraphics[width=5.1cm]{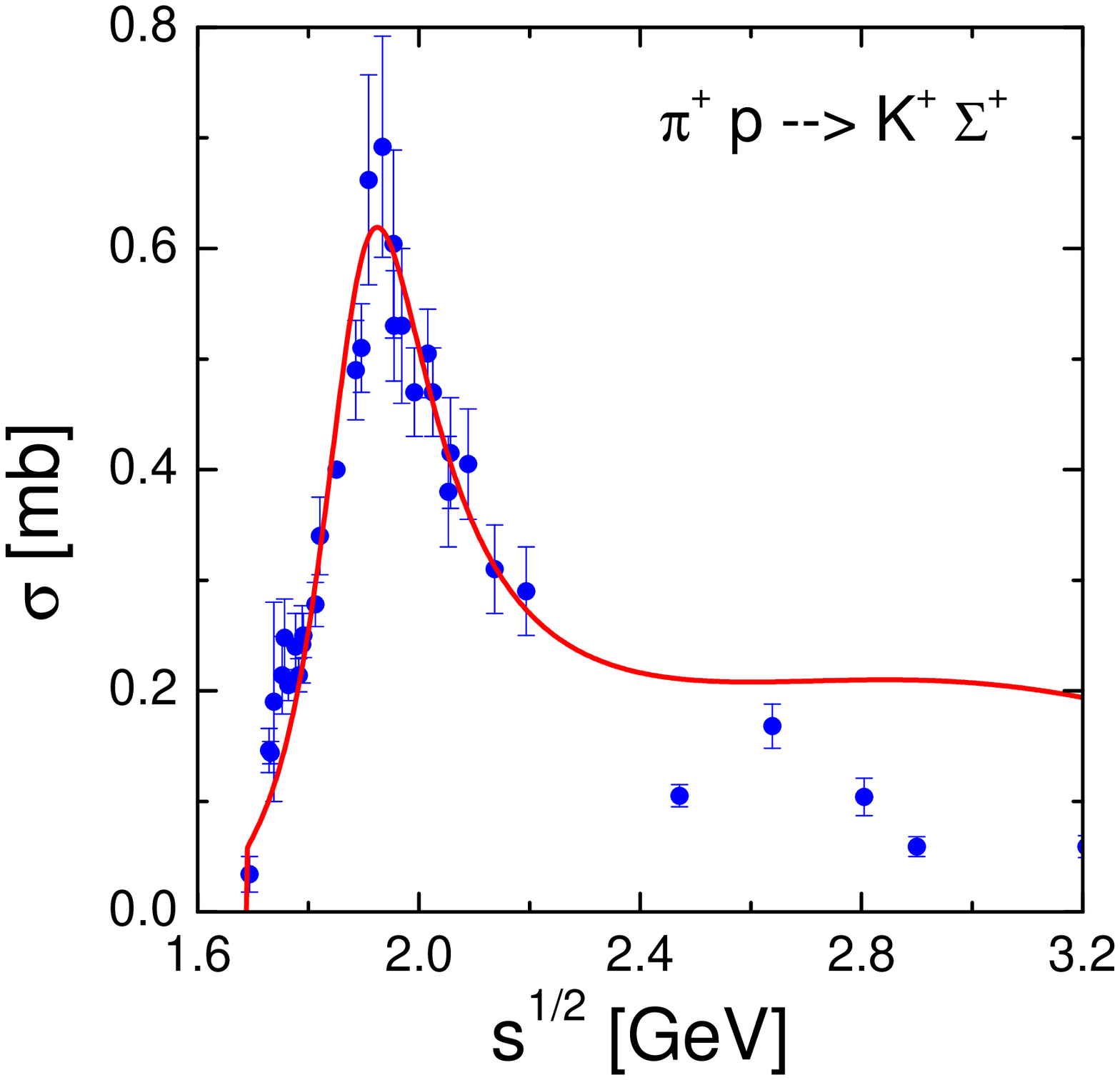}}\,
\parbox{5cm}{\includegraphics[width=5cm]{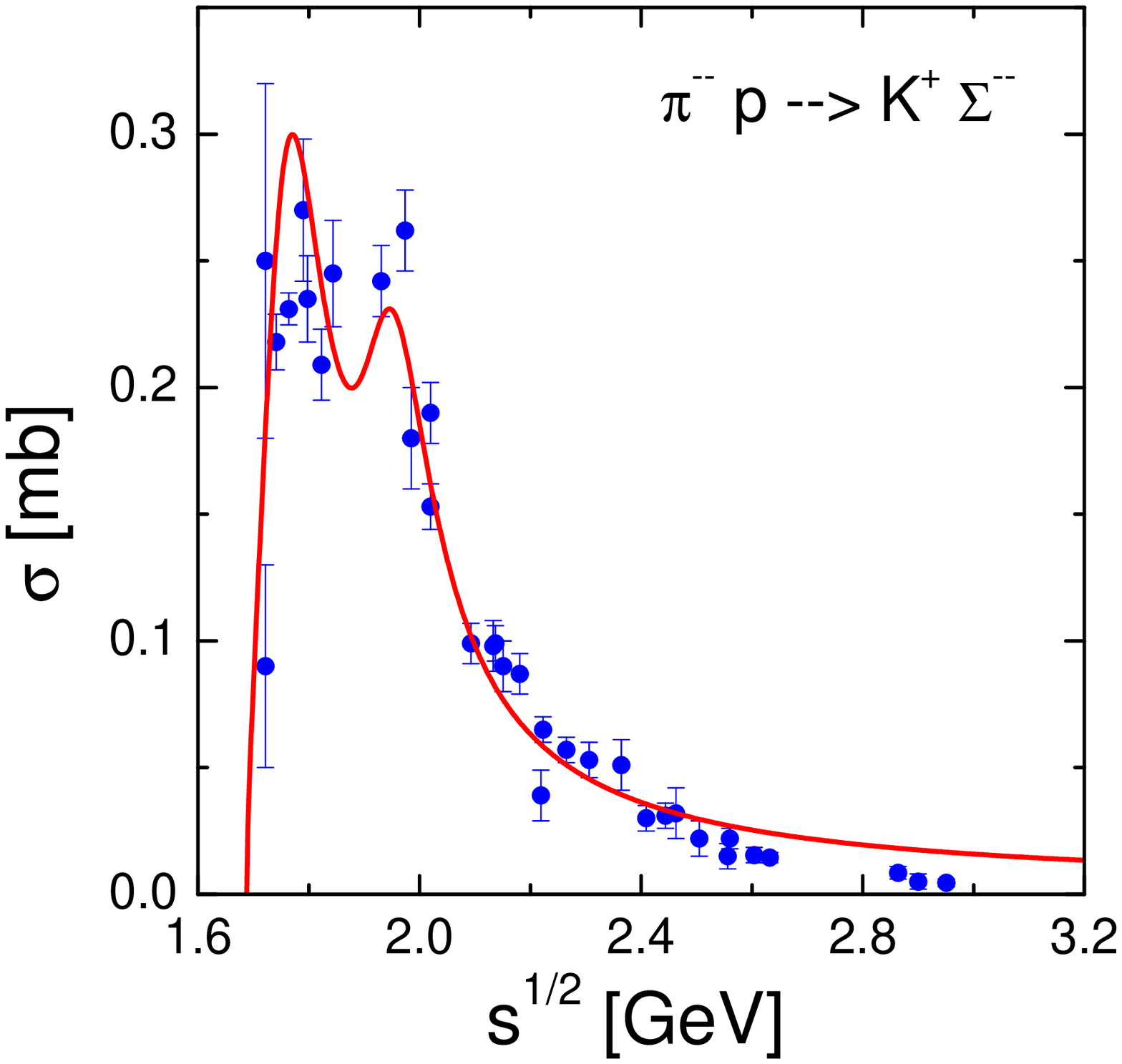}}
\caption{Total cross section of the $\pip p \to K^0 \La$,
$\pip p \to K^+ \Sigma^+$, and $\pim p \to K^+ \Sigma^-$ reactions: solid line
 calculated with eq.~(\ref{i1}), dots are the experimental data from Ref.~\cite{expdata} }
\label{fig:pin2yk}
\end{figure*}
>From \cite{tsu} we get the $\pi N \to \Sigma \Kp$ cross sections
\be
\label{i3}
\si(\pip  p \to \Kp  \Sp) =\, \si(\pim n \to \Kn \Sm)  &=&
\frac{0.03591(\sqs - 1.688)^{0.9541}}{(\sqs - 1.890)^2 + 0.01548} +
\frac{0.1594(\sqs - 1.688)^{0.01056}}{(\sqs - 3.000)^2 +
0.9412}\,,
\\ \label{i4}
\si(\,\pin  p \to \Kp  \Sn\,) =\, \si(\,\pin n \to \Kn \,\Sn\,)  &=&
\frac{0.003978(\sqs-1.688)^{0.5848}}{(\sqs - 1.740)^2 + 0.00667} +
\frac{0.04709 (\sqs-1.688)^{2.1650}}{(\sqs - 1.905)^2 +
0.006358}\,,
\\ \label{i6}
\si(\pim  p \to \Kp  \Sm) =\, \si(\pip n \to \Kn \Sp)  &=&
\frac{0.009803(\sqs - 1.688)^{0.6021}}{(\sqs - 1.742)^2 + 0.006583} +
\frac{0.006521(\sqs - 1.688)^{1.4728}}{(\sqs - 1.940)^2 +
0.006248}\,,
\ee
\begin{equation}
\label{i5}
\si(\pip   n \to \Kp  \Sn) =
\si(\pim p \to \Kn \Sn) =\si(\pin  n \to \Kp  \Sm) =
\si(\pin p \to \Kn \Sp) =
\frac{0.05014(\sqs - 1.688)^{1.2878}}{(\sqs - 1.730)^2 +
0.006455}\,.
\end{equation}
\end{widetext}
The quality of these parameterizations is demonstrated in
Fig.~\ref{fig:pin2yk} for $\pip p \to K^+ \Sigma^+$, and $\pim p \to K^+ \Sigma^-$
reactions.

\begin{figure*}
\parbox{4.1cm}{\includegraphics[width=4.1cm]{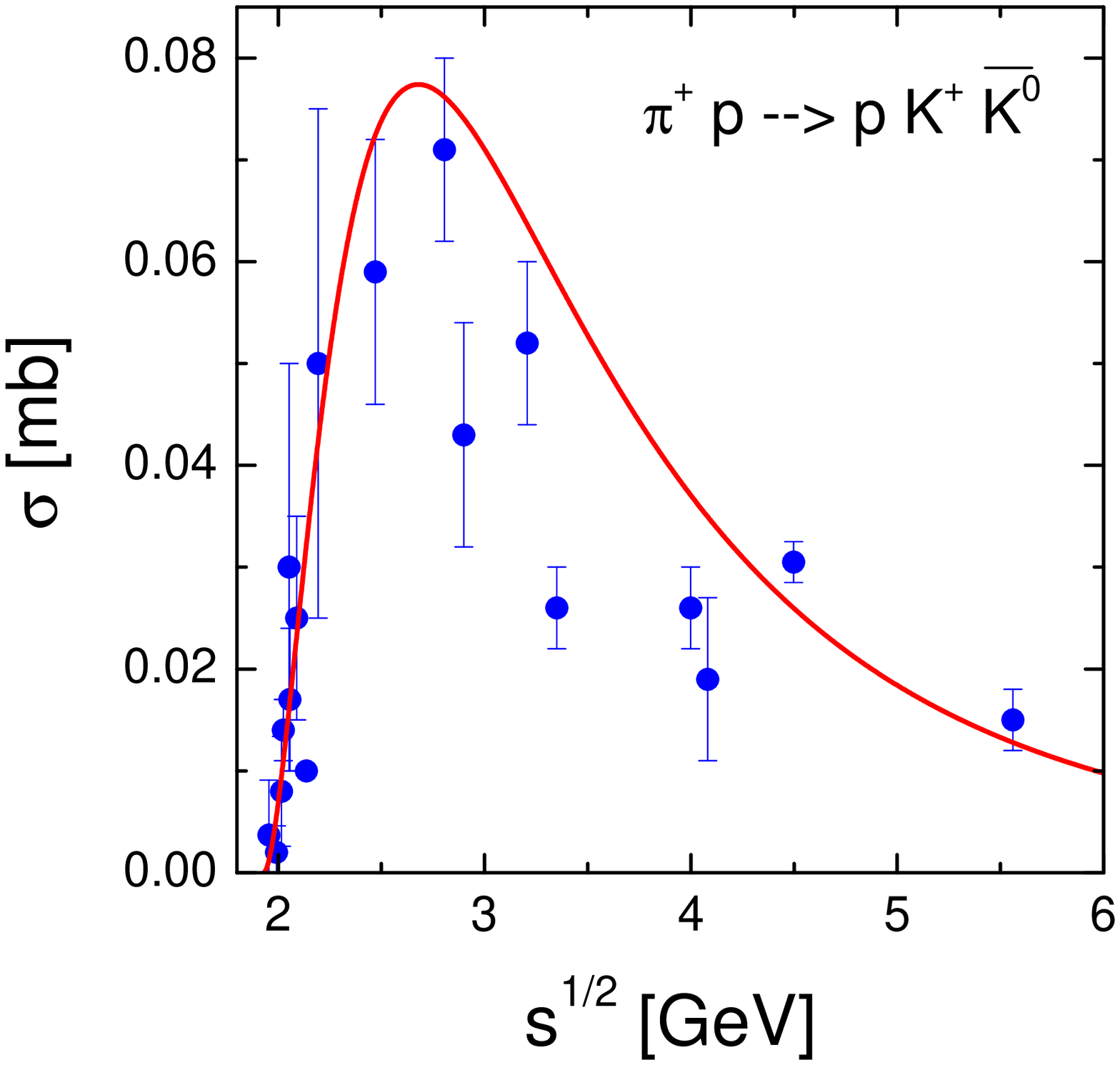}}\,
\parbox{4cm}{\includegraphics[width=4cm]{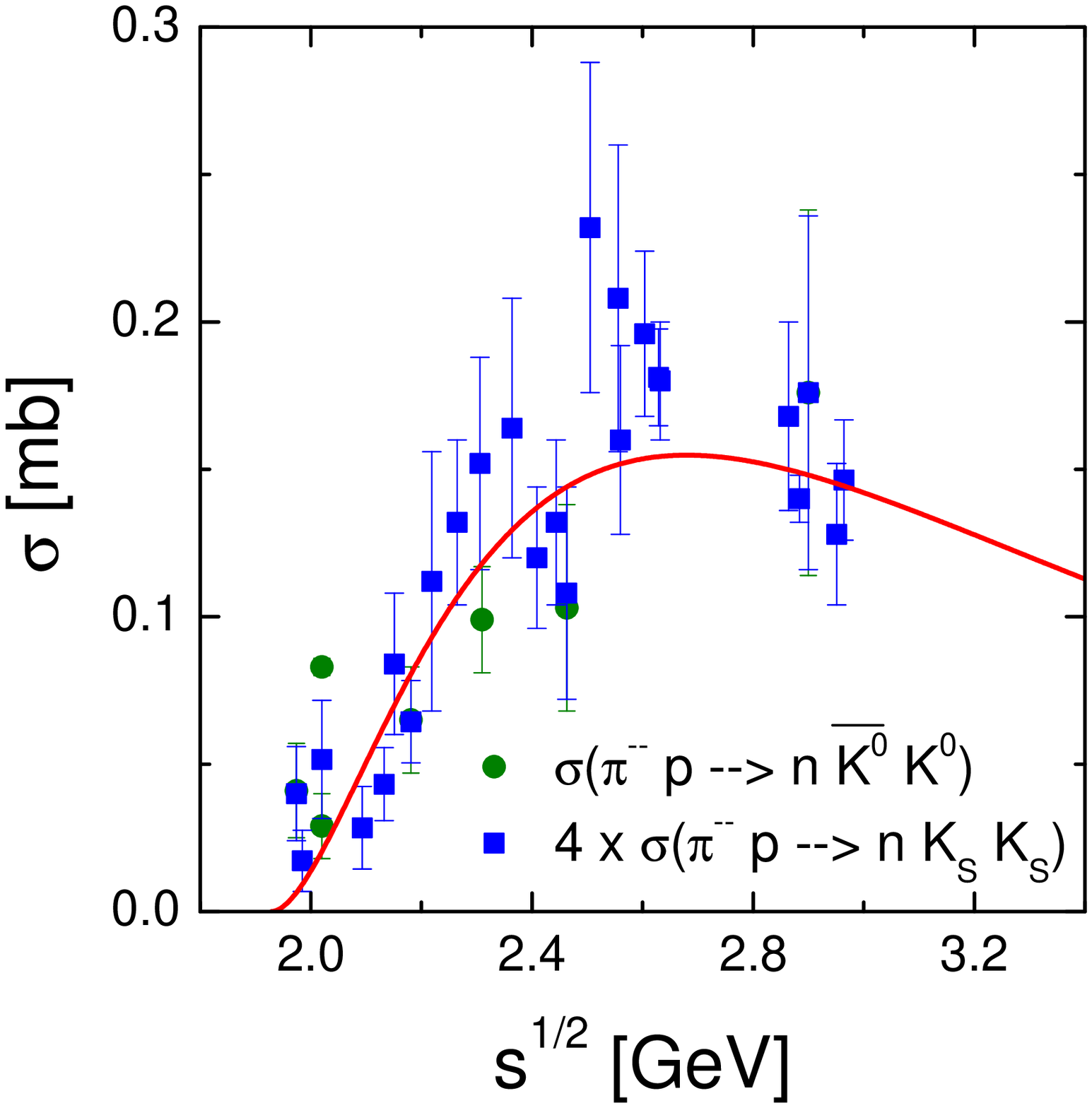}}\,
\parbox{4cm}{\includegraphics[width=4cm]{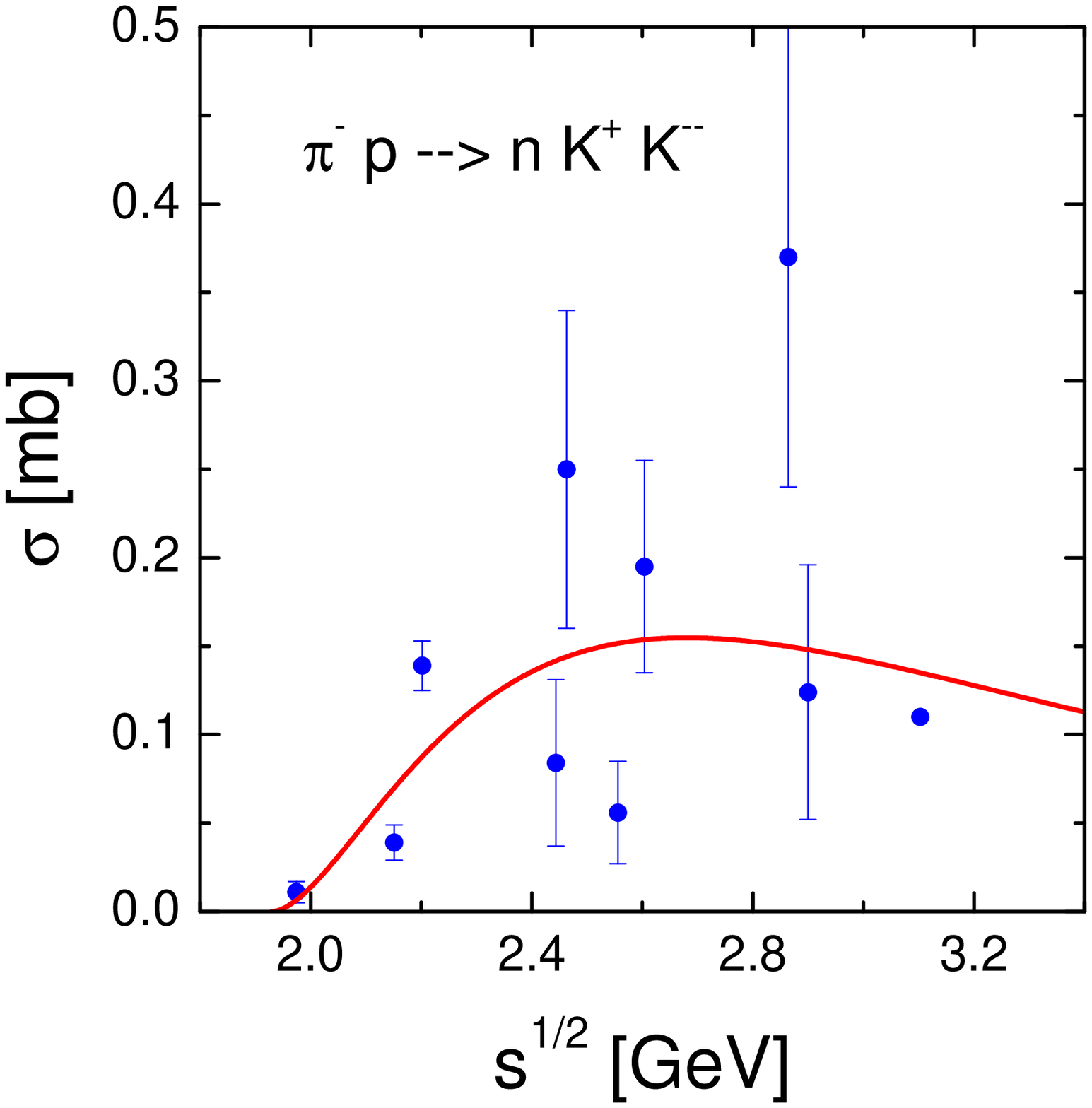}}\,
\parbox{4.1cm}{\includegraphics[width=4.1cm]{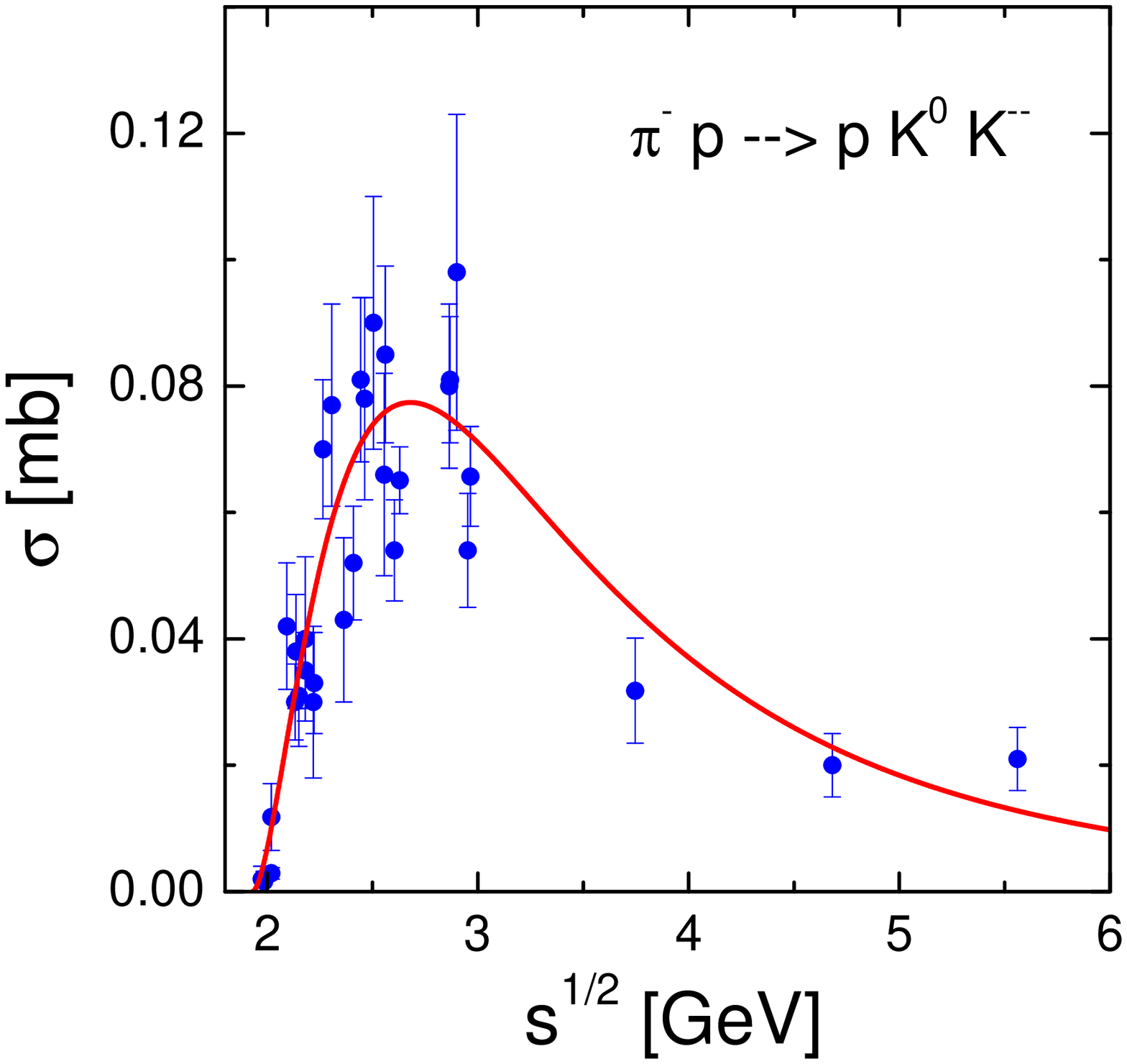}}
\caption{Total cross section of the $\pip p \to p K^+ \bar {K^0}$,
$\pim p \to n K^0 \bar {K}^0$, $\pim p \to n K^+ K^-$,
and $\pim p \to p K^0 K^-$ reactions: solid line
calculated with eq.~(\ref{pin2nkk}), dots are the experimental data from Ref.~\cite{expdata} }
\label{fig:pin2nkk}
\end{figure*}

\begin{figure}
\parbox{4cm}{\includegraphics[width=4cm]{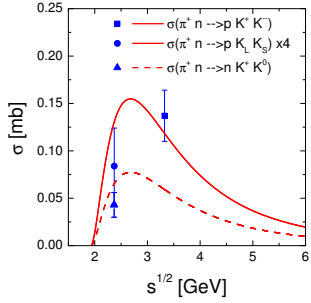}}
\caption{Total cross section of the $\pip n \to p K^+ K^-$,
$\pip n \to p K^0 \bar K^0 $, and $\pip n \to n K^+ \bar K^0$ reactions: solid line
calculated with eq.~(\ref{pin2nkk}), dots are the experimental data from Ref.~\cite{expdata}.
Note that isospin symmetry of the diagram (\ref{diag:pin2nkk}) predicts
that cross sections for $\pip n \to p K^+ K^-$ and the one for
$\pip n \to p K^0 \bar K^0 $ multiplied by 4 are same.}
\label{fig:pin2nkk2}
\end{figure}

Reactions $\pi N \to N K \bar K$ were studied in
Ref.~\cite{cassing}. It was found that under assumption that main
contribution to the $\pi N \to N K \bar K$ reaction is given by
the diagram
\be
\parbox{25mm}{\includegraphics[width=25mm]{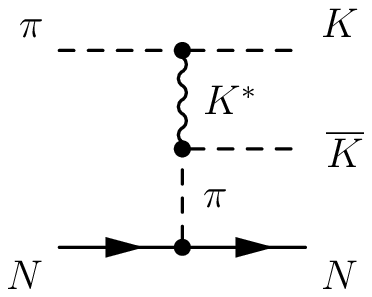}}
\label{diag:pin2nkk}
\ee
all different isospin channels can be
parameterized as
\be
\bar\si(\pi_a N_b \to N_c K_d \bar K_e)  = c\,
1.121 \, \left[ 1 - \frac{s_0}{s} \right]^{1.86}\!
\left[ \frac{s_0}{s} \right ]^2\!,
\label{pin2nkk}
\ee
where $\sqrt{s_0} = m_N + 2 m_K=1.929$~GeV and $c$ is a channel
dependent isospin coefficient. The latter ones are
collected Table~\ref{tab:pin2nkk}. In Figs.~\ref{fig:pin2nkk} and \ref{fig:pin2nkk2} we confront these
parameterizations with the available experimental data.
\begin{table}[b]
\caption{Isospin coefficient in the parameterization (\ref{pin2nkk})
of $\pi N \to N K \bar K$ reactions}
\begin{tabular}{ccccc}
\hline\hline
reaction & $c$ &\phantom{xxx}& reaction & $c$\\
\hline\hline
$\pi^+ p \to p K^+ \bar K^0$ & 1         && $\pi^- p \to p K^0 K^- $     & 1\\
$\pi^+ n \to p K^0 \bar K^0$ & 2         && $\pi^- p \to n K^0 \bar K^0$ & 2\\
$\pi^+ n \to p K^+ K^-$      & 2         && $\pi^- p \to n K^+ K^- $     & 2\\
$\pi^+ n \to n K^+ \bar K^0$ & 1         && $\pi^0 n \to p K^0 K^- $     & 2\\
$\pi^0 p \to p K^0 \bar K^0$ & $\frac12$ && $\pi^0 n \to n K^0 \bar K^0$ & $\frac12$\\
$\pi^0 p \to p K^+ K^-$      & $\frac12$ && $\pi^0 n \to n K^+ K^- $     & $\frac12$\\
$\pi^0 p \to n K^+ \bar K^0$ & 2         && $\pi^- n \to n K^0 K^-$      & 1\\
\hline\hline
\end{tabular}
\label{tab:pin2nkk}
\end{table}

\subsection{Reactions of $\pi + \Delta$}

The cross sections for $\pi \Delta \to Y \Kp$ reactions we take from Ref.~\cite{tsau}
\be
&&\si(\pim   \Dpp \to \Kp   \La)
=\frac{0.00983(\sqs - 1.613)^{0.7866}}{(\sqs - 1.720)^2 + 0.004852} \,.
\nonumber\\ \label{i15}
\ee
The other channels leading to $\La$ production are obtained from isospin
symmetry
\be
&\si(\pim   \Dpp \to \Kp   \La)& = \si(\pip \Dm \to \Kn \La),
\nonumber\\
&\si(\pin   \Dp\,\, \to \Kp   \La)& = \si(\pin \Dn \to \Kn \La)
\nonumber\\
&&=\frac{2}{3}\si(\pim   \Dpp \to \Kp   \La),
\nonumber\\
&\si(\pip   \Dn \to \Kp   \La)& = \si(\pim \Dp \to \Kn \La)
\nonumber\\
&&=\frac{1}{3}\si(\pim   \Dpp \to \Kp   \La),
\nonumber\\
\label{i17}\ee

For $\pi \Delta \to \Sigma K$ reactions
one assumes in~\cite{tsug} that the cross section
has a contribution from the isospin $1/2$ channel only.
The $I=3/2$ contribution was found small.
Then all reactions are related to four
isospin combinations parameterized in \cite{tsug} as:
\be
\si(\pim   \Dpp \to \Kp   \Sn) =
\frac{0.004959 (\sqs - 1.688)^{0.7785}}{(\sqs-1.725)^2 + 0.008147},
\nonumber\\
\si(\pin   \Dn \to \Kp \Sm) =
\frac{0.006964 (\sqs - 1.688)^{0.8140}}{(\sqs-1.725)^2 + 0.007713},
\nonumber\\
\si(\pip   \Dn \to \Kp   \Sn) =
\frac{0.002053(\sqs-1.688)^{0.9853}}{(\sqs-1.725)^2 + 0.005414},
\nonumber\\
+\frac{0.3179(\sqs-1.688)^{0.9025}}{(\sqs-2.675)^2 + 44.88},
\nonumber\\
\si(\pip \Dm \to \Kp   \Sm) =
\frac{0.01741(\sqs - 1.688)^{1.2078}}{(\sqs - 1.725)^2 + 0.003777},
\nonumber
\ee
For other reactions one finds
\be
&\si(\pin   \Dpp \to \Kp   \Sp)& = \si(\pin \Dm \to \Kn \Sm) = 0,
\nonumber\\
&\si(\pip   \Dp \to \Kp   \Sp) &= \si(\pim \Dn \to \Kn \Sm) = 0,
\nonumber\\
& \si(\pip \Dm \to \Kn \Sn) &=\si(\pim   \Dpp \to \Kp   \Sn),
\nonumber\\
&\si(\pin   \Dp \to \Kp   \Sn)& = \si(\pin \Dn \to \Kn \Sn)
\nonumber\\ &&=
\frac{1}{3} \si(\pip \Dm \to \Kp   \Sm),
\nonumber\\
& \si(\pim \Dp \to \Kn \Sn)& = \si(\pip   \Dn \to \Kp   \Sn),
\nonumber\\
&\si(\pim   \Dp \to \Kp   \Sm)& = \si(\pip \Dn \to \Kn \Sp)
\nonumber\\&&=
\frac{2}{3}\si(\pim   \Dpp \to \Kp   \Sn),
\nonumber\\
& \si(\pin \Dp \to \Kn \Sp)& = \si(\pin   \Dn \to \Kp \Sm),
\nonumber\\
&\si(\pim \Dpp \to \Kn \Sp)& = \si(\pip \Dm \to \Kp   \Sm),
\nonumber\\
\ee

We also include reactions $\pi \Delta \to N K \bar K$. These are
taken with the cross section parameterized as in (\ref{pin2nkk}).
The corresponding isospin constant $c$
follows from the analysis of the diagram (\ref{diag:pin2nkk}),
where the incoming nucleon line is replaced by the $\Delta$, and
depends on
incoming and outgoing isospin via Clebsch-Gordan coefficients (cf.
\cite{eff}).
The coefficients are collected in Table~\ref{tab:pid2nkk}.
\begin{table}[h]
\caption{Isospin coefficient in the parameterization (\ref{pin2nkk})
of $\pi \Delta \to N K \bar K$ reactions}
\begin{tabular}{lcclc}
\hline\hline
reaction & $c$ &\phantom{xx}& reaction & $c$\\
\hline\hline
$\pi^+ \Delta^+ \,\,\,\,\to p K^+ \bar K^0$    & 2 && $\pi^+ \Delta^- \to n K^+ K^- $         & 3\\
$\pi^0\, \Delta^{++} \to p K^+ \bar K^0$       & 3 && $\pi^+ \Delta^- \to n K^0 \bar K^0 $    & 3\\
$\pi^+ \Delta^0\, \,\,\,\,\to p K^0 \bar K^0$  & 1 && $\pi^+ \Delta^- \to p K^0 K^- $         & 0\\
$\pi^+ \Delta^0\, \,\,\,\,\to p K^+ K^-$       & 1 && $\pi^0\, \Delta^0\, \to p K^0 K^- $     & 1\\
$\pi^+ \Delta^0\, \,\,\,\,\to n K^+ \bar K^0$  & 2 && $\pi^0\, \Delta^0\, \to n K^0 \bar K^0$ & 3\\
$\pi^0 \,\Delta^+ \,\,\,\,\to p K^0 \bar K^0$  & 1 && $\pi^0\, \Delta^0\, \to n K^+ K^- $     & 1\\
$\pi^0 \,\Delta^+ \,\,\,\,\to p K^+ K^-$       & 3 && $\pi^- \Delta^+ \to n K^+ K^- $         & 1\\
$\pi^0 \,\Delta^+ \,\,\,\,\to n K^+ \bar K^0$  & 1 && $\pi^- \Delta^+ \to n K^0 \bar K^0 $    & 1\\
$\pi^- \Delta^{++} \to p K^0 \bar K^0$         & 3 && $\pi^- \Delta^+ \to p K^0 K^- $         & 2\\
$\pi^- \Delta^{++} \to p K^+ K^-$              & 3 && $\pi^0\, \Delta^- \to n K^0 K^-$        & 3\\
$\pi^- \Delta^{++} \to n K^+ \bar K^0$         & 0 && $\pi^- \Delta^0\, \to n K^0 K^-$        & 2\\
\hline\hline
\end{tabular}
\label{tab:pid2nkk}
\end{table}


\subsection{Reactions of $N + N$}
\label{NNre}

\begin{figure}
\parbox{4cm}{\includegraphics[width=4cm]{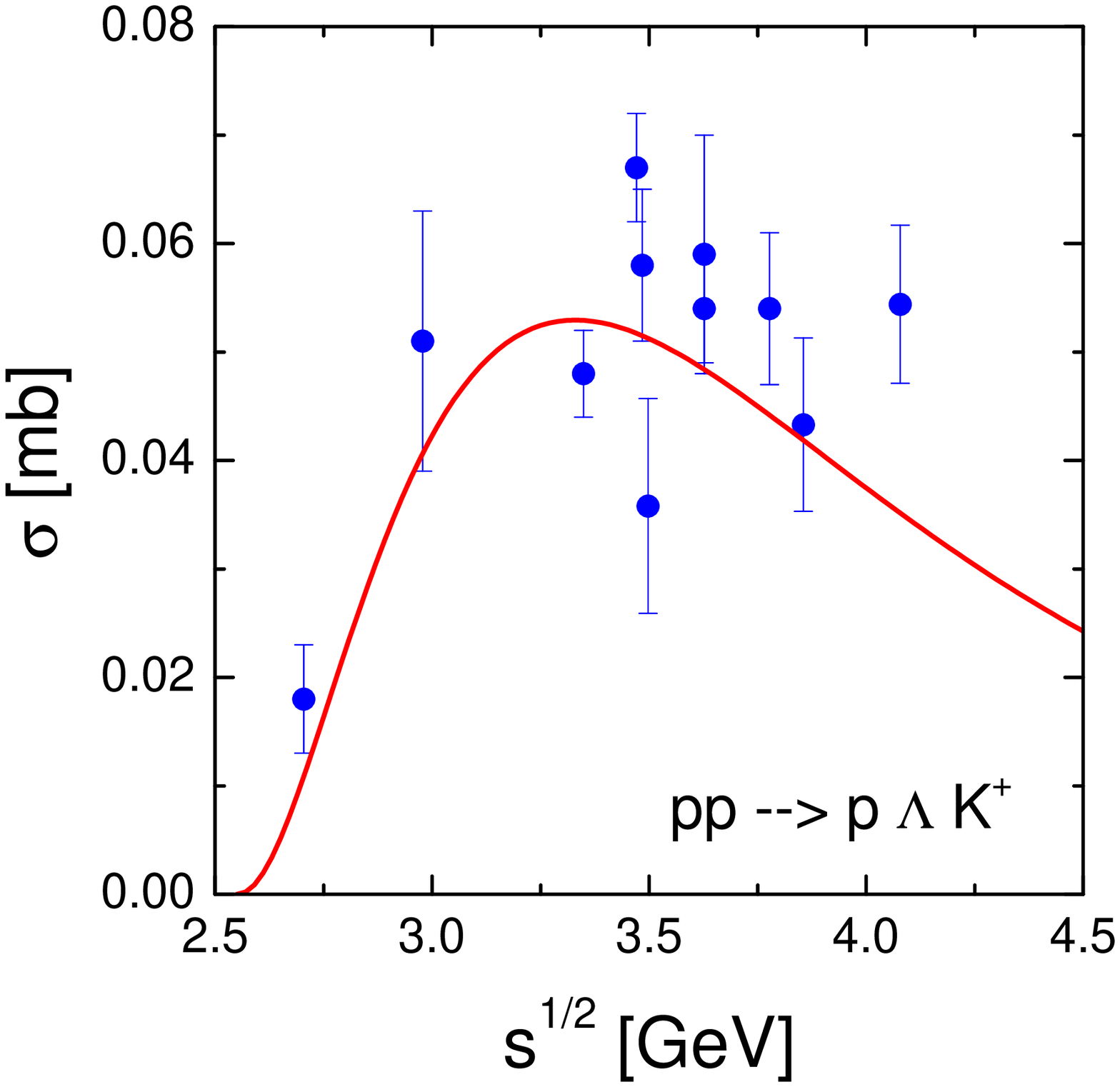}}
\parbox{4cm}{\includegraphics[width=4cm]{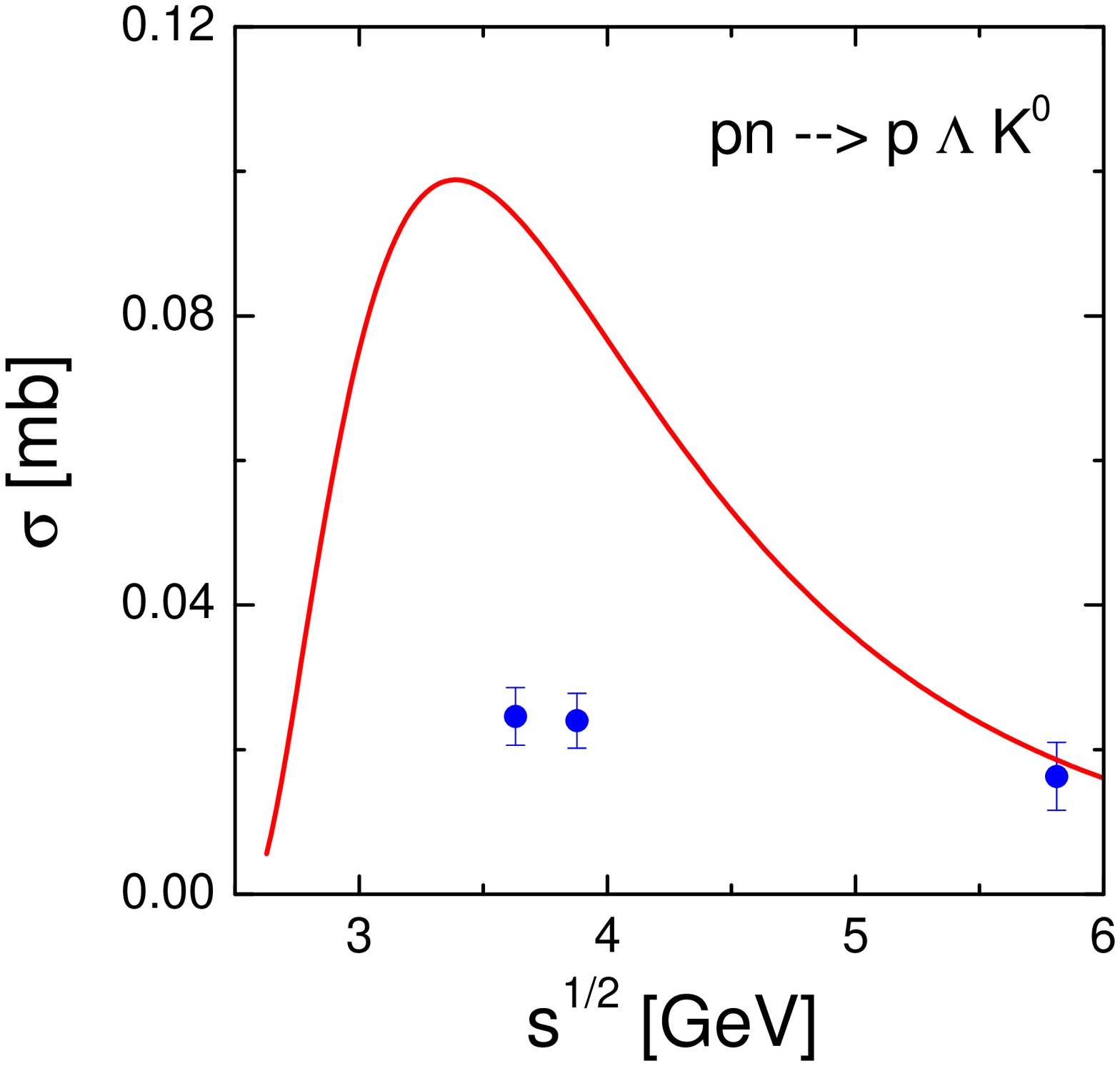}}
\caption{Total cross section of the $p p \to p \La K^+$,
$p n \to p \La K^0$ reactions: solid line
calculated with eq.~(\ref{NN2NLK}), dots are the experimental data from Ref.~\cite{expdata}}
\label{fig:NN2NLK}
\end{figure}

Cross sections for
reactions of the type
$NN\to N \La K$ and $NN \to N \Sigma K$ are parameterized in
\cite{tstl}.
For $NN\to N \La K$ reactions we have
\be
\nonumber
&&\si(p   p \to   p   \La \Kp) = \si(n n \to n \La \Kn)
\\ \nonumber
&&\qquad=
1.879 \left (\frac{s}{s_{KN\Lambda}} -1  \right )^{2.176}
\left ( \frac{s_{KN\Lambda}}{s} \right )^{5.264}
\\ \nonumber
&&\si(p   n \to   n   \La \Kp) = \si(n p \to p \La \Kn)
\\
&&\qquad=
2.812 \left (\frac{s}{s_{KN\Lambda}} -1  \right )^{2.121}
\left ( \frac{s_{KN\Lambda}}{s} \right )^{4.893}
\label{NN2NLK}
\ee
where $s_{KN\Lambda} = (m_\Lambda + m_N + m_K)^2$\,.
These parameterizations are confronted with the available
experimental data in Fig.~\ref{fig:NN2NLK}.
For $NN\to N \Sigma K$ reactions we use
\be \nonumber
&&\si(p   p \to  p   \Sn \Kp ) = \si(n n \to n \Sn \Kn)
\\ \nonumber
&&\qquad=
5.321 \left (\frac{s}{s_{KN\Sigma}} -1  \right )^{2.753}
\left (\frac{s_{KN\Sigma}}{s} \right )^{8.510},
\\ \nonumber
&&\si(p   p \to   n   \Sp \Kp) = \si(n n \to p \Sm \Kn)
\\ \nonumber
&&\qquad=
1.466
\left (\frac{s}{s_{KN\Sigma}} -1  \right )^{2.743}
\left (\frac{s_{KN\Sigma}}{s} \right )^{3.271},
\\ \nonumber
&&\si(p   n \to   p   \Sm \Kp) = \si( n p \to n \Sp \Kn)
\\ \nonumber
&&\qquad=
11.02
\left (\frac{s}{s_{KN\Sigma}} -1  \right )^{2.782}
\left (\frac{s_{KN\Sigma}}{s} \right )^{7.674},
\\ \nonumber
&&\si(p   n \to   n   \Sn \Kp) = \si(n p \to p \Sn \Kn)
\\ \nonumber
&&\qquad=
6.310\left (\frac{s}{s_{KN\Sigma}} -1  \right )^{2.773}
\left (\frac{s_{KN\Sigma}}{s} \right )^{7.820},
\\ \nonumber
&&\si(n   n \to n \Sm \Kp) = \si(p p \to p \Sp \Kn)
\\ \nonumber
&&\qquad=
7.079
\left (\frac{s}{s_{KN\Sigma}} -1  \right )^{2.760}
\left (\frac{s_{KN\Sigma}}{s} \right )^{8.164},
\\ \label{NN2NSK}
\ee
where  $s_{KN\Sigma}=(m_\Sigma + m_N + m_K)^2$.
\begin{figure}
\parbox{4cm}{\includegraphics[width=4cm]{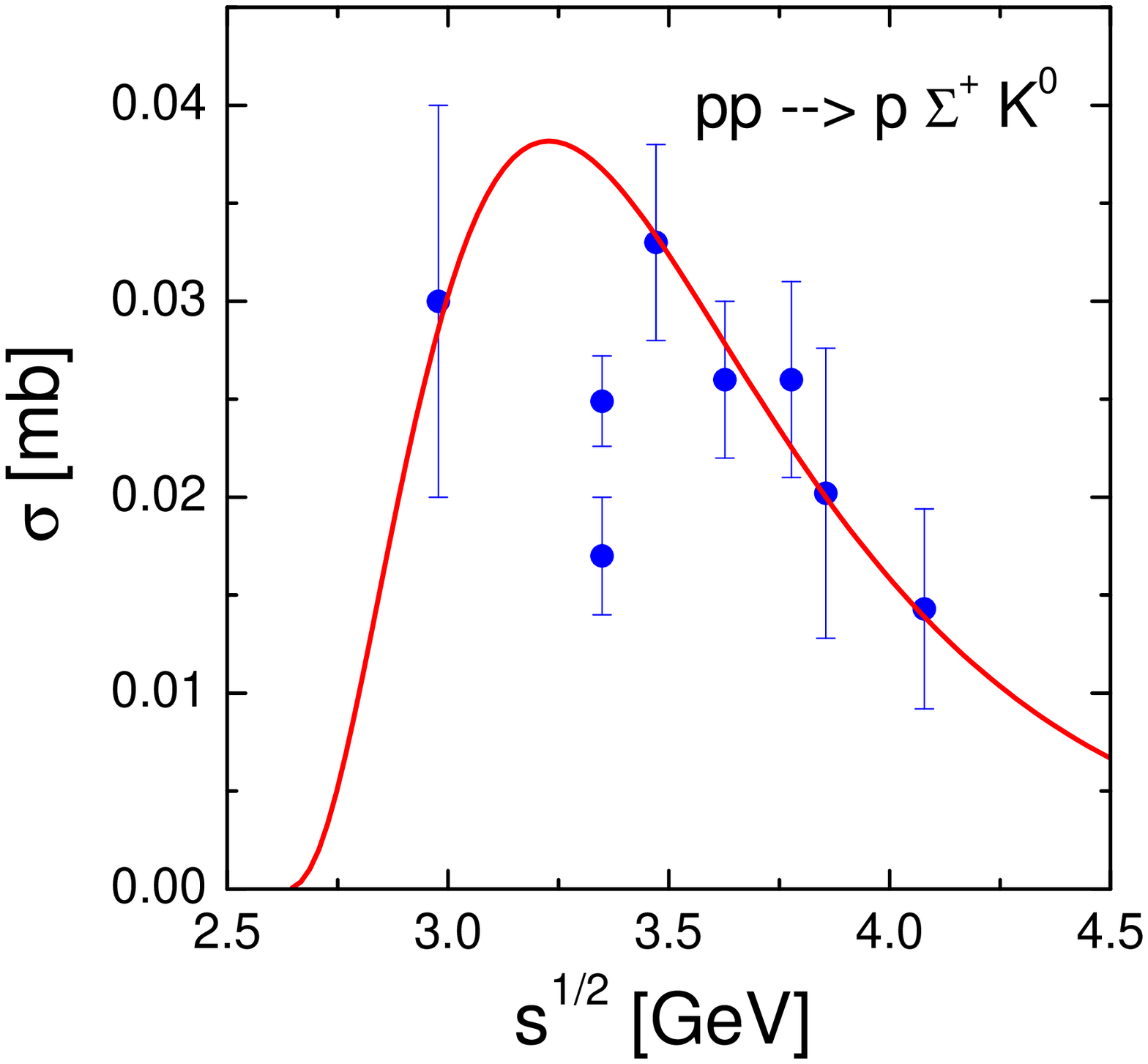}}
\parbox{4cm}{\includegraphics[width=4cm]{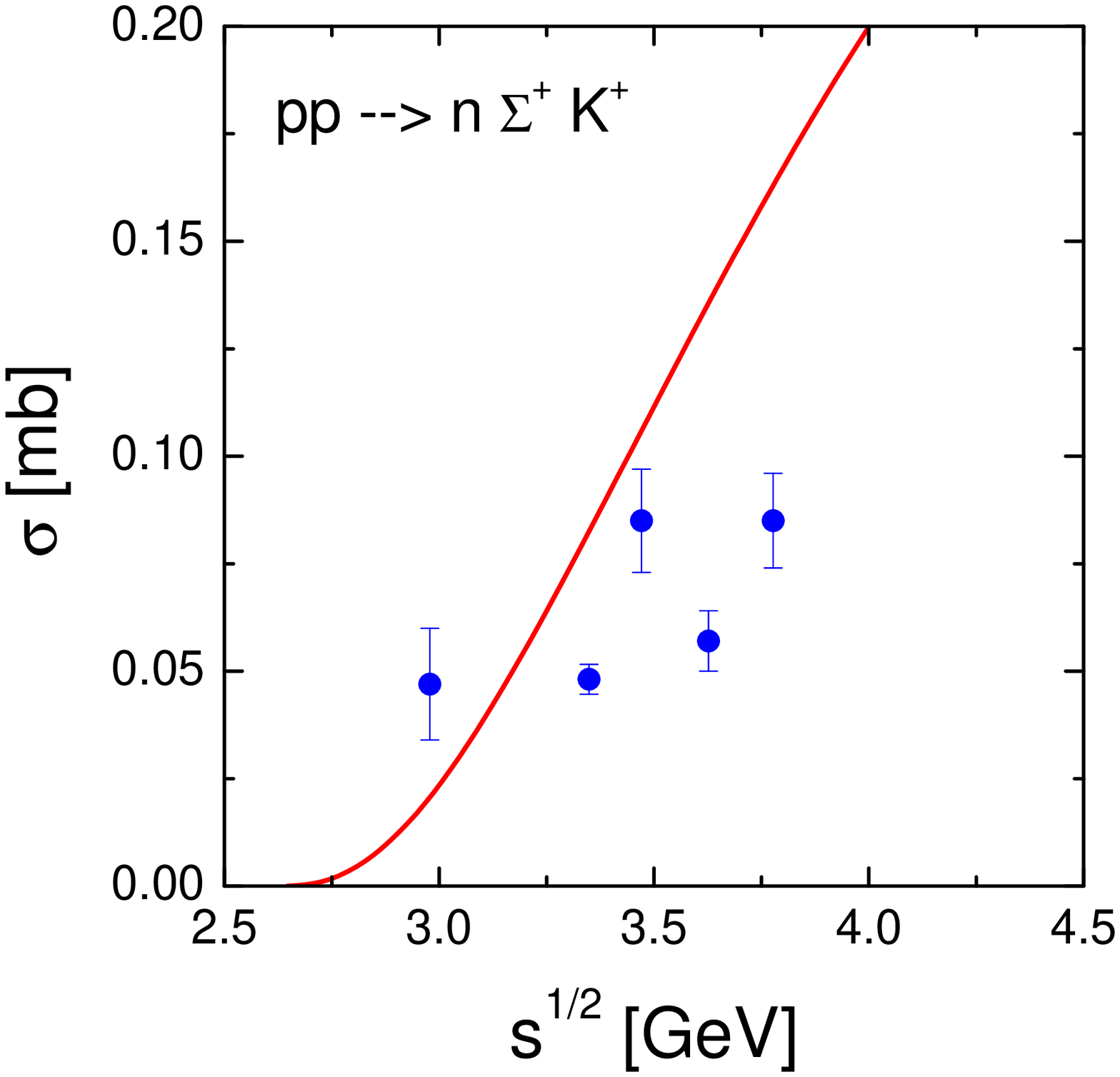}}
\caption{Total cross section of the $p p \to p \Sp  K^0$,
$p p \to n \Sp K^+$ reactions: solid line
calculated with eq.~(\ref{NN2NSK}), dots are the experimental data from Ref.~\cite{expdata}}
\label{fig:NN2NSK-1}
\end{figure}
\begin{figure}
\parbox{4cm}{\includegraphics[width=4cm]{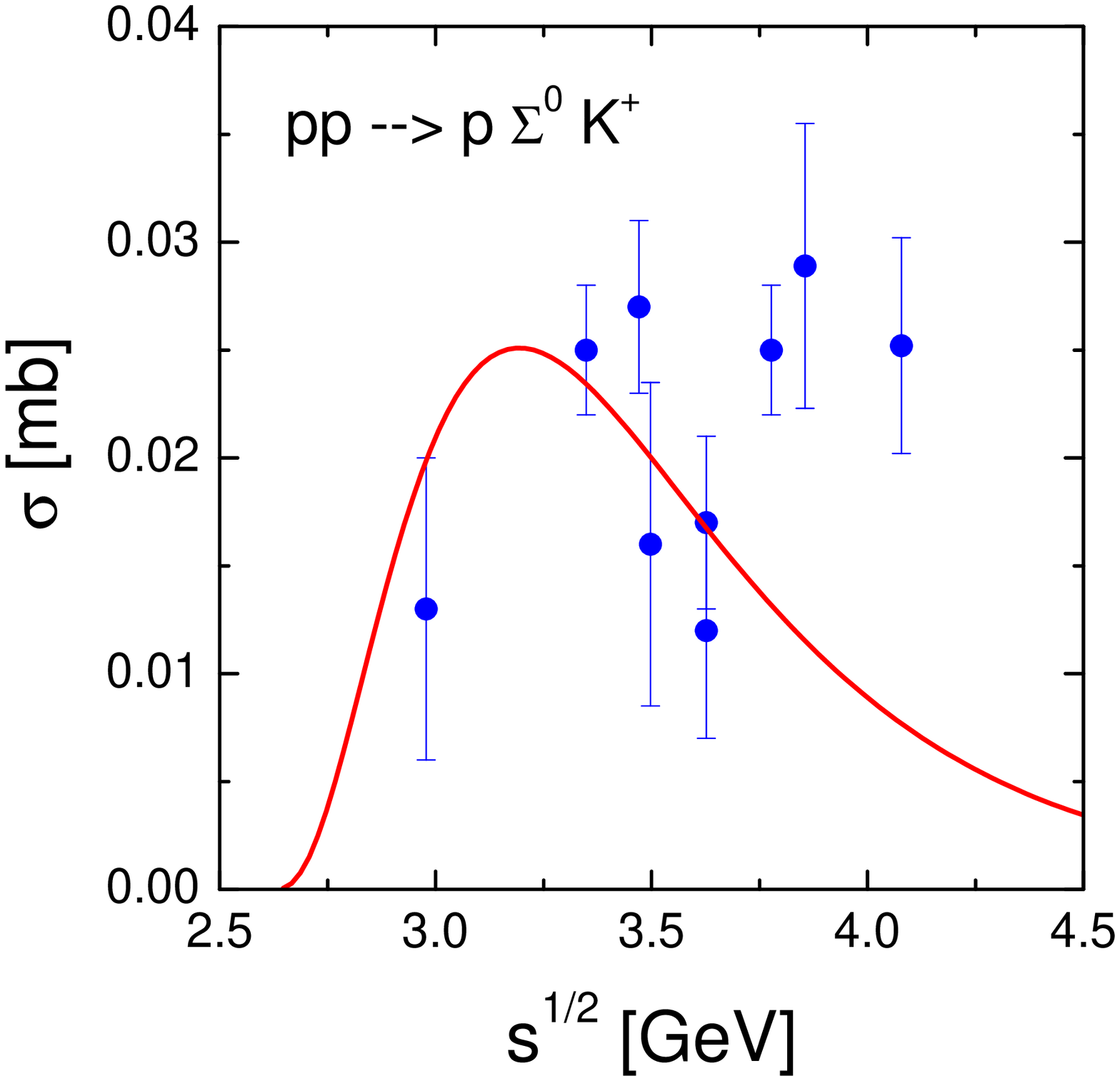}}
\parbox{4cm}{\includegraphics[width=4cm]{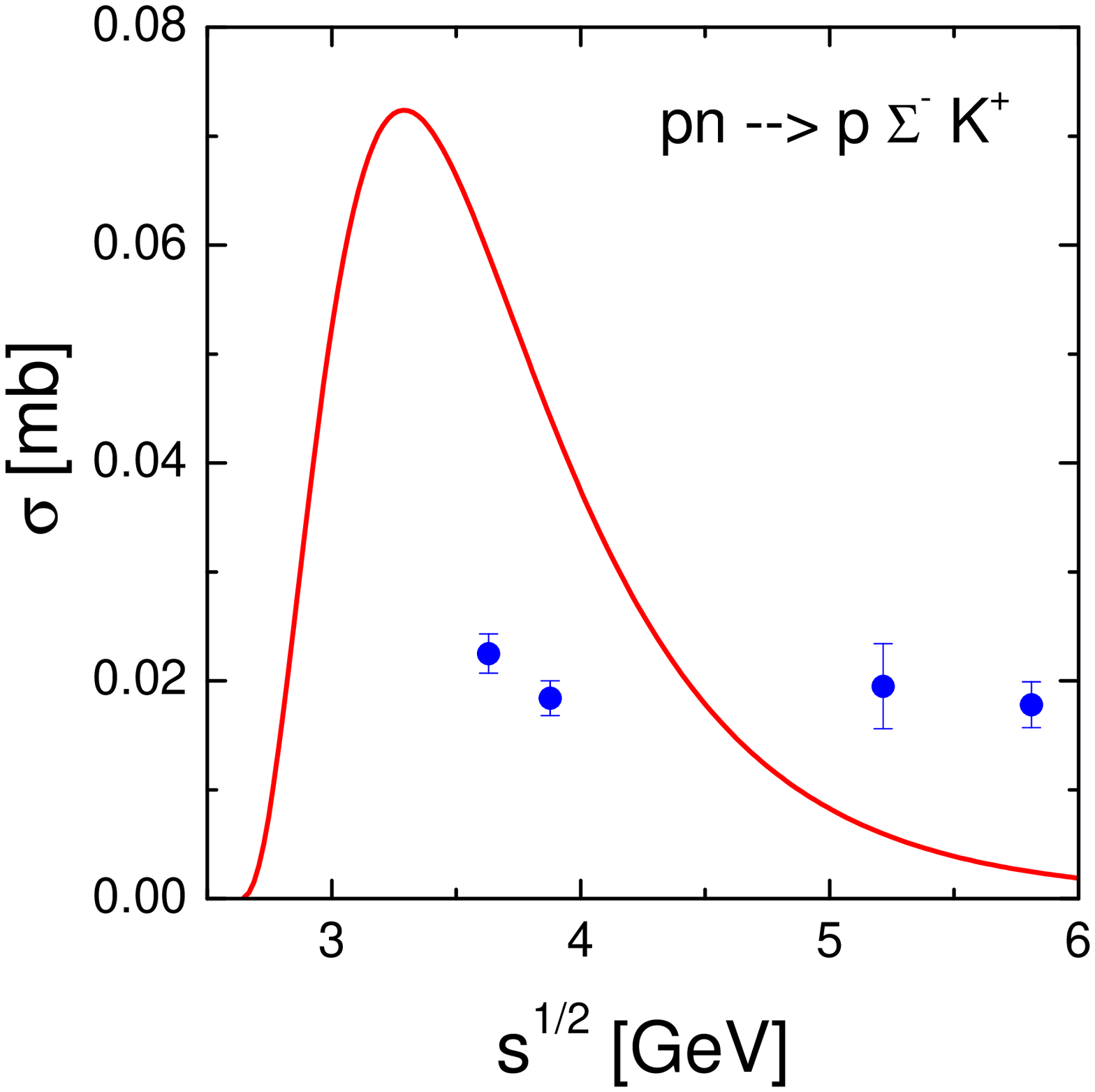}}
\caption{Total cross section of the $p p \to p \Sn K^+$,
$p n \to p \Sm K^+$ reactions: solid line
calculated with eq.~(\ref{NN2NSK}), dots are the experimental data from Ref.~\cite{expdata}}
\label{fig:NN2NSK-2}
\end{figure}
\begin{figure}
\parbox{4cm}{\includegraphics[width=4cm]{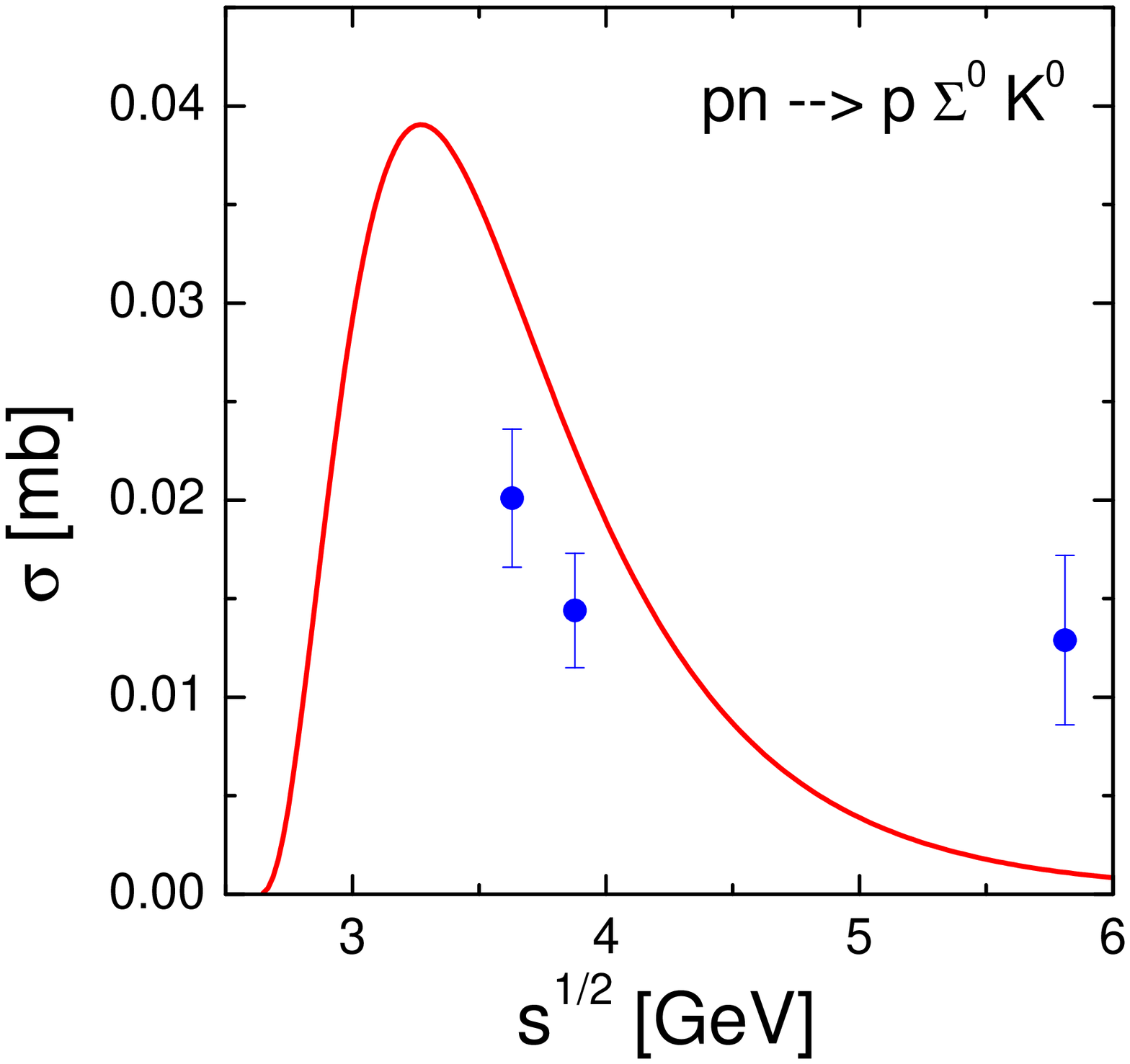}}
\caption{Total cross section of the $p n \to p \Sn K^0$ reaction: solid line
calculated with eq.~(\ref{NN2NSK}), dots are the experimental data from Ref.~\cite{expdata}}
\label{fig:NN2NSK-3}
\end{figure}
These parameterizations are compared with the experimental data in
Figs.~\ref{fig:NN2NSK-1}, \ref{fig:NN2NSK-2}, \ref{fig:NN2NSK-3}.

Next we proceed to
reactions of the type $NN\to NNK\bar K$.
Following \cite{eff} we will use the same isospin-averaged cross sections
\cite{cas}
\be
\bar\si(NN\to NNK\bar K) = 1.5 \left( 1 - \frac{s_1}{s} \right )^{3.17}
\left ( \frac{s_1}{s}\right )^{1.96}
\label{nn2nnkk}
\ee
with $s_1 = (2 m_N + 2 m_K)^2$
for all possible isospin channels
\be
\begin{tabular}{lcl}
$pp \to pp K^+K^-$       &\phantom{xx}& $pn \to pp K^0 K^-$\\
$pp \to pp K^0 \bar K^0$ && $pn \to nn K^+ \bar K^0$\\
$pp \to pn K^+ \bar K^0$ && $nn \to nn K^+K^-$\\
$pn \to pn K^+K^-$       && $nn \to nn K^0 \bar K^0$\\
$pn \to pn K^0 \bar K^0$ && $nn \to pn K^0 K^-$\\
\end{tabular}
\ee

Cross sections for processes
$NN \to \Delta \Lambda K$ and $NN \to \Delta \Sigma K$
were calculated in Ref.~\cite{tstl}.
The cross sections of reactions $p p \to \Dpp \La \Kn$ and
$pp \to \Dpp \Sm \Kp$ are parameterized as follows
\be
\si(p p \to \Dpp\, \La\,\, \Kn\,) \!\!\!&=&\!\!\!
6.166 \left[ \frac{s}{s_{\Delta\La K}} -1 \right]^{2.842}
\left[ \frac{s_{\Delta\La K}}{s} \right]^{1.96}\!\!\!\!,
\nonumber\\
\label{pp2DLk}\\
\si(pp \to \Dpp \Sm \Kp)\!\!\! &=&\!\!\!
10.00 \left[ \frac{s}{s_{\Delta\Sigma K}} -1 \right]^{2.874}
\left[ \frac{s_{\Delta\Sigma K}}{s} \right]^{2.543}\!\!\!\!,
\nonumber\\
\label{pp2DSk}
\ee
where
$s_{\Delta\La K} = (m_\Delta + m_\Lambda + m_K)^2$ and
$s_{\Delta\Sigma K} = (m_\Delta + m_\Sigma + m_K)^2$\,.
The other reactions channels are related to these two by isospin
coefficients $c_\Lambda$ and $c_\Sigma$
\be
\sigma(N_aN_b \to \Delta_c \Lambda K_d)\!\!\! &=&\!\!\! c_\Lambda\,\si(p p \to \Dpp \La \Kn)
\nonumber\\
\sigma(N_aN_b \to \Delta_c \Sigma_d K_e)\!\!\!&=&\!\!\! c_\Sigma\,\si(pp \to \Dpp \Sm \Kp)
\nonumber\\
\label{clcs}
\ee
This coefficients are collected in Table~\ref{tab:clcs}.
\begin{table}
\fontsize{8}{2}
\caption{The isospin coefficients in parameterizations of
the reactions $NN \to \Delta \Lambda K$ and
$NN \to \Delta \Sigma K$~(\ref{pp2DLk},\ref{pp2DSk},\ref{clcs}) }
\begin{tabular}{lclclc}
\hline\hline
\multicolumn{1}{c}{reaction} & $c_\Lambda$ & \multicolumn{1}{c}{reaction} & $c_\Sigma$ & \multicolumn{1}{c}{reaction} & $c_\Sigma$ \\
\hline\hline
$pp\to \Dp\La K^+$   & $\frac13$ & $pp\to \Dp\,\,\,\Sn K^+$  & $\frac16$ & $pn\to \Dp\,\,\,\Sn\, K^0$  & $\frac16$\\
$pn\to \Dp \La K^0$  & $\frac13$ & $pp\to \Dp\,\,\,\Sp K^0$  & $\frac13$ & $pn\to \Dn\,\,\,\,\Sp K^0$  & $\frac13$\\
$pn\to \Dn\,\La K^0$ & $\frac13$ & $pp\to \Dpp\Sn K^0$       & $\frac12$ & $pn\to \Dpp\Sm K^0$         & $0$      \\
$nn\to \Dn\,\La K^0$ & $\frac13$ & $pp\to \Dn\,\,\,\,\Sp K^+$& $0$       & $nn\to \Dn\,\,\,\,\Sm K^+$  & $\frac13$\\
$nn\to \Dm \La K^+$  & $1$       & $pn\to \Dn\,\,\,\,\Sn K^+$& $\frac16$ & $nn\to \Dm\,\,\,\Sn K^+$    & $\frac12$\\
                     &           & $pn\to \Dp\,\,\,\Sm K^+$  & $\frac13$ & $nn\to \Dn\,\,\,\,\Sn\, K^0$& $\frac16$\\
                     &           & $pn\to \Dm\,\,\,\Sp K^+$  & $0$       & $nn\to \Dm\,\,\,\Sp K^0$    & $1$      \\
                     &           &                           &           & $nn\to \Dp\,\,\,\Sm K^0$    & $0$      \\
\hline\hline
\end{tabular}
\label{tab:clcs}
\end{table}


\subsection{Reactions of $\Delta + N$}
Reactions $\Delta N \to N Y K$ ($Y = \Lambda,\, \Sigma$) are
parameterized in \cite{tstl} in terms of two cross sections
\be
\si(\Dpp n \to p \La \Kp) =
8.337 \left[\textstyle \frac{s}{s_{K N\La}} -1 \right]^{2.227}\!\!
\left[\textstyle \frac{s_{K N\La}}{s} \right]^{2.511}
\nonumber\\
\label{DN2LK}
\\
\si(\Dm p \to n \Sm \Kp) = 52.72
\left[\textstyle \frac{s}{s_{K N\Sigma}} -1 \right]^{2.799}\!\!
\left[\textstyle \frac{s_{K N\Sigma}}{s} \right]^{6.303}
\nonumber\\
\label{DN2SK}
\ee
The other reaction channels are related to these two by the
isospin coefficients $\bar{c}_\Lambda$ and $\bar{c}_\Sigma$
\be
\sigma(\Delta_a N_b \to N_c \Lambda K_d)\!\!\! &=&\!\!\! c_\Lambda\,\si(\Dpp n \to p \La \Kp)
\nonumber\\
\sigma(\Delta_a N_b \to N_c \Sigma_d K_e)\!\!\!&=&\!\!\! c_\Sigma\,\si(\Dm p \to n \Sm \Kp)
\nonumber\\
\label{bclbcs}
\ee
\begin{table}
\fontsize{8}{2}
\caption{The isospin coefficients in parameterizations of
the reactions $\Delta N \to N Y K$ ~(\ref{DN2LK},\ref{DN2SK},\ref{bclbcs}) }
\begin{tabular}{lclclc}
reaction & $\bar{c}_\Lambda$ & reaction & $\bar{c}_\Sigma$ & reaction & $\bar{c}_\Sigma$\\
\hline\hline
$\Delta^{++} n\to p\La K^+$       &    $1$    & $\Delta^{++} p \to p\Sp K^+$     &$0$      & $\Delta^{+} n \to p\Sm K^+$  &$\frac13$\\
$\Delta^{+}\,\,\,p\to p\La K^+$   &$\frac13$& $\Delta^{+}\,\,\, p \to p\Sp K^0$  &$\frac13$& $\Delta^{0}\, p \to p\Sm K^+$&$\frac13$\\
$\Delta^{+}\,\,\,n\to p\La K^0$   &$\frac13$& $\Delta^{++} n \to p\Sp K^0$       &$1$      & $\Delta^{+} n \to n\Sn K^+$  &$\frac16$\\
$\Delta^{0}\,\,\,\,p\to p\La K^0$ &$\frac13$& $\Delta^{+}\,\,\, p \to p\Sn K^+$  &$\frac16$& $\Delta^{0}\, p \to n\Sn K^+$&$\frac16$\\
$\Delta^{+}\,\,\, n\to n\La K^+$  &$\frac13$& $\Delta^{++} n \to p\Sn K^+$       &$\frac12$& $\Delta^{0}\, n \to n\Sn K^0$&$\frac16$\\
$\Delta^{0}\,\,\,\,p\to n\La K^+$ &$\frac13$& $\Delta^{+}\,\,\, p \to n\Sp K^+$  &$0$      & $\Delta^{-} p \to n\Sn K^0$  &$\frac12$\\
$\Delta^{-}\,\,\,p\to n\La K^0$   &$1$      & $\Delta^{++} n \to n\Sp K^+$       &$0$      & $\Delta^{0}\, n \to p\Sm K^0$&$0$      \\
$\Delta^{0}\,\,\,n\to n\La K^0$   &$\frac13$& $\Delta^{+}\,\,\, n \to p\Sn K^0$  &$\frac16$& $\Delta^{-} p \to p\Sm K^0$  &$0$      \\
                                  &         & $\Delta^{0}\,\,\,\, p \to p\Sn K^0$&$\frac16$& $\Delta^{0}\, n \to n\Sm K^+$&$\frac13$\\
                                  &         & $\Delta^{+}\,\,\, n \to n\Sp K^0$  &$\frac13$& $\Delta^{-} p \to n\Sm K^+$  &$1$\\
                                  &         & $\Delta^{0}\,\,\,\, p \to n\Sp K^0$&$\frac13$& $\Delta^{-} n \to n\Sm K^0$  &$0$\\
\hline\hline
\end{tabular}
\end{table}

For reactions of the type $\Delta N \to NN K \bar K$, all channels
\be
\begin{tabular}{ll}
$\Dpp p \to pp K^+\bar K^0$ &  $\Dn p \to pn K^0\bar K^0$\\
$\Dpp n \to pp K^0\bar K^0$ &  $\Dn p \to nn K^+\bar K^0$\\
$\Dpp n \to pp K^+ K^-$     &  $\Dn p \to pp K^0     K^-$\\
$\Dpp n \to pn K^+\bar K^0$ &  $\Dn p \to pn K^+     K^-$\\
$\Dp\,\,\, p \to pp K^0\bar K^0$  &  $\Dn n \to nn K^0\bar K^0$\\
$\Dp\,\,\, p \to pp K^+ K^-$      &  $\Dn n \to pn K^0     K^-$\\
$\Dp\,\,\, p \to pn K^+\bar K^0$  &  $\Dn n \to nn K^+     K^-$\\
$\Dp\,\,\, n \to pn K^0\bar K^0$  &  $\Dm p \to nn K^0\bar K^0$\\
$\Dp\,\,\, n \to nn K^+\bar K^0$  &  $\Dm p \to pn K^0     K^-$\\
$\Dp\,\,\, n \to pp K^0     K^-$  &  $\Dm p \to nn K^+     K^-$\\
$\Dp\,\,\, n \to pn K^+     K^-$  &  $\Dm n \to nn K^0     K^-$\\
\end{tabular}
\label{dn2nnkk}
\ee
will be taken with the averaged cross section for
$NN\to N N K \bar K$ \cite{eff}.

Next, we consider the family of processes
$\Delta N \to \Delta Y K$ ($Y=\Lambda,\Sigma$). In
Ref.~\cite{tstl} the following
ten reactions were parameterized:
\be
\sigma_{\Delta\Delta}^{(1)}&=&\si(\Dpp p \to \Dpp \La \Kp)
\nonumber\\
&=&2.704
\left[\textstyle \frac{s}{s_{\Delta\La K}} -1 \right]^{2.303}
\left[\textstyle \frac{s_{\Delta\La K}}{s} \right]^{5.551},
\nonumber\\
\sigma_{\Delta\Delta}^{(2)}&=&\si(\Dp p \to \Dp \La \Kp)
\nonumber\\
&=&2.917
\left[\textstyle \frac{s}{s_{\Delta\La K}} -1 \right]^{2.350}
\left[\textstyle \frac{s_{\Delta\La K}}{s} \right]^{6.557},
\nonumber\\
\sigma_{\Delta\Delta}^{(3)}&=&\si(\Dp n \to \Dn \La \Kp)
\nonumber\\
&=&0.312
\left[\textstyle \frac{s}{s_{\Delta\La K}} -1 \right]^{2.110}
\left[\textstyle \frac{s_{\Delta\La K}}{s} \right]^{2.165},
\label{dn2dlk}
\ee
\be
\sigma_{\Delta\Delta}^{(4)}&=&\si(\Dpp p \to \Dpp \Sn \Kp)
\nonumber\\
&=&
10.30
\left[\textstyle \frac{s}{s_{\Delta\Sigma K}} -1 \right]^{2.748}
\left[\textstyle \frac{s_{\Delta\Sigma K}}{s} \right]^{9.321},
\nonumber\\
\sigma_{\Delta\Delta}^{(5)}&=&\si(\Dpp n \to \Dpp \Sm \Kp)
\nonumber\\
&=&
10.33
\left[\textstyle \frac{s}{s_{\Delta\Sigma K}} -1 \right]^{2.743}
\left[\textstyle \frac{s_{\Delta\Sigma K}}{s} \right]^{8.915},
\nonumber\\
\sigma_{\Delta\Delta}^{(6)}&=&\si(\Dp p\to \Dp \Sn \Kp)
\nonumber\\
&=&
10.62
\left[\textstyle \frac{s}{s_{\Delta\Sigma K}} -1 \right]^{2.759}
\left[\textstyle \frac{s_{\Delta\Sigma K}}{s} \right]^{10.20},
\nonumber\\
\sigma_{\Delta\Delta}^{(7)}&=&\si(\Dp p \to \Dn \Sp \Kp)
\nonumber\\
&=&
0.647
\left[\textstyle \frac{s}{s_{\Delta\Sigma K}} -1 \right]^{2.830}
\left[\textstyle \frac{s_{\Delta\Sigma K}}{s} \right]^{3.862},
\nonumber\\
\sigma_{\Delta\Delta}^{(8)}&=&\si(\Dn p \to \Dp \Sm \Kp)
\nonumber\\
&=&
2.128
\left[\textstyle \frac{s}{s_{\Delta\Sigma K}} -1 \right]^{2.843}
\left[\textstyle \frac{s_{\Delta\Sigma K}}{s} \right]^{5.986},
\nonumber\\
\sigma_{\Delta\Delta}^{(9)}&=&\si(\Dp n \to \Dp \Sm \Kp)
\nonumber\\
&=&
10.57
\left[\textstyle \frac{s}{s_{\Delta\Sigma K}} -1 \right]^{2.757}
\left[\textstyle \frac{s_{\Delta\Sigma K}}{s} \right]^{10.11},
\nonumber\ee\be
\sigma_{\Delta\Delta}^{(10)}&=&\si(\Dp n \to \Dn \Sn \Kp)
\nonumber\\
&=&
1.112
\left[\textstyle \frac{s}{s_{\Delta\Sigma K}} -1 \right]^{2.846}
\left[\textstyle \frac{s_{\Delta\Sigma K}}{s} \right]^{5.943}
\label{dn2dsk}
\ee
The other isospin channels can be expressed in terms of these ten cross
sections as it is given in Table~\ref{tab:dn2dyk}.

\begin{table*}\fontsize{8}{2}
\caption{Cross sections of reactions $\Delta N \to \Delta Y K$ ($Y=\Lambda,\Sigma$)
in terms of cross sections (\ref{dn2dlk},\ref{dn2dsk})}
\label{tab:dn2dyk}
\begin{tabular}{lrlrlrlrlr}
\hline\hline
\multicolumn{10}{c}{$\Delta N \to\Delta \Lambda K$}\\
\hline\hline
reaction & $\si$ & reaction & $\si$ & reaction & $\si$ &  reaction & $\si$ & reaction & $\si$ \\
\hline\hline
$\Delta^{++} p \to \Delta^{++} \Lambda K^+$ & $\si^{(1)}_{\Delta\Delta}$ &
$\Delta^{++} n \to \Delta^{++} \Lambda K^0$ & $\si^{(1)}_{\Delta\Delta}$ &
$\Delta^{++} n \to \Delta^+ \Lambda K^+$ & $\frac43\si^{(3)}_{\Delta\Delta}$ &
$\Delta^+ p \to \Delta^{++} \Lambda K^0$ & $\frac43\si^{(3)}_{\Delta\Delta}$ &
$\Delta^+ p \to \Delta^+ \Lambda K^+$ & $\si^{(2)}_{\Delta\Delta}$ \\
$\Delta^+\,\,\, n \to \Delta^+\,\,\, \Lambda K^0$ & $\si^{(2)}_{\Delta\Delta}$ &
$\Delta^+\,\,\, n \to \Delta^0\,\,\,\, \Lambda K^+$ & $\si^{(3)}_{\Delta\Delta}$ &
$\Delta^0\,\,\,\, p \to \Delta^+ \Lambda K^0$ & $\si^{(3)}_{\Delta\Delta}$ &
$\Delta^0\, p \to \Delta^0\,\,\,\,\, \Lambda K^+$ & $\si^{(2)}_{\Delta\Delta}$ &
$\Delta^0\, n \to \Delta^0 \Lambda K^0$ & $\si^{(2)}_{\Delta\Delta}$ \\
$\Delta^0\,\,\,\, n \to \Delta^-\,\,\, \Lambda K^+$ & $\frac43\si^{(3)}_{\Delta\Delta}$ &
$\Delta^-\,\,\, p \to \Delta^0\,\,\,\, \Lambda K^0$ & $\frac43\si^{(3)}_{\Delta\Delta}$ &
$\Delta^-\,\,\, p \to \Delta^- \Lambda K^+$ & $\si^{(1)}_{\Delta\Delta}$ &
$\Delta^- n \to \Delta^-\,\,\, \Lambda K^0$ & $\si^{(1)}_{\Delta\Delta}$\\
\hline\hline
\end{tabular}
\fontsize{7}{2}
\begin{tabular}{lrlrlrlrlr}
\multicolumn{10}{c}{$\Delta N \to\Delta \Sigma K$}\\
\hline\hline
reaction & $\si$ & reaction & $\si$ & reaction & $\si$ &  reaction & $\si$ & reaction & $\si$ \\
\hline\hline
$\Delta^{++} p \to \Delta^{++} \Sigma^0 K^+$ &$\si_{\Delta\Delta}^{(4)}$          & $\Delta^+ p \to \Delta^{++} \Sigma^0 K^0$    &$\frac34\si_{\Delta\Delta}^{(10)}$  & $\Delta^+ n \to \Delta^0 \Sigma^0 K^+$       &$\si_{\Delta\Delta}^{(10)}$         & $\Delta^0 p \to \Delta^{-} \Sigma^+ K^+$     &$\frac34\si_{\Delta\Delta}^{(7)}$   & $\Delta^- p \to \Delta^+ \Sigma^- K^0$       &$0\,\,\,$                           \\
$\Delta^{++} p \to \Delta^{++} \Sigma^+ K^0$ &$\si_{\Delta\Delta}^{(5)}$          & $\Delta^+ p \to \Delta^+ \Sigma^0 K^+$       &$\si_{\Delta\Delta}^{(6)}$          & $\Delta^+ n \to \Delta^{++} \Sigma^- K^0$    &$\frac34\si_{\Delta\Delta}^{(7)}$   & $\Delta^0 n \to \Delta^0 \Sigma^0 K^0$       &$\si_{\Delta\Delta}^{(6)}$          & $\Delta^- p \to \Delta^- \Sigma^+ K^0$       &$\si_{\Delta\Delta}^{(5)}$          \\
$\Delta^{++} p \to \Delta^+ \Sigma^+ K^+$    &$\frac34\si_{\Delta\Delta}^{(7)}$   & $\Delta^+ p \to \Delta^+ \Sigma^+ K^0$       &$\si_{\Delta\Delta}^{(9)}$          & $\Delta^+ n \to \Delta^{-} \Sigma^+ K^+$     &$0\,\,\,$                           & $\Delta^0 n \to \Delta^0 \Sigma^- K^+$       &$\si_{\Delta\Delta}^{(9)}$          & $\Delta^- p \to \Delta^- \Sigma^0 K^+$       &$\si_{\Delta\Delta}^{(4)}$          \\
$\Delta^{++} n \to \Delta^{++} \Sigma^0 K^0$ &$\si_{\Delta\Delta}^{(4)}$          & $\Delta^+ p \to \Delta^{++} \Sigma^- K^+$    &$\frac34\si_{\Delta\Delta}^{(8)}$   & $\Delta^0 p \to \Delta^+ \Sigma^0 K^0$       &$\si_{\Delta\Delta}^{(10)}$         & $\Delta^0 n \to \Delta^+ \Sigma^- K^0$       &$\si_{\Delta\Delta}^{(7)}$          & $\Delta^- n \to \Delta^- \Sigma^0 K^0$       &$\si_{\Delta\Delta}^{(4)}$          \\
$\Delta^{++} n \to \Delta^+ \Sigma^0 K^+$    &$\frac34\si_{\Delta\Delta}^{(10)}$  & $\Delta^+ p \to \Delta^{0} \Sigma^+ K^+$     &$\si_{\Delta\Delta}^{(7)}$          & $\Delta^0 p \to \Delta^+ \Sigma^- K^+$       &$\si_{\Delta\Delta}^{(8)}$          & $\Delta^0 n \to \Delta^- \Sigma^+ K^0$       &$\frac34\si_{\Delta\Delta}^{(8)}$   & $\Delta^- n \to \Delta^- \Sigma^- K^+$       &$\si_{\Delta\Delta}^{(5)}$          \\
$\Delta^{++} n \to \Delta^+ \Sigma^+ K^0$    &$\frac34\si_{\Delta\Delta}^{(8)}$   & $\Delta^+ n \to \Delta^+ \Sigma^0 K^0$       &$\si_{\Delta\Delta}^{(6)}$          & $\Delta^0 p \to \Delta^0 \Sigma^+ K^0$       &$\si_{\Delta\Delta}^{(9)}$          & $\Delta^0 n \to \Delta^- \Sigma^0 K^+$       &$\frac34\si_{\Delta\Delta}^{(10)}$  & $\Delta^- n \to \Delta^0 \Sigma^- K^0$       &$\frac34\si_{\Delta\Delta}^{(7)}$   \\
$\Delta^{++} n \to \Delta^{++} \Sigma^- K^+$ &$\si_{\Delta\Delta}^{(5)}$          & $\Delta^+ n \to \Delta^+ \Sigma^- K^+$       &$\si_{\Delta\Delta}^{(9)}$          & $\Delta^0 p \to \Delta^0 \Sigma^0 K^+$       &$\si_{\Delta\Delta}^{(6)}$          & $\Delta^- p \to \Delta^0 \Sigma^0 K^0$       &$\frac34\si_{\Delta\Delta}^{(10)}$  &                                              &                                    \\
$\Delta^{++} n \to \Delta^{0} \Sigma^+ K^+$  &$0\,\,\,$                           & $\Delta^+ n \to \Delta^0 \Sigma^+ K^0$       &$\si_{\Delta\Delta}^{(8)}$          & $\Delta^0 p \to \Delta^{++} \Sigma^- K^0$    &$0\,\,\,$                           & $\Delta^- p \to \Delta^0 \Sigma^- K^+$       &$\frac34\si_{\Delta\Delta}^{(8)}$   &                                              &                                    \\
\hline\hline
\end{tabular}
\end{table*}


\subsection{Reactions $\Delta + \Delta$}

Reactions $\Delta \Delta \to \Delta Y K$ have small cross sections
and thus contribute marginally to $\Kp$ production. Nevertheless, they
are included in the calculation with parameterizations from \cite{tstl}.
\be
\bar\si^{(1)}_{\Delta\Delta}&=& \si(\Dp \Dpp \to \Dpp \La \Kp)
\nonumber\\
&=&
1.154
\left( \frac{s}{s_{\Delta\La K}} -1 \right )^{2.149}
\left( \frac{s_{\Delta\La K}}{s} \right )^{7.969},
\nonumber\\
\bar\si^{(2)}_{\Delta\Delta}&=&\si(\Dn \Dpp \to \Dp \La \Kp)
\nonumber\\
&=&
0.881
\left ( \frac{s}{s_{\Delta\La K}} -1 \right )^{2.150}
\left ( \frac{s_{\Delta\La K}}{s} \right )^{7.977},
\nonumber\\
\bar\si^{(3)}_{\Delta\Delta}&=&\si(\Dn \Dp \to \Dn \La \Kp)
\nonumber\\
&=&
0.291
\left ( \frac{s}{s_{\Delta\La K}} -1 \right )^{2.148}
\left ( \frac{s_{\Delta\La K}}{s} \right )^{7.934},
\label{dd2dlk}
\ee
\be
\bar\si^{(4)}_{\Delta\Delta}&=&\si(\Dpp \Dn \to \Dpp \Sm \Kp)
\nonumber\\
&=&
3.532
\left ( \frac{s}{s_{\Delta\Sigma K}} -1 \right )^{2.953}
\left ( \frac{s_{\Delta\Sigma K}}{s} \right )^{12.06},
\nonumber\\
\bar\si^{(5)}_{\Delta\Delta}&=&\si(\Dpp \Dn \to \Dp \Sn \Kp)
\nonumber\\
&=&
2.931
\left ( \frac{s}{s_{\Delta\Sigma K}} -1 \right )^{2.952}
\left ( \frac{s_{\Delta\Sigma K}}{s} \right )^{12.03},
\nonumber\ee\be
\bar\si^{(6)}_{\Delta\Delta}&=&\si(\Dm \Dp \to \Dn \Sm \Kp)
\nonumber\\
&=&
5.861
\left ( \frac{s}{s_{\Delta\Sigma K}} -1 \right )^{2.952}
\left ( \frac{s_{\Delta\Sigma K}}{s} \right )^{12.04},
\nonumber\\
\bar\si^{(7)}_{\Delta\Delta}&=&\si(\Dm \Dn \to \Dm \Sm \Kp)
\nonumber\\
&=&
7.047
\left ( \frac{s}{s_{\Delta\Sigma K}} -1 \right )^{2.952}
\left ( \frac{s_{\Delta\Sigma K}}{s} \right )^{12.05},
\label{dd2dsk}
\ee
The other isospin channels can be expressed in term of these seven cross
sections as it is given in Table~\ref{tab:dd2dyk}.
\begin{table*}\fontsize{7}{2}
\caption{Cross sections of reactions $\Delta \Delta \to \Delta Y K$ ($Y=\Lambda,\Sigma$)
in terms of cross sections (\ref{dd2dlk},\ref{dd2dsk})}
\label{tab:dd2dyk}
\begin{tabular}{lrlrlrlrlr}
\hline\hline
\multicolumn{10}{c}{$\Delta \Delta \to\Delta \Lambda K$}\\
\hline\hline
reaction & $\si$ & reaction & $\si$ & reaction & $\si$ &  reaction & $\si$ & reaction & $\si$ \\
\hline\hline
$\Delta^{++} \Delta^+ \to \Delta^{++} \Lambda K^+$ & $\bar\si^{(1)}_{\Delta\Delta}$ &
$\Delta^{++} \Delta^0 \to \Delta^{++} \Lambda K^0$ & $0\,\,\,$ &
$\Delta^{++} \Delta^0 \to \Delta^+ \Lambda K^+$ & $\bar\si^{(2)}_{\Delta\Delta}$ &
$\Delta^+\,\,\,\, \Delta^+ \to \Delta^{++} \Lambda K^0$ & $\frac13\bar\si^{(1)}_{\Delta\Delta}$ &
$\Delta^+ \Delta^+ \to \Delta^+ \Lambda K^+$ & $\frac19\bar\si^{(1)}_{\Delta\Delta}$ \\
$\Delta^+\,\,\, \Delta^0 \to \Delta^+\,\,\, \Lambda K^0$ & $\bar\si^{(3)}_{\Delta\Delta}$ &
$\Delta^+\,\,\, \Delta^0 \to \Delta^0\,\,\,\, \Lambda K^+$ & $\bar\si^{(3)}_{\Delta\Delta}$ &
$\Delta^{++} \Delta^- \to \Delta^+ \Lambda K^0$ & $0\,\,\,$ &
$\Delta^{++} \Delta^- \to \Delta^0\,\,\,\,\, \Lambda K^+$ & $0\,\,\,$ &
$\Delta^0\, \Delta^0 \to \Delta^0 \Lambda K^0$ & $\frac19\bar\si^{(1)}_{\Delta\Delta}$ \\
$\Delta^0\,\,\,\, \Delta^0 \to \Delta^-\,\,\, \Lambda K^+$ & $\frac13\bar\si^{(1)}_{\Delta\Delta}$ &
$\Delta^-\,\,\, \Delta^+ \to \Delta^0\,\,\,\, \Lambda K^0$ & $\bar\si^{(2)}_{\Delta\Delta}$ &
$\Delta^-\,\,\, \Delta^+ \to \Delta^- \Lambda K^+$ & $0\,\,\,$ &
$\Delta^- \,\,\,\Delta^0 \to \Delta^-\,\,\, \Lambda K^0$ & $\si^{(1)}_{\Delta\Delta}$\\
\hline\hline
\end{tabular}
\fontsize{9}{2}
\begin{tabular}{lrlrlrlr}
\multicolumn{8}{c}{$\Delta \Delta \to\Delta \Sigma K$}\\
\hline\hline
reaction & $\si$ & reaction & $\si$ & reaction & $\si$ &  reaction & $\si$  \\
\hline\hline
$\Delta^{++} \Delta^+ \to \Delta^{++} \Sigma^0 K^+$ &$\frac12\bar\si_{\Delta\Delta}^{(7)}$   & $\Delta^+ \Delta^+ \to \Delta^+ \Sigma^+ K^0$    &$\frac19\bar\si_{\Delta\Delta}^{(7)}$& $\Delta^- \Delta^{++}\to \Delta^+ \Sigma^- K^+$        &$\bar\si_{\Delta\Delta}^{(4)}$          & $\Delta^- \Delta^+ \to \Delta^0 \Sigma^0 K^0$        &$\bar\si_{\Delta\Delta}^{(5)}$       \\
$\Delta^{++} \Delta^+ \to \Delta^{++} \Sigma^+ K^0$ &$\bar\si_{\Delta\Delta}^{(7)}$          & $\Delta^+ \Delta^+ \to \Delta^{++} \Sigma^- K^+$ &$\frac13\bar\si_{\Delta\Delta}^{(7)}$& $\Delta^- \Delta^{++}\to \Delta^0 \Sigma^+ K^0$        &$\bar\si_{\Delta\Delta}^{(4)}$          & $\Delta^- \Delta^+ \to \Delta^0 \Sigma^- K^+$        &$\bar\si_{\Delta\Delta}^{(6)}$       \\
$\Delta^{++} \Delta^+ \to \Delta^+ \Sigma^+ K^+$    & $0\,\,\,$                              & $\Delta^+ \Delta^+ \to \Delta^{0} \Sigma^+ K^+$  &  $0\,\,\,$                             & $\Delta^- \Delta^{++}\to \Delta^0 \Sigma^0 K^+$        &$\bar\si_{\Delta\Delta}^{(4)}$          & $\Delta^- \Delta^+ \to \Delta^+ \Sigma^- K^0$        &  $0\,\,\,$                              \\
$\Delta^{++} \Delta^0\,\to \Delta^{++} \Sigma^0 K^0$&$\bar\si_{\Delta\Delta}^{(4)}$          & $\Delta^+ \Delta^0\,\to \Delta^+ \Sigma^0 K^0$   &$\frac16\bar\si_{\Delta\Delta}^{(6)}$& $\Delta^- \Delta^{++}\to \Delta^{++} \Sigma^- K^0$     & $0\,\,\,$                                 & $\Delta^- \Delta^+ \to \Delta^- \Sigma^+ K^0$        &$\bar\si_{\Delta\Delta}^{(4)}$       \\
$\Delta^{++} \Delta^0\,\to \Delta^+ \Sigma^0 K^+$   &$\bar\si_{\Delta\Delta}^{(5)}$          & $\Delta^+ \Delta^0\,\to \Delta^+ \Sigma^- K^+$   &$\frac23\bar\si_{\Delta\Delta}^{(5)}$& $\Delta^- \Delta^{++}\to \Delta^{-} \Sigma^+ K^+$      &  $0\,\,\,$                                & $\Delta^- \Delta^+ \to \Delta^- \Sigma^0 K^+$        &$\bar\si_{\Delta\Delta}^{(4)}$       \\
$\Delta^{++} \Delta^0\,\to \Delta^+ \Sigma^+ K^0$   &$\bar\si_{\Delta\Delta}^{(6)}$          & $\Delta^+ \Delta^0\,\to \Delta^0 \Sigma^+ K^0$   &$\frac23\bar\si_{\Delta\Delta}^{(5)}$& $\Delta^0\,\Delta^0\,\,\,\,\,\to \Delta^0 \Sigma^0 K^0$&$\frac1{18}\bar\si_{\Delta\Delta}^{(7)}$& $\Delta^- \Delta^0 \to \Delta^- \Sigma^0 K^0$        &$\frac12\bar\si_{\Delta\Delta}^{(7)}$\\
$\Delta^{++} \Delta^0\,\to \Delta^{++} \Sigma^- K^+$&$\bar\si_{\Delta\Delta}^{(4)}$          & $\Delta^+ \Delta^0\to \Delta^0 \Sigma^0 K^+$     &$\frac16\bar\si_{\Delta\Delta}^{(6)}$& $\Delta^0\,\Delta^0\,\,\,\,\,\to \Delta^0 \Sigma^- K^+$&$\frac19\bar\si_{\Delta\Delta}^{(7)}$   & $\Delta^- \Delta^0 \to \Delta^- \Sigma^- K^+$        &$\bar\si_{\Delta\Delta}^{(7)}$       \\
$\Delta^{++} \Delta^0\,\to \Delta^{0} \Sigma^+ K^+$ &  $0\,\,\,$                              & $\Delta^+ \Delta^0\to \Delta^{++} \Sigma^- K^0$  &  $0\,\,\,$                             & $\Delta^0\,\Delta^0\,\,\,\,\,\to \Delta^+ \Sigma^- K^0$& $0\,\,\,$                                & $\Delta^- \Delta^0 \to \Delta^0 \Sigma^- K^0$        &  $0\,\,\,$                              \\
$\Delta^+\,\,\, \Delta^+ \to \Delta^{++} \Sigma^0 K^0$ &$\frac16\bar\si_{\Delta\Delta}^{(7)}$& $\Delta^+ \Delta^0\to \Delta^{-} \Sigma^+ K^+$   &  $0\,\,\,$                               & $\Delta^0\,\Delta^0\,\,\,\,\,\to \Delta^- \Sigma^+ K^0$&$\frac13\bar\si_{\Delta\Delta}^{(7)}$   & $\Delta^{++} \Delta^{++} \to \Delta^{++}\Sigma^+ K^+$&  $0\,\,\,$                            \\
$\Delta^+\,\,\ \Delta^+ \to \Delta^+ \Sigma^0 K^+$    &$\frac1{18}\bar\si_{\Delta\Delta}^{(7)}$ & $\Delta^- \Delta^{++}\to \Delta^+ \Sigma^0 K^0$  &$\bar\si_{\Delta\Delta}^{(4)}$       & $\Delta^0\,\Delta^0\,\,\,\,\,\to \Delta^- \Sigma^0 K^+$&$\frac16\bar\si_{\Delta\Delta}^{(7)}$   &   $\Delta^{-}\,\,\, \Delta^{-} \to \Delta^{-}\Sigma^- K^0$  &   $0\,\,\,$                                   \\
\hline\hline
\end{tabular}
\end{table*}

Another group of reactions are $\Delta \Delta \to NN K \bar K$.
We use the same parameterization (\ref{nn2nnkk}) for all isospin
channels
\be
\begin{tabular}{ll}
$\Dpp \Dp \to pp K^+\bar K^0$       &  $\Dm  \Dpp \to pn K^0\bar K^0$\\
$\Dpp \Dn \to pp K^0\bar K^0$       &  $\Dm  \Dpp \to nn K^+\bar K^0$\\
$\Dpp \Dn \to pp K^+ K^-$           &  $\Dm  \Dpp \to pp K^0     K^-$\\
$\Dpp \Dn \to pn K^+\bar K^0$       &  $\Dm  \Dpp \to pn K^+     K^-$\\
$\Dp\,\,\, \Dp \to pp K^0\bar K^0$  &  $\Dn\,\Dn\,\,\,\, \to nn K^0\bar K^0$\\
$\Dp\,\,\, \Dp \to pp K^+ K^-$      &  $\Dn\,\Dn\,\,\,\, \to pn K^0     K^-$\\
$\Dp\,\,\, \Dp \to pn K^+\bar K^0$  &  $\Dn\,\Dn\,\,\,\, \to nn K^+     K^-$\\
$\Dp\,\,\, \Dn \to pn K^0\bar K^0$  &  $\Dm  \Dp\,\,\, \to nn K^0\bar K^0$\\
$\Dp\,\,\, \Dn \to nn K^+\bar K^0$  &  $\Dm  \Dp\,\,\, \to pn K^0     K^-$\\
$\Dp\,\,\, \Dn \to pp K^0     K^-$  &  $\Dm  \Dp\,\,\, \to nn K^+     K^-$\\
$\Dp\,\,\, \Dn \to pn K^+     K^-$  &  $\Dm  \Dn\,\,\,\, \to nn K^0     K^-$\\
\end{tabular}
\label{dd2nnkk}
\ee


\subsection{Reactions of $\pi + \pi$}

\begin{figure}
\includegraphics[width=4cm]{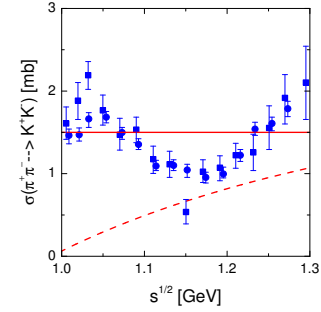}
\caption{The cross section of the $\pi^+\pi^- \to K^+K^-$
reaction: solid line represents parameterization (\ref{kppi38}), dashed
line is calculated as in Refs.~\cite{mosel,cas}. The experimental points
are from \cite{pipikk,petersen} }
\label{fig:pp2kk}
\end{figure}

The source of information about $\pi\pi\to K \bar K$ reactions is
the inelastic $\pi\pi$ scattering which was intensively studied
experimentally~\cite{exppipi}.
A characteristic feature of the $\pi\pi$ scattering is  very rapid
rise of the phase shift and the inelasticity parameter at energies
about 1~GeV.
Such a strong energy dependence is caused by the $f_0$
resonance with a pole around~$980-i14$~MeV (cf. Ref.~\cite{oset}).
The inelasticity parameter translated into the cross section of
the $\pi^+\pi^- \to K^+K^-$ reaction~\cite{pipikk,petersen} is depicted in
Fig.~\ref{fig:pp2kk}.
We checked, that the velocity-averaged cross section is reproduced almost
exactly if we used a constant for a $\pi\pi\to K\bar K$
\begin{equation}
\label{kppi38}
\si(\pip   \pim \to \Kp   \Km) = \si(\pim \pip \to \Kn \Kb) = 1.6\,
\mbox{mb}\,,
\end{equation}
shown in Fig.~\ref{fig:pp2kk} by the solid line.
>From isospin symmetry, the other $\pi\pi$ channels have cross sections
\be
\label{kppi39}
&&\si(\pin   \pin \to \Kp   \Km) = \si(\pin \pin \to \Kn \Kb)
\\ \nonumber
&&\qquad\qquad =
\frac{2}{5}\, \si(\pip   \pim \to \Kp   \Km) = 0.64\, \mbox{mb}\, ,
\\ \label{kppi40}
&&\si(\pip \pin \to \Kp \Kb) = \si(\pim \pin\to \Kn \Km)
\\ \nonumber
&&\qquad\qquad =
\frac{6}{5}\, \si(\pip   \pim \to \Kp   \Km) = 1.92\, \mbox{mb} \, .
\ee
Note that this implementation is different from the $\pi\pi$ cross sections
used in~\cite{mosel,cas}. The latter is shown in
Fig.~\ref{fig:pp2kk} by the dashed line.


\subsection{Reactions of $\rho+\rho$}
\label{rhorho}

The cross section for these reactions is not known, because there
are no data to which calculations could be compared. It has been
calculated in \cite{broko} as a kaon exchange process. We adopt this
result here and parameterize the cross section as
\be
&&\si(\rop\rom \to \Kp \Km) = \si(\rop\rom \to \Kn\Kb)
\label{rr1}\\ \nonumber
&&\qquad\qquad =
\frac{1}{32\pi}\, \frac{p^K_{cm}}{p^\rho_{cm}}\,
\exp\left ( - \frac{\sqrt{s} - 2 m_\rho}{1\, \mbox{GeV}}\right )\,
29.11\, \mbox{mb}\, ,
\ee
where $p^K_{cm}(s)$, and $p^\rho_{cm}(s)$ are the momenta of kaons and
rhos in the centre of mass frame, respectively. The exponential form is
chosen such that it reproduces very well the matrix element calculated in
\cite{broko}.

Other $\rho\rho$ channels are determined from isospin symmetry, in analogy
to $\pi\pi$ scattering
\be
&&\si(\ron   \ron \to \Kp   \Km) = \si(\ron \ron \to \Kn \Kb)
\nonumber\\ \label{rr2}
&&\qquad\qquad
=\frac{2}{5}\, \si(\rop   \rom \to \Kp   \Km) \, ,
\\
&&\si(\rop \ron \to \Kp \Kb) = \si(\rom \ron\to \Kn \Km)
\nonumber\\\label{rrr3}
&&\qquad\qquad=
\frac{6}{5}\, \si(\rop   \rom \to \Kp   \Km) \, .
\ee


\subsection{Reactions of $\pi +\rho$}
\label{pirho}

Production of a $K\bar K$ pair in this reaction is suppressed
because the $s$-wave component of cross section is forbidden.
The reaction is dominated by $\phi$ resonance \cite{HSD}
\begin{equation}
\si(\pi\rho \to \phi \to K\bar K) = \frac{4\, \pi \, s}{p_i^2}\,
\frac{\Gamma_{\phi\to \pi\rho}\, \left (\frac{k}{k_0}\right )^3\,
\Gamma_{\phi\to K\bar K} \, \left (\frac{q}{q_0}\right )^3 }%
{(s - M^2_\phi)^2 + s \, \Gamma_{\rm tot}^2(\sqrt{s})}\, ,
\end{equation}
where $p_i$ is the momentum of incoming particles in the centre-of-mass
system and
\be
&&q_0 =p_{\rm cm}(M_\phi^2,m_K,m_K)\,,\quad  q = p_{\rm
cm}(s,m_K,m_K)\,,
\nonumber\\
&&k_0=p_{\rm cm}(M_\phi^2,m_\pi,m_\rho)\,,\quad\,\,\,
k=p_{\rm cm}(s,m_\pi,m_\rho)\,,
\nonumber
\ee
and the total width is determined from the partial widths
\begin{equation}
\Gamma_{\rm tot}(\sqrt{s})
= \Gamma_{\phi\to \pi\rho}\frac{k^3}{k_0^3}
+ \left(\Gamma_{\phi\to K^+ K^-} + \Gamma_{\phi\to K^0\bar
K^0}\right)\frac{q^3}{q_0^3}\, .
\end{equation}
The partial widths for decays of $\phi$ into $K^+K^-$ and $K^0_L K^0_S$
are different, but since we keep isospin symmetry in all other reactions
throughout this work, we shall use
\begin{equation}
\Gamma_{\phi\to K\bar K} = \frac{1}{2} \,
\left ( \Gamma_{\phi\to K^+ K^-} + \Gamma_{\phi\to K^0\bar K^0}\right )
\end{equation}
for all $I=0$ channels.

\begin{figure}
\includegraphics[width=5cm]{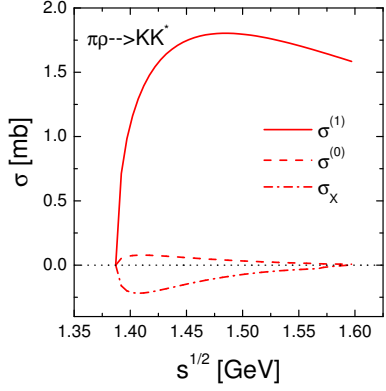}
\caption{Cross sections for reactions $\pi\rho\to K\bar K^*$ with
given isospin $\sigma^{(0)}$, $\sigma^{(1)}$ and the mixed terms
$\sigma_{\rm X}$}
\label{fig:pirho2kvk}
\end{figure}

For the reaction of $\pi\rho$ into $K\bar K^*$,
the leading order diagram in chiral counting is the contact
Weinberg-Tomozawa term \cite{LK}. We calculated the cross section
\be
&\sigma^{(I)}&=\frac{9}{64 \pi s}\,
\frac{p_{\rm cm}(s,m_K,m_{K*})}{p_{\rm cm}(s,m_\pi,m_\rho)}\,
|M^{(I,0)}|^2\,\quad I=0,1
\nonumber\\
&\sigma_{\rm X}&=\frac{9}{64 \pi s}\,
\frac{p_{\rm cm}(s,m_K,m_{K*})}{p_{\rm cm}(s,m_\pi,m_\rho)}\,
\Re\Big(M^{(0,0)}\,\big[M^{(1,0)}\big]^*\Big)\,,
\nonumber
\ee
using unitarized amplitudes $M^{(I,S)}$, from~\cite{LK}, where
$I$ and $S$ are isospin and strangeness quantum numbers
indicating a particular reaction channel. The cross sections are
shown in Fig.~\ref{fig:pirho2kvk}.
We realized that, to a very good approximation,
only the $I=1$ component of the cross section effectively contributes and
can be replaced by a constant. From combining the involved isospin states
we obtain relations
\be
&&\si(\rop\pim \to \Kp K^{*-} ) = \si(\rom \pip \to \Kn \bar K^{*0} )=
0.2\, \mbox{mb},
\nonumber\\
&&\si(\rom \pip \to \Kp K^{*-}) = \si(\rop \pim \to \Kn \bar K^{*0}) =
0.2\, \mbox{mb},
\nonumber\\
&&\si(\ron \pin \to \Kp K^{*-}) = \si(\ron \pin \to \Kn \bar K^{*0}) = 0
\nonumber\\
&&\si(\rop\pin \to \Kp \bar K^{*0}) = \si(\ron\pin \to \Kn \bar K^{*0})
=0.4\, \mbox{mb},
\nonumber\\
&&\si(\ron \pip \to \Kp \bar K^{*0}) = \si(\ron \pim \to \Kn K^{*-}) =
0.4\, \mbox{mb}\, .
\nonumber\\
\ee


\subsection{Decays of $K^*$}

Kaons can be also produced in decays of $K^*$. The decay rate is simply given
by $\Gamma_{K^*} \rho_{K^*}$ where $\Gamma_{K^*} = 50.8\, \mbox{MeV}$ is
the width and $\rho_{K^*}$ the density. Individual channels must be
multiplied by appropriate Clebsch-Gordan coefficients. These are
\be\begin{array}{lll}
K^{*+} \to \Kp \pin \, ,&  K^{*0} \to \Kn \pin \quad & \to \frac{1}{3}\\
K^{*+} \to \Kn \pip \, ,&  K^{*0} \to \Kp \pim \quad & \to \frac{2}{3}
\end{array}
\ee
The inverse reaction destroys kaons. The cross section is derived as
\cite{DanBer}
\begin{equation}
\sigma(K\pi \to K^*) = \frac{\pi\,\Gamma_{K^*}\, A_{K^*}}{2\, p^2_{\rm
cm}(m_{K^*}^2,m_\pi,m_K)}\,,
\end{equation}
where
\begin{equation}
A_{K^*}(\sqrt{s}) =
\frac{4\, m_{K^*}^2 \Gamma_{K^*}}{(s - m_{K^*}^2)^2 + m_{K^*}^2 \Gamma_{K^*}^2}
\,.
\end{equation}


\subsection{Reactions of $Y+\pi$}

These reactions produce kaon and a cascade. The cross sections are unknown,
so we just use a parameterization
\[
\sigma = C\, \frac{p_{\rm cm}(s,m_{\Xi},m_K)}{p_{\rm cm}(s,m_Y,m_\pi)}\,
10\, \mbox{mb}\, ,
\]
where the constant $C$
results from the  Clebsch-Gordon coefficients of isospin
adding. In the following
we just list the parameterizations and the corresponding constants $C$:
\be\begin{array}{lllc}
\La \pip \to \Kp \Xn \, ,&  \La \pim \to \Kn \Xm \quad &\to &1 \\
\La \pin \to \Kp \Xm \, ,&  \La \pin \to \Kn \Xn \quad &\to &\frac{1}{2}
\end{array}
\ee
\be
\begin{array}{lllc}
\Sp   \pin\,\to \Kp \Xn \,, & \Sm  \pin\,\to \Kn \Xm \quad &\to &\frac{1}{2}\\
\Sp   \pim  \to \Kp \Xm \,, & \Sm  \pip  \to \Kn \Xn \quad &\to &\frac{5}{12} \\
\Sn\, \pip  \to \Kp \Xn \,, & \Sn\,\pim  \to \Kn \Xm \quad &\to &\frac{1}{2}\\
\Sn\, \pin\,\to \Kp \Xm \,, & \Sn\,\pin\,\to \Kn \Xn \quad &\to &\frac{1}{3}\\
\Sm   \pip  \to \Kp \Xm \,, & \Sp  \pim  \to \Kn \Xn \quad &\to &\frac{5}{12}
\end{array}
\ee


\subsection{Total cross sections of $KN$ reactions}

\begin{figure}
\includegraphics[width=5cm]{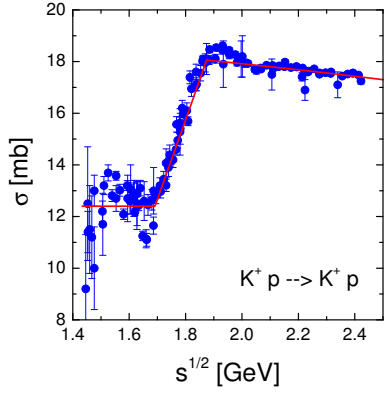}
\caption{The total $K^+ p$ cross section as given by (\ref{kpxs}) in comparison
with the experimental data from \cite{expdata} }
\label{fig:Kptot}
\end{figure}

These parameterizations are taken from \cite{casibi}.
\begin{equation}
\si_{K^+p} =
\begin{cases}
12.4 & : \quad p_K \le 0.78  \\
1.09 + 14.5 p_K & : \quad 0.78  < p_K \le 1.17 \\
18.64 - 0.5 p_K & : \quad 1.17  < p_K \le 2.92
\end{cases}
\label{kpxs}
\end{equation}
\begin{equation}
\si_{K^+n} =
\begin{cases}
15.5 & : \quad p_K \le 0.78  \\
4.19 + 14.5 p_K & : \quad 0.78  < p_K \le 1.1 \\
22.78 - 2.4 p_K & : \quad 1.1  < p_K \le 2.3
\end{cases}
\label{knxs}
\end{equation}
where $p_K$ is kaon momentum in the lab frame.
In Fig.~\ref{fig:Kptot} we illustrate the quality of the
paremeterization~(\ref{kpxs})\,.



\section{Feed-down from resonance decays}
\label{feeddown}

We list in Table~\ref{t:pifd} the average numbers of pions which
are obtained by decays of individual resonances.
\begin{table}
\caption{\label{t:pifd}
List of resonances decaying into pions together with the average numbers
of pion species produced by their decays.
}
\begin{tabular}{c|cc||c|cc}
\hline\hline
&
$\langle \pi^+\rangle_R$ & $\langle \pi^- \rangle_R$ &
&
$\langle \pi^+\rangle_R$ & $\langle \pi^- \rangle_R$ \\
\hline
$\eta$ & 0.28 & 0.28 & $N^0(1440)$ & 0.175 & 0.725 \\
$\rho^+$ & 1 & 0 & $\bar N^+(1440)$ & 0.175 & 0.725 \\
$\rho^0$ & 1 & 1 & $\bar N^0(1440)$ & 0.725 & 0.175 \\
$\rho^-$ & 0 & 1 & $K^{*+}$ & 2/3 & 0 \\
$\omega$ & 0.91 & 0.91 & $K^{*0}$ & 0 & 2/3 \\
$\Delta^{++}$ & 1 & 0 & $K^-$ & 0 & 2/3 \\
$\Delta^{+}$ & 1/3 & 0 & $\bar K^{*0}$ & 2/3 & 0 \\
$\Delta^0$ & 0 & 1/3 & $\La(1405)$ & 1/3 & 1/3 \\
$\Delta^-$ & 0 & 1 & $\La(1520)$ & 0.217 & 0.217 \\
$\bar \Delta^{++}$ & 0 & 1 & $\Sp(1385)$ & 0.94 & 0 \\
$\bar \Delta^+$ & 0 & 1/3 & $\Sn(1385)$ & 0.06 & 0.06 \\
$\bar \Delta^0$ & 1/3 & 0 & $\Sm(1385)$ & 0 & 0.94 \\
$\bar \Delta^-$ & 1 & 0 & $\Xi^-$ & 0 & 1 \\
$N^+(1440)$ & 0.725 & 0.175 & $\Omega$ & 0 & 0.322\\
\hline\hline
\end{tabular}
\end{table}
Resonances which feed-down into kaon, $\La$ and $\Sigma$ production are
listed in Table~\ref{t:kfd}.
\begin{table}
\caption{\label{t:kfd}
List of resonances decaying into kaons, lambdas or sigmas together
with average numbers of decay products.
}
\begin{tabular}{c|ccccc}
\hline\hline
&
$\langle K^+\rangle_R$ & $\langle K^- \rangle_R$ &
$\langle \La \rangle_R$ &
$\langle \Sp \rangle_R$ & $\langle \Sm \rangle_R$ \\
\hline
$K^{*+}$ & 1/3 & & & & \\
$K^{*0}$ & 2/3 & & & & \\
$K^{*-}$ &   & 1/3 & & & \\
$\bar K^{*0}$ &  & 2/3 & & & \\
$\Sn$ & & & 1 & & \\
$\La (1405)$ &   &  & 1/3 & 1/3 & 1/3 \\
$\La(1520)$  &   &  & 0.11 & 0.143 & 0.143 \\
$\Sp(1385)$ &  &  & 0.94 & 0.06 & \\
$\Sn(1385)$ &  &  & 0.88 & 0.06 & 0.06 \\
$\Sm(1385)$ &  &  & 0.94 &  & 0.06 \\
$\Xi^0$ &   &  & 1 & & \\
$\Xi^-$ &   &  & 1 & & \\
$\Omega$ &  & 0.678 & 1 &   & \\
\hline\hline
\end{tabular}
\end{table}


\section{Strangeness content of the initial state}
\label{istate}

\subsection{Strangeness multiplicity in $NN$ collisions}

Strange particles are also produced in the initial nucleon-nucleon
interactions. We shall assume that their initial multiplicities
are given by production of kaons in $NN$ reactions in vacuum.
Since in our approach kaon production is calculated explicitly and species
with $S<0$ are populated statistically,
we will only need input on primordial $K^+$ production. Data on kaon
multiplicities in $pp$ collisions were summarized in \cite{grs}.
In order to obtain the {\em average} multiplicity in {\em nucleon-nucleon}
collisions, we also need multiplicities in $nn$ and $pn$ collisions.
It was observed that in $NN$ collisions multiplicity of charged
kaons does not depend on isospin of the projectile/target and we
can represent $\langle K^+ \rangle_{NN}$ with $\langle K^+ \rangle_{pp}$
\cite{ondrej,na49a10}
\begin{equation}
\langle K^+ \rangle_{NN} =
\langle K^+ \rangle_{pp} = \langle K^+\rangle_{nn} = \langle K^+\rangle_{pn}
\, .
\end{equation}

\subsection{Density from multiplicity}

What we actually need for the simulation is the {\em density}, not the
{\em multiplicity}. We will estimate the former by equating
\begin{equation}
\frac{\rho_{K^+}}{\rho_{h^-}} =
\frac{\langle K^+ \rangle_{NN}}{\langle h^- \rangle_{NN}}
\label{ra}
\end{equation}
Data on multiplicity of negatively charged hadrons from $NN$
interactions were collected by Ga\'zdzicki and R\"ohrich
in \cite{grns}. They collected $h^-$ yields from $pp$ and $pn$ collisions
at different energies and
estimated them for $nn$ collisions. The {\em primordial}
yield (per nucleon-nucleon pair)
for {\em nucleus-nucleus} collisions is then obtained as \cite{grns}
\begin{multline}
\label{getNN}
\langle h^- \rangle_{NN} = \left ( \frac{Z}{A} \right )^2
\langle h^-\rangle_{pp} \\
+ 2 \frac{Z}{A} \left ( 1 - \frac{Z}{A} \right )
\langle h^- \rangle_{pn} + \left ( 1 - \frac{Z}{A} \right )^2
\langle h^- \rangle_{nn}\, .
\end{multline}
This relation is formulated for nuclei with atomic number $Z$
which contain $A$ nucleons.

The values for $h^-$ yields are not measured at all energies we need.
Thus we use data collected in \cite{grs,grns} and linearly interpolate
as a function of $\sqrt{s}$. The compilation is displayed in
Table~\ref{t:hmmult}.
\begin{table}
\caption{\label{t:hmmult}
Multiplicities of negative hadrons in $pp$, $pn$, and $nn$ collisions
interpolated from data compiled in \cite{grs,grns}. The averaged
values for nucleon-nucleon collisions are calculated according to
eq.~\eqref{getNN}.}
\begin{tabular}{c|ccc|c} \hline\hline
 $ $ & $\langle h^-\rangle_{pp}$ & $\langle h^-\rangle_{pn}$ &
$\langle h^-\rangle_{nn}$ & $\langle h^-\rangle_{NN}$ \\
\hline
Au+Au @ 11.6 AGeV & 0.697 & 0.900 & 1.390 & 1.044 \\
Pb+Pb @ 30 AGeV   & 1.265 & 1.728 & 1.965 & 1.741 \\
Pb+Pb @ 40 AGeV   & 1.482 & 1.899 & 2.182 & 1.937 \\
Pb+Pb @ 80 AGeV   & 2.025 & 2.434 & 2.725 & 2.477 \\
Pb+Pb @ 158 AGeV  & 2.611 & 3.058 & 3.311 & 3.081 \\
\hline\hline
\end{tabular}
\end{table}
The resulting ratios $\langle K^+\rangle_{NN}/\langle h^-\rangle_{NN}$
are listed in Table~\ref{t:inis}.

\subsection{Composition of multiplicity data and feed-down}

The major part of the $h^-$ multiplicity are pions, a smaller fraction
comes from negative kaons. Very small contribution is due to
antiprotons. There is feed-down from resonance decays mainly into
pion multiplicity. Feed-down to antiprotons is neglected because
antibaryons are populated scarcely in general at these energies. Feed-down to
$K^-$ and $\bar K^0$ from $K^*$ is taken into account.

This is implemented by using the {\em effective} densities
of pions, kaons and antiprotons when determining the $h^-$ density
at the left-hand-side of eq.~\eqref{ra}
\begin{equation}
\rho_{h^-} = \rho_{\tilde \pi^-} + \rho_{\tilde K^-} + \rho_{\bar p}\, .
\end{equation}
The effective densities, denoted by tildes over subscripts, include
the actual densities and pions (kaons) which can be produced by decays of all
resonances in the system. These additions are calculated according
to the same prescription as the feed-down in final state, see
Section~\ref{fistate} and Appendix~\ref{feeddown}. By
using the same strategy we keep the simulation internally consistent.

\subsection{Algorithm to obtain the initial state}

Technically, the initial state is obtained iteratively:

\begin{enumerate}
\item
Set densities of all strange particles to zero. Set
temperature and chemical potential to some ``reasonable'' value.
\item \label{bdva}
Determine energy density and $I_3$ density contributions stored in kaons,
$\eden_K$ and $\rho_{3K}$, respectively.
This is done by first determining phase space occupancies $\gamma_j$ via
\begin{equation}
\gamma_j = \frac{2\, \pi^2\, \rho_j}{%
\exp \left ( \frac{I_{3,j}\, \mu_3}{T}
\right )\, m_K^2\, T \, K_2 \left ( \frac{m_K}{T} \right )}\, ,
\end{equation}
where $j=K^+,\, K^0$. Then
\begin{multline}
\label{edk}
\eden_K = \frac{1}{2\pi^2}\,
m_K^2\, T^2\,
\left \{ \frac{m_K}{T} K_1 \left ( \frac{m_K}{T} \right )
+ 3 K_2 \left (\frac{m_K}{T} \right ) \right \} \\
\times \left ( \gamma_{K^+}\, \exp(\mu_3/2T)
+ \gamma_{K^0}\, \exp (-\mu_3/2T) \right )  \, ,
\end{multline}
and
\begin{equation}
\rho_{3K} = \frac{1}{2} ( \rho_{K^+} - \rho_{K^0} ) \, .
\label{idk}
\end{equation}
Subtract them from the total $\eden$ and $\rho_3$, i.e.,
calculate
\begin{eqnarray}
\tilde \eden & = & \eden - \eden_K\\
\tilde \rho_3 & = & \rho_3 - \rho_{3K}
\end{eqnarray}
\item\label{sbar}
Calculate $\rho_S = \rho_{K^+} + \rho_{K^0}$.
\item
>From the new $\tilde\eden,\, \rho_B,\, \tilde \rho_3,\, \rho_S$
calculate temperature, chemical potentials and the
suppression factor $\gamma_S$.
\item From $T,\, \mu_B,\, \mu_3,\, \gamma_S$
calculate $\rho_{\tilde \pi^-} + \rho_{\tilde K^-} + \rho_{\bar p}$.
This also includes resonance feed-down to pions and kaons according to
the given temperature and chemical potentials.
\item Get $K^+$ density as
\begin{multline}
\rho_{K^+} =
\frac{\langle K^+\rangle_{NN}}{\langle h^- \rangle_{NN}}\,
( \rho_{\tilde \pi^-}(T,\mu_B,\mu_3,\gamma_S) \\
+ \rho_{\bar p}(T,\mu_B,\mu_3,\gamma_S)  +
\rho_{\tilde K^-}(T,\mu_B,\mu_3,\gamma_S) )
\end{multline}
\item Determine
\begin{equation}
\rho_{K^0} = \rho_{K^+} \, \exp (-\mu_3/T)
\end{equation}
\item Calculate again $\rho_{S} = \rho_{K^+} + \rho_{K^0}$.
\item If the obtained $\rho_{S}$ differs from that determined in
step \ref{sbar}, go to step \ref{bdva}.
If they agree, the routine has converged.
\end{enumerate}


\end{document}